\documentclass[a4paper,12pt]{article}
\pdfoutput=1
\usepackage[DIV=14,BCOR=0mm,headinclude=true,footinclude=false]{typearea}
\usepackage[utf8]{inputenc}
\usepackage[T1]{fontenc}

\usepackage{style}

\usepackage{subcaption}
\usepackage{tipa}
\usepackage{graphicx}
\usepackage[bottom]{footmisc}
\usepackage[titletoc]{appendix}
\usepackage{hyperref}
\usepackage{listings}
\usepackage{float}
\usepackage{amsmath}
\usepackage{mathrsfs}
\usepackage{amssymb}
\usepackage{mathtools}
\usepackage{amsfonts}
\usepackage{braket}
\usepackage{slashed}
\usepackage{dashbox}
\usepackage[colorinlistoftodos]{todonotes}
\usepackage{tikz-feynman}   % For Feynman diagrams
\usetikzlibrary{snakes}     % For snake drawing line
\usepackage[normalem]{ulem}

\usepackage{scalerel}
\usepackage{multirow}  % For merged rows in table
\usepackage{makecell}  % For adjusting height of a single row in table

\newcommand\reallywidehat[1]{\arraycolsep=0pt\relax%
	\begin{array}{c}
		\stretchto{
			\scaleto{
				\scalerel*[\widthof{\ensuremath{#1}}]{\kern-.5pt\bigwedge\kern-.5pt}
				{\rule[-\textheight/2]{1ex}{\textheight}} %WIDTH-LIMITED BIG WEDGE
			}{\textheight} % 
		}{0.5ex}\\           % THIS SQUEEZES THE WEDGE TO 0.5ex HEIGHT
		#1\\                 % THIS STACKS THE WEDGE ATOP THE ARGUMENT
		\rule{-1ex}{0ex}
	\end{array}
}

\newcommand{\vev}[1]{\ensuremath{\left\langle #1 \right\rangle}}
\def\be{\begin{equation}}
\def\ee{\end{equation}}
\def\ba{\begin{aligned}}
\def\ea{\end{aligned}}

\newcommand{\SL}{\text{SL}\left(2,\mathbb{R}\right)}

\numberwithin{equation}{section}
\numberwithin{table}{section}

\title{Scalar Love numbers and Love symmetries of 5-dimensional Myers-Perry black holes}
\author[a]{Panagiotis Charalambous}%\footnote{\texttt{pc2560@nyu.edu}}}
\author[b,c]{Mikhail M. Ivanov}%\footnote{\texttt{ivanov@ias.edu}}}

\affiliation[a]{Center for Cosmology and Particle Physics, Department of Physics,
New York University,\\
726 Broadway, New York, NY 10003, USA}
\affiliation[b]{Center for Theoretical Physics, Massachusetts Institute of Technology, \\ Cambridge, MA 02139, USA}
\affiliation[c]{School of Natural Sciences, Institute for Advanced Study, 1 Einstein Drive, \\ Princeton, NJ 08540, USA}

\emailAdd{pc2560@nyu.edu}
\emailAdd{ivanov@ias.edu}
\date{}

\abstract{
	The near-zone ``Love'' symmetry resolves the naturalness issue of black hole Love number vanishing with $\SL$ representation theory. Here, we generalize this proposal to $5$-dimensional asymptotically flat and doubly spinning (Myers-Perry) black holes. We consider the scalar response of Myers-Perry black holes and extract its static scalar Love numbers. In agreement with the naturalness arguments, these Love numbers are, in general, non-zero and exhibit logarithmic running unless certain resonant conditions are met; these conditions include new cases with no previously known analogs. We show that there exist two near-zone truncations of the equations of motion that exhibit enhanced $\SL$ Love symmetries that explain the vanishing of the static scalar Love numbers in the resonant cases. These Love symmetries can be interpreted as local $\SL\times\SL$ near-zone symmetries spontaneously broken down to global $\SL\times U\left(1\right)$ symmetries by the periodic identification of the azimuthal angles. We also discover an infinite-dimensional extension of the Love symmetry into $\SL\ltimes\hat{U}\left(1\right)_{\mathcal{V}}^2$ that contains both Love symmetries as particular subalgebras, along with a family of $\SL$ subalgebras that reduce to the exact near-horizon Myers-Perry black hole isometries in the extremal limit. Finally, we show that the Love symmetries acquire a geometric interpretation as isometries of subtracted (effective) black hole geometries that preserve the internal structure of the black hole and interpret these non-extremal $\SL$ structures as remnants of the enhanced isometry of the near-horizon extremal geometries.
}

\begin{document}

\maketitle

\section{Introduction}
\label{sec:Intro}

The engulfment of gravitational wave astronomy into the area of precision astronomy~\cite{LIGOScientific:2016aoc} has attracted growing attention towards the tidal response problem of a compact body. During the early stages of the inspiraling phase of a binary system, the tidal perturbations of the bodies are parameterized by the static Love numbers~\cite{Love1912,PoissonWill2014}. More importantly, Love numbers are able to probe the equation of state of the involved relativistic configurations~\cite{Flanagan:2007ix,Yagi:2013bca,LIGOScientific:2017vwq,Raithel:2018ncd,Chatziioannou:2020pqz}, while their measurement has been proposed as a testing arena for strong-field gravity~\cite{Cardoso:2017cfl}.

Love numbers' imprint on gravitational wave signals can be calculated by means of the worldline Effective Field Theory (EFT), 
a toolkit for the construction of gravitational waveform templates
during the inspiral phase~\cite{Goldberger:2004jt,Porto:2005ac,Porto:2016pyg,Levi:2015msa,Levi:2018nxp,Goldberger:2022ebt}. Within the worldline EFT, a compact body is approximated by its large-distance universal appearance as a point-particle evolving along a worldline. Finite size effects are systematically captured by non-minimal couplings of the worldline with various curvature operators. The Love numbers, in particular, appear as Wilson coefficients for operators quadratic in the curvature and their computation reduces to a matching condition. %\cite{Kol:2011vg,Charalambous:2021mea,Ivanov:2022hlo,Ivanov:2022qqt}.

The static Love numbers for general-relativistic black holes have recently been gaining a growing theoretical interest. In four spacetime dimensions, both static black holes~\cite{Fang:2005qq,Damour:2009vw,Binnington:2009bb,Gurlebeck:2015xpa}, as well as spinning black holes~\cite{Poisson:2014gka,Landry:2015zfa,Pani:2015hfa,LeTiec:2020spy,LeTiec:2020bos,Chia:2020yla,Charalambous:2021mea,Ivanov:2022hlo,Ivanov:2022qqt} have been shown to posses vanishing static Love numbers. This fact raises naturalness concerns from the EFT point of view, calling upon the existence of an enhanced symmetry structure~\cite{Porto:2016zng,tHooft:1979rat}.

Black holes have been dubbed ``the hydrogen atom of the 21st century''~\cite{tHooft:2016sdu,EHT2019}. This statement has been given some rigor for asymptotically flat black holes in General Relativity~\cite{Bertini:2011ga,Kim:2012mh,Charalambous:2021kcz,Charalambous:2022rre}. It is ought to persisting hidden conformal structures of asymptotically flat black holes, with the classic paradigms being the extremal~\cite{Bardeen:1999px,Guica:2008mu,Lu:2008jk} and the non-extremal~\cite{Castro:2010fd,Krishnan:2010pv} Kerr/CFT conjectures, while more efforts have recently been put forward to constructing holographic-like dictionaries between black hole geometries and conformal field theories~\cite{Bonelli:2021uvf,Consoli:2022eey}. It has also been suggested that conformal structures associated with black holes can leave distinct signatures on polarimteric observations, see e.g.~\cite{Johnson:2019ljv,Himwich:2020msm}.

We have recently proposed that one new form of such conformal structures of black holes can be used to explain the vanishing of static Love numbers for general-relativistic black holes in four spacetime dimensions~\cite{Charalambous:2021kcz,Charalambous:2022rre}, see also~\cite{Cvetic:2021vxa} of its presence in SUGRA black holes. This ``Love'' symmetry is an $\SL$ symmetry manifesting itself in the near-zone region, where perturbations have small frequencies compared to the inverse distance from the black hole. The Love symmetry outputs the vanishing of the static Love numbers as a selection rule following from the fact that the relevant perturbation solution belongs to a highest-weight representation. Geometrically, it can be realized as an approximate isometry of the black hole geometry, in the sense that it is an exact isometry of an effective black hole geometry which preserves the thermodynamic properties of the black hole but approximates (''subtracts'') its environment. Such geometries have been introduced in~\cite{Cvetic:2011hp,Cvetic:2011dn} and go by the name ``subtracted geometries''. Interestingly, the Love symmetry appears to be a cousin of another well-known $\SL$ symmetry associated with degenerate black holes: the enhanced isometry of the near-horizon geometry for extremal black holes~\cite{Bardeen:1999px,Kunduri:2007vf}. This hints at the interpretation of the Love symmetry as a remnant of this enhanced isometry for extremal black holes. It should be remarked here that there have been other attempts of explaining the vanishing of black holes Love numbers via symmetry arguments, notably, the ``ladder symmetries'' proposal~\cite{Hui:2021vcv,Hui:2022vbh} (see also~\cite{BenAchour:2022uqo,Katagiri:2022vyz}), which bares resemblance to the earlier notion of ``mass ladder operators'' for spacetimes admitting closed conformal Killing vectors~\cite{Cardoso:2017qmj}.

The static Love numbers have also been studied for higher-dimensional general-relativistic black holes in~\cite{Kol:2011vg,Hui:2020xxx,Pereniguez:2021xcj}, all of which have been spherically symmetric. In these examples, the static Love numbers are in general non-zero and exhibit no running. However, there exist some resonant conditions between the multipolar order $\ell$ and the spacetime dimensionality $d$, for which the static Love numbers vanish again, which hints again on a symmetry explanation.
Specifically, for the case of massless scalar perturbations,
this happens when
the generalized 
angular momentum 
$\hat{\ell}=\ell/\left(d-3\right)$ is an integer. 
Despite this more intricate structure of the black hole static Love numbers, the Love symmetry exists for any multipolar order and in any spacetime dimension, and explains these puzzling results~\cite{Charalambous:2021kcz,Charalambous:2022rre}. Here, we extend this analysis to higher-dimensional asymptotically flat, axisymmetric spinning (Myers-Perry) black holes~\cite{Myers:1986un}. We will devote the present work to setting up the arena for this investigation by focusing to the $d=5$ scalar Love numbers.

The structure of this paper is as follows. We begin by reviewing the definition of the response coefficients in Newtonian gravity~\cite{PoissonWill2014} and the subsequent generalization to relativistic configurations within the worldline EFT formalism~\cite{Goldberger:2004jt,Porto:2005ac} in Section~\ref{sec:TLNsDefinition}. We also describe how it is natural to distinguish between the conservative and dissipative parts of the response and we identify the former as the ``Love'' part of the response~\cite{Chia:2020yla,Charalambous:2021mea,Ivanov:2022hlo}. This prescription is then applied to axisymmetric spinning bodies in $5$ spacetime dimensions whose background symmetries do not allow the mixing between different multipolar orders, a case relevant for the Myers-Perry black hole, to extract a simple matching formula for the extraction of the Love numbers from the microscopic computation.

In Section \ref{sec:TLNsComputation}, we treat massless scalar perturbations of the $5$-d Myers-Perry black hole with the scope of extracting the static scalar Love numbers. We introduce a near-zone approximation that allows us to analytically solve the equations of motion in terms of hypergeometric functions and extract the scalar response coefficients that are exact in the static limit. We find some qualitatively new features compared to the case of higher-dimensional non-spinning black holes, namely, that the static scalar Love numbers are always non-zero and exhibit running for generic black hole spin and angular modes of perturbations. We also explore the resonant conditions for which the Love numbers extracted from the near-zone approximation acquire some seemingly fine-tuned properties in the form of ``magic zeroes.''

These resonant conditions are addressed in Section~\ref{sec:SL2R} by introducing two Love symmetries for the $5$-d Myers-Perry black hole, that is, by two enhanced $\SL$ symmetries associated with the two near-zone truncations of the equations of motion that had been employed. Similar to the Love symmetry argument for $4$-dimensional black holes~\cite{Charalambous:2021kcz,Charalambous:2022rre}, these resonant conditions are in one-to-one correspondence with the states spanning the highest-weight representations of the Love symmetries. The highest-weight property then outputs the vanishing of Love numbers, while the absence of running is also realized from algebraic local criteria.

In Section~\ref{sec:Properties}, we begin investigating the properties of the Love symmetries. We start by their geometric interpretation as exact isometries of effective black hole geometries which are identified as relatives to subtracted geometries~\cite{Cvetic:2011dn,Cvetic:2011hp}. We then reveal an infinite-dimensional extension of the symmetry structure into \mbox{$\SL\ltimes\hat{U}\left(1\right)_{\mathcal{V}}^2$} that contains both Love symmetries as particular subalgebras, similar to the infinite extension for $4$-dimensional spinning black holes~\cite{Charalambous:2021kcz,Charalambous:2022rre}.

This larger symmetry also contains a family of $\SL$ subalgebras which are closely related to the enhanced isometries of the near-horizon geometries in the extremal limit~\cite{Bardeen:1999px,Kunduri:2007vf,Galajinsky:2012vh,Galajinsky:2013mla,Hakobyan:2017qee,Figueras:2008qh}. This is explicitly studied in Section~\ref{sec:NHE} where we show that appropriately taking the extremal limit of these $\SL$ subalgebras of the infinite extension precisely recovers the Killing vectors of the near-horizon $\text{AdS}_2$ throat. This further hints towards the interpretation of the Love symmetries as remnants of this enhanced isometry associated with the extremal black holes.

We close with a summary and discussion in Section \ref{sec:Discussion}. We also supplement with a number of appendices. In Appendix \ref{sec:Ap5dMPGeometry}, we review the geometry of the $5$-d Myers-Perry black hole. In Appendix \ref{sec:ApSphericalHarmonics}, we introduce a modified spherical harmonics basis which proves more natural to employ for the study of axisymmetric bodies in $5$ spacetime dimensions. In Appendix \ref{sec:ApSchwarzschildLimit}, we collect the formulas for the source/response split of the black hole perturbations as dictated by the matching with the worldline EFT. We also compare our
findings with known results for Schwarzschild black holes~\cite{Kol:2011vg,Hui:2020xxx} by taking the spinless limit of the relevant expressions extracted in the current work. Last, in Appendix \ref{sec:ApSL2RGenerators}, we sketch the derivation of the $\SL$ symmetries associated with truncations of the equations of motion that preserve the characteristic exponents of the black hole near the event horizon, a subset of which contains near-zone truncations and the associated Love symmetries.

\textit{Notation and conventions}: We will employ geometrized units with $c=1$ and work with the mostly-positive metric Lorentzian signature, $\left(\eta_{\mu\nu}\right) = \text{diag}\left(-1,+1,+1,\dots\right)$. Small Latin letters from the beginning of the alphabet will denote spatial indices running from $1$ to $d-1$ for a $d$-dimensional spacetime, while small Greek letters will denote spacetime indices running from $0$ to $d-1$, with $x^0$ the temporal coordinate, and repeated indices will be summed over. We will also be employing the multi-index notation $a_1\dots a_{\ell}\equiv L$, within which $x^{a_1}\dots x^{a_{\ell}}\equiv x^{L}$ and $\partial_{a_1}\dots\partial_{a_{\ell}}\equiv\partial_{L}$. Last, we will denote the symmetric trace-free (STF) part of a tensor with respect to a set of indices $\left\{a_1,a_2,\dots\right\}$ by enclosing the indices within angular brackets (``$\left\langle a_1a_2 \dots \right\rangle$'').
\section{Response coefficients and Love numbers}
\label{sec:TLNsDefinition}

In this section, we present the formulation of the response problem for general compact bodies in terms of response tensors in any number of spatial dimensions with the eventual goal to define the Love numbers in $5$ spacetime dimensions.

\subsection{Newtonian definition}
Let us start with the standard formulation of the tidal response problem for a compact body in Newtonian gravity~\cite{PoissonWill2014}. This consists of solving the Poisson equation\footnote{Newton's gravitational constant $G$ is identified as the coupling constant appearing in the Einstein-Hilbert action $S=\frac{1}{16\pi G}\int d^{d}x\,\sqrt{-g}\,R$ such that the Einstein field equations $G_{\mu\nu}=8\pi G T_{\mu\nu}$ preserve their form in any number of spacetime dimensions.} for the Newtonian gravitation potential $\Phi_{\text{N}}$,
\be
	\nabla^2\Phi_{\text{N}} = \frac{d-3}{d-2}8\pi G\rho \,,
\ee
with $\rho=\rho\left(t,\mathbf{x}\right)$ the mass density of the mass configuration. In practice, this is accompanied by two field equations; the continuity equation and Euler's equations. Once supplemented with an equation of state, the problem is well posed.

The setup for introducing the tidal response coefficients begins with an unperturbed mass configuration at equilibrium, practically being hydrostatic equilibrium, which is perturbed by a weak distant mass configuration sourcing tidal forces parameterized by its tidal moments $\bar{\mathcal{E}}_{L}\left(t\right)$. In response, the mass distribution of the body rearranges until a new equilibrium state is reached. The response is encoded in induced mass multipole moments $\delta Q_{L}\left(t\right)$~\cite{PoissonWill2014}. The perturbation in the Newtonian gravitational potential in the exterior of the body is then given by, in frequency space,
\be
	\delta\Phi_{\text{N}}\left(\omega,\mathbf{x}\right) = \sum_{\ell=2}^{\infty}\frac{\left(\ell-2\right)!}{\ell!}\left[\bar{\mathcal{E}}_{L}\left(\omega\right) - N_{\ell}G\frac{\delta Q_{L}\left(\omega\right)}{r^{2\ell+d-3}}\right]x^{L} \,,
\ee
where we have defined the dimensionless constants
\be\label{eq:NnormDimLess}
	N_{\ell} \equiv  \frac{8\pi}{\left(d-2\right)\Omega_{d-2}}\frac{\left(2\ell+d-5\right)!!}{\left(\ell-2\right)!\left(d-5\right)!!} \,,
\ee
with $\Omega_{d-2}=2\pi^{\left(d-1\right)/2}/\Gamma\left(\frac{d-1}{2}\right)$ the surface area of the unit $\left(d-2\right)$-sphere $\mathbb{S}^{d-2}$. Both the tidal moments $\bar{\mathcal{E}}_{L}$ and the induced mass multipole moments $\delta Q_{L}$ are rank-$\ell$ STF spatial tensors by virtue of the Laplace equation in the exterior\footnote{If $V\left(\mathbf{x}\right)$ is a harmonic function, then $\partial_{L}V$ are rank-$\ell$ STF spatial tensors. In the same way, $\partial_{L}\frac{1}{r^{d-3}}$, which are involved in the multipole expansion defining the mass multipole moments, are STF tensors of rank-$\ell$, since $\left|\mathbf{x}-\mathbf{x}^{\prime}\right|^{-\left(d-3\right)}$ is a harmonic function of $\mathbf{x}$ in the exterior.}. We also remark here that we are working in the body-centered frame where the induced dipole moment vanishes identically and all the sums start from $\ell=2$.

Assuming that the body exhibits no gravitational hysteresis, i.e. that it develops no new permanent multipole moments after the tidal source is switched off, we can apply linear response theory and define the dimensionful tidal response tensor $\lambda_{LL^{\prime}}$ in frequency space as the corresponding retarded Green's function\footnote{Lower and upper spatial indices in Newtonian gravity are raised and lowered with the flat space metric.},
\be
	\delta Q_{L}\left(\omega\right) = -\sum_{\ell^{\prime}=2}^{\infty}\lambda_{LL^{\prime}}\left(\omega\right)\bar{\mathcal{E}}^{L^{\prime}}\left(\omega\right) \,,
\ee
where the mixing of different $\ell$-modes is a necessary implementation for rotating and non-spherically symmetric bodies and we are suppressing non-linear corrections. The tidal response tensor $\lambda_{LL^{\prime}}$ is STF with respect to the first multi-index $L$, while only $\lambda_{L\left\langle L^{\prime} \right\rangle}$ is physically relevant. Although we have only displayed an $\omega$-dependence, $\lambda_{LL^{\prime}}$ will strongly depend on other properties associated with the internal structure of the body, e.g. its background multipole moments, including its mass and angular momentum, as well as other parameters entering its equation of state. As a last conventional step, we introduce the computationally favorable dimensionless tidal response tensor $k_{LL^{\prime}}\left(\omega\right)$, defined according to
\be\label{eq:DimensionlessResponse}
	\lambda_{LL^{\prime}}\left(\omega\right) \equiv k_{LL^{\prime}}\left(\omega\right)\frac{\mathcal{R}^{2\ell+d-3}}{N_{\ell}G} \,,
\ee
with $\mathcal{R}$ a scale associated to the unperturbed body's size, e.g. its radius if it is spherically symmetric, such that the frequency space gravitational potential perturbation in the exterior takes the form
\be
	\delta\Phi_{\text{N}}\left(\omega,\mathbf{x}\right) = \sum_{\ell,\ell^{\prime}=2}^{\infty}\frac{\left(\ell-2\right)!}{\ell!}\left[\delta_{L,L^{\prime}} + k_{LL^{\prime}}\left(\omega\right)\left(\frac{\mathcal{R}}{r}\right)^{2\ell+d-3}\right]\bar{\mathcal{E}}^{L^{\prime}}\left(\omega\right)x^{L} \,.
\ee
Although we have presented an analysis for the tidal deformation of a self-gravitating body, it can be extended to compact bodies supported by other types of long-range forces as well. In particular, it can be extended to systems responding to spin-$1$ and spin-$0$ forces and define the corresponding response tensors. Then, the $\ell$-sums will start from $\ell=1$ or $\ell=0$ respectively, but the general prescription outlined above can be carried away unaffected, up to the conventional overall $\left(\ell-s\right)!/\ell!$ factor for a spin-$s$ force system, and define the spin-$s$ response tensors $k_{LL^{\prime}}^{\left(s\right)}$,
\be\label{eq:NewtonianResponseTensros}
	\delta\Phi^{\left(s\right)}\left(\omega,\mathbf{x}\right) = \sum_{\ell,\ell^{\prime}=s}^{\infty}\frac{\left(\ell-s\right)!}{\ell!}\left[\delta_{L,L^{\prime}} + k_{LL^{\prime}}^{\left(s\right)}\left(\omega\right)\left(\frac{\mathcal{R}}{r}\right)^{2\ell+d-3}\right]\bar{\mathcal{E}}^{\left(s\right)L^{\prime}}\left(\omega\right)x^{L} \,.
\ee
In this language, the above analysis of the tidal response of a gravitational system corresponds to the Newtonian definition of the $k_{LL^{\prime}}^{\left(2\right)}$ response tensor. For $s=1$, $\Phi^{\left(1\right)}$ is the electrostatic potential and $k_{LL^{\prime}}^{\left(1\right)}$ defines the electric susceptibility tensor of the body, while there is an analogous definition of the magnetic susceptibility tensor associated with induced electric current multipole moments in the vector potential profile which we do not write down here. Last, for $s=0$, $\Phi^{\left(0\right)}$ is a scalar field for the potential of a system interacting via scalar forces and $k_{LL}^{\left(0\right)}$ defines the scalar susceptibility tensor of the body.

\subsection{General relativistic EFT definition and Newtonian matching}
When relativistic effects are taken into account, the definition of the associated response tensors is more subtle. To begin with, the response tensors should be gauge invariant under diffeomorphisms. In addition, the growing mode in the profiles of the potentials, the ``source'' part of the field, acquires relativistic corrections and results in an overlapping with the decaying mode, the ``response'' part of the field, thus, raising concerns for a source/response ambiguity~\cite{Gralla:2017djj}.

These concerns are all addressed within the framework of the worldline EFT~\cite{Goldberger:2004jt,Porto:2005ac} whose starting point is the universal point-particle appearance of compact bodies from large distances. We will only briefly review the EFT definition of Love numbers here. A more complete review fitted to the tidal response problem can be found in~\cite{Charalambous:2021mea,Ivanov:2022hlo}, while comprehensive reviews of the worldline EFT formalism can be found in~\cite{Porto:2016pyg,Levi:2018nxp}.
One is effectively integrating out the short-scale modes associated with the internal structure of the body, leaving an effective action whose degrees of freedom are the worldline position $x_{\text{cm}}\left(\lambda\right)$ along which the center of the mass of the body propagates and parameterized by an affine parameter $\lambda$, a set of vielbein vectors $e_{a}^{\mu}\left(\lambda\right)$ localized on the worldline and capturing rotational degrees of freedom in the case the body is spinning\footnote{If the body is spinning, the notion of a ``center of mass'' is not invariant and one needs to supplement with a spin gauge symmetry~\cite{Porto:2005ac} to compensate for this. The center of mass is then fixed via a ``Spin Supplementary Condition'' (SSC) corresponding to fixing the time-like vector of the worldline vielbein $e_{a=0}^{\mu}$~\cite{Porto:2016pyg,Levi:2018nxp}.}, and the long distance metric perturbations with respect to the Minkowski background $h_{\mu\nu}=g_{\mu\nu}-\eta_{\mu\nu}$. These are supplemented with other types of bulk fields and symmetries in the case of non-pure gravity, e.g. with a $U\left(1\right)$ gauge field $A_{\mu}$ for the Einstein-Maxwell theory or a real scalar field $\Phi$ for systems interacting via scalar forces. The effective action then contains a ``minimal'' point-particle action, while finite-size effects are captured by non-minimal couplings of the worldline with higher-derivative operators,
\be
	S_{\text{EFT}}\left[x_{\text{cm}},e,h,A,\Phi\right] = S_{\text{bulk}}\left[\eta+h,A,\Phi\right] + S_{\text{pp}}\left[x_{\text{cm}},e,h\right] + S_{\text{finite-size}}\left[x_{\text{cm}},e,h,A,\Phi\right] \,.
\ee

Love numbers are defined as particular Wilson coefficients in front of quadratic couplings of the worldline with field strength tensors. For the simplest case of a spherically symmetric, non-rotating body, for example, the static Love numbers are defined from
\be
	S_{\text{finite-size}} \supset S_{\text{Love}} = \sum_{s=0}^{2}\sum_{\ell=s}^{\infty}\frac{C_{\text{el},\ell}^{\left(s\right)}}{2\ell!}\int d\tau\,\mathcal{E}_{L}^{\left(s\right)}\left(x_{\text{cm}}\left(\tau\right)\right)\mathcal{E}^{\left(s\right)L}\left(x_{\text{cm}}\left(\tau\right)\right) + \left(\mathcal{E}\leftrightarrow \mathcal{B},\mathcal{T}\right) \,,
\ee
where we have chosen the affine parameter to be equal to the proper time along the worldline and $\mathcal{E}_{L}^{\left(s\right)}\equiv \mathcal{E}_{a_1\dots a_{\ell}}^{\left(s\right)}$ are the multipole moments of the electric-type field strength tensors projected onto spatial slices orthogonal to the $d$-velocity $u^{\mu}=\frac{dx_{\text{cm}}^{\mu}}{d\tau}$ of the body defined via a set of local vielbein vectors $e_{a}^{\mu}$ satisfying $u_{\mu}e_{a}^{\mu}=0$. For $s=0,1,2$,
\be\ba
	\mathcal{E}_{L}^{\left(s=0\right)} &= e_{a_1}^{\mu_1}\dots e_{a_{\ell}}^{\mu_{\ell}}\nabla_{\langle \mu_1}\dots \nabla_{\mu_{\ell}\rangle}\Phi \,, \\
	\mathcal{E}_{L}^{\left(s=1\right)} &= e_{a_1}^{\mu_1}\dots e_{a_{\ell}}^{\mu_{\ell}}\nabla_{\langle \mu_1}\dots \nabla_{\mu_{\ell-1}}E_{\mu_{\ell}\rangle} \,,\quad E_{\mu} = u^{\nu}F_{\mu\nu} \,, \\
	\mathcal{E}_{L}^{\left(s=2\right)} &= e_{a_1}^{\mu_1}\dots e_{a_{\ell}}^{\mu_{\ell}}\nabla_{\langle \mu_1}\dots \nabla_{\mu_{\ell-2}}E_{\mu_{\ell-1}\mu_{\ell}\rangle} \,,\quad E_{\mu\nu} = u^{\rho}u^{\sigma}C_{\mu\rho\nu\sigma} \,,
\ea\ee
and they are by construction rank-$\ell$ STF spatial tensors. The Wilson coefficients $C_{\text{el},\ell}^{\left(s\right)}$ define the spin-$s$ electric-type static Love numbers. Although not written here explicitly, there is also a magnetic version of this interaction term defining the magnetic-type static Love numbers, while, for $s=2$ in $d>4$, one should furthermore take into account tensor-type gravitational perturbations which define the tensor-type tidal Love numbers, see e.g.~\cite{Hui:2020xxx}. In the rest of this section, we will focus to the electric-type responses for economy but all the analysis below can be straightforwardly applied for magnetic-type and tensor-type responses as well.

For a generic compact body, the Wilson coefficients $C_{\text{el},\ell}^{\left(s\right)}$ become ``Wilson tensors'' $C_{\text{el},LL^{\prime}}^{\left(s\right)}$ defining the static Love tensors\footnote{We are also assuming parity invariance of the geometry of the unperturbed body here, i.e. we are omitting mixing between electric and magnetic components which would otherwise be allowed.},
\be
	S_{\text{Love}}^{\text{electric}} = \sum_{s=0}^{2}\sum_{\ell,\ell^{\prime}=s}^{\infty}\int d\tau\,\frac{C_{\text{el},LL^{\prime}}^{\left(s\right)}}{2\ell!}\,\mathcal{E}^{\left(s\right)L}\left(x_{\text{cm}}\left(\tau\right)\right)\mathcal{E}^{\left(s\right)L^{\prime}}\left(x_{\text{cm}}\left(\tau\right)\right) \,.
\ee
The time dependent conservative responses can 
be captured by operators like $D\mathcal{E}D\mathcal{E}$
on the world line, where $D\equiv (d x^{\mu}_{\rm cm}/d\tau)\partial_\mu$. It is more practical, however, 
to switch to frequency space in this case, 
and consider the ``dynamical'' Love number
action of the form, 
\be
	S_{\text{dynamical Love}}^{\text{electric}} = \sum_{s=0}^{2}\sum_{\ell,\ell^{\prime}=s}^{\infty}\int\frac{d\omega}{2\pi}\,\frac{C_{\text{el},LL^{\prime}}^{\left(s\right)}\left(\omega\right)}{2\ell!}\,\mathcal{E}^{\left(s\right)L}\left(-\omega\right)\mathcal{E}^{\left(s\right)L^{\prime}}\left(\omega\right) \,.
\ee

To compute the Love tensors, one matches onto observables of the full theory. We will employ here the ``Newtonian matching'' condition consisting of inserting a pure $2^{\ell}$-pole background Newtonian source at large distances,
\be
	\Phi^{\left(s\right)}\left(\omega,\mathbf{x}\right) = \bar{\Phi}^{\left(s\right)}\left(\omega,\mathbf{x}\right) + \delta\Phi^{\left(s\right)}\left(\omega,\mathbf{x}\right) \,,\quad \bar{\Phi}^{\left(s\right)}\left(\omega,\mathbf{x}\right) = \frac{\left(\ell-s\right)!}{\ell!}\bar{\mathcal{E}}_{L}^{\left(s\right)}\left(\omega\right)x^{L} \,,
\ee
and matching EFT $1$-point functions onto microscopic computations of perturbation theory~\cite{Kol:2011vg,Charalambous:2021mea,Ivanov:2022hlo}. Diagrammatically\footnote{We are employing the dimensional regularization scheme here which sets all the unphysical power-law divergences to zero.},
\be\label{eq:1ptNewtonianMatching}
	\vev{\delta\Phi^{\left(s\right)}\left(\omega,\mathbf{x}\right)} = \vcenter{\hbox{\begin{tikzpicture}
				\begin{feynman}
					\vertex (a0);
					\vertex[right=0.6cm of a0] (gblobaux);
					\vertex[left=0.00cm of gblobaux, blob] (gblob){};
					\vertex[below=1cm of a0] (p1);
					\vertex[above=1cm of a0] (p2);
					\vertex[right=1cm of p1] (a1);
					\vertex[right=1cm of p2] (a2);%{$\times$};
					\vertex[right=0.69cm of p2] (a22){$\times$};
					\diagram*{
						(p1) -- [double,double distance=0.5ex] (p2),
						(a1) -- (gblob) -- (a2),
					};
				\end{feynman}
	\end{tikzpicture}}} =
	\underbrace{\vcenter{\hbox{\begin{tikzpicture}
					\begin{feynman}
						\vertex[dot] (a0);
						\vertex[below=1cm of a0] (p1);
						\vertex[above=1cm of a0] (p2);
						\vertex[right=0.4cm of a0, blob] (gblob){};							
						\vertex[right=1.5cm of p1] (b1);
						\vertex[right=1.5cm of p2] (b2);
						\vertex[right=1.19cm of p2] (b22){$\times$};
						\vertex[above=0.7cm of a0] (g1);
						\vertex[above=0.4cm of a0] (g2);
						\vertex[right=0.05cm of a0] (gdtos){$\vdots$};
						\vertex[below=0.7cm of a0] (gN);
						\diagram*{
							(p1) -- [double,double distance=0.5ex] (p2),
							(g1) -- [photon] (gblob),
							(g2) -- [photon] (gblob),
							(gN) -- [photon] (gblob),
							(b1) -- (gblob) -- (b2),
						};
					\end{feynman}
	\end{tikzpicture}}}}_{\text{``source''}} + 
	\underbrace{\vcenter{\hbox{\begin{tikzpicture}
					\begin{feynman}
						\vertex[dot] (a0);
						\vertex[left=0.00cm of a0] (lambda){$C_{\text{el},LL^{\prime}}^{\left(s\right)}\left(\omega\right)$};
						\vertex[below=1.6cm of a0] (p1);
						\vertex[above=0.4cm of a0] (p2);
						\vertex[right=1.5cm of p1] (b1);
						\vertex[right=1.5cm of p2] (b2);
						\vertex[right=1.19cm of p2] (b22){$\times$};
						\vertex[below=0.5cm of a0] (g1);
						\vertex[below=0.3cm of g1] (g2);
						\vertex[below=0.15cm of g2] (gdotsaux);
						\vertex[right=0.00cm of gdotsaux] (gdtos){$\vdots$};
						\vertex[below=0.5cm of gdotsaux] (gN);
						\vertex[below=0.7cm of a0] (gblobaux);
						\vertex[right=0.4cm of gblobaux, blob] (gblob){};
						\diagram*{
							(p1) -- [double,double distance=0.5ex] (p2),
							(b2) -- (a0) -- (gblob) -- (b1),
							(g1) -- [photon] (gblob),
							(g2) -- [photon] (gblob),
							(gN) -- [photon] (gblob),
						};
					\end{feynman}
	\end{tikzpicture}}}}_{\text{``response''}} \,,
\ee
where the double line represents the worldline, straight lines indicate propagators of the fields $\delta\Phi^{\left(s\right)}$, a ``$\times$'' represents a $\bar{\Phi}^{\left(s\right)}$ insertion and wavy lines correspond to interactions of the worldline with the graviton arising from the minimal point-particle action. We note here that we are not including dissipative effects which will be addressed shortly. In the above diagrammatic representation we have also demonstrated how the worldline EFT definition allows to unambiguously separate relativistic corrections in the ``source'' part of the field profile from tidal effects~\cite{Charalambous:2021mea,Ivanov:2022hlo}. This splitting is in fact equivalent to the method of analytically continuing the spacetime dimensionality $d$~\cite{Kol:2011vg} or the multipolar order $\ell$~\cite{LeTiec:2020bos,Charalambous:2021mea,Ivanov:2022hlo,Creci:2021rkz} as the ``source'' and ``response'' diagrams have indicial powers $r^{\alpha}$ with $\alpha=\ell$ and $\alpha=-\left(\ell+d-3\right)$ respectively. These receive PN corrections from the interaction of the graviton with the worldline which have the form $r^{\alpha-n}$ with positive integer $n$.

In the Newtonian limit, in a gauge where the fields $\delta\Phi^{\left(s\right)}$ are canonical variables up to an overall normalization constant $N^{\left(s\right)}_{\text{prop}}$ in momentum space,
\be
	\vev{\delta\Phi^{\left(s\right)}\delta\Phi^{\left(s\right)}}\left(p\right) = N^{\left(s\right)}_{\text{prop}}\frac{-i}{p^2} \,,
\ee
and in the body centered frame where $x_{\text{cm}}=\left(t,\mathbf{0}\right)$ and $u^{\mu}=\left(1,\mathbf{0}\right)$, this gives
\be\ba
	{}&\vev{\delta\Phi^{\left(s\right)}\left(\omega,\mathbf{x}\right)} \rightarrow
		\vcenter{\hbox{\begin{tikzpicture}
						\begin{feynman}
							\vertex[dot] (a0);
							\vertex[below=1cm of a0] (p1);
							\vertex[above=1cm of a0] (p2);						
							\vertex[right=1cm of p1] (b1);
							\vertex[right=1cm of p2] (b2);
							\vertex[right=0.69cm of p2] (b22){$\times$};
							\diagram*{
								(p1) -- [double,double distance=0.5ex] (p2),
								(b1) -- (b2),
							};
						\end{feynman}
		\end{tikzpicture}}} + 
		\vcenter{\hbox{\begin{tikzpicture}
						\begin{feynman}
							\vertex[dot] (a0);
							\vertex[left=0.00cm of a0] (lambda){$C_{\text{el},LL^{\prime}}^{\left(s\right)}\left(\omega\right)$};
							\vertex[below=1cm of a0] (p1);
							\vertex[above=1cm of a0] (p2);
							\vertex[right=1.2cm of p1] (b1);
							\vertex[right=1.2cm of p2] (b2);
							\vertex[right=0.89cm of p2] (b22){$\times$};
							\diagram*{
								(p1) -- [double,double distance=0.5ex] (p2),
								(b2) -- (a0) -- (b1),
							};
						\end{feynman}
		\end{tikzpicture}}} \\
		&= \frac{\left(\ell-s\right)!}{\ell!} \sum_{\ell^{\prime}=s}^{\infty}\left[\delta_{L,L^{\prime}} + \frac{2^{\ell-2}\Gamma\left(\ell+\frac{d-3}{2}\right)}{\pi^{\left(d-1\right)/2}}N^{\left(s\right)}_{\text{prop}}\frac{\left[C_{\text{el},LL^{\prime}}^{\left(s\right)}\left(\omega\right)\right]_{\text{TRS}}}{r^{2\ell+d-3}}\right]\bar{\mathcal{E}}^{\left(s\right)L^{\prime}}\left(\omega\right)x^{L} \,,
\ea\ee
where,
\be
	\left[C_{\text{el},LL^{\prime}}^{\left(s\right)}\left(\omega\right)\right]_{\text{TRS}}\equiv\frac{1}{2}\left(C_{\text{el},LL^{\prime}}^{\left(s\right)}\left(\omega\right)+C_{\text{el},L^{\prime}L}^{\left(s\right)}\left(-\omega\right)\right) \,,
\ee
with ``TRS'' standing for time-reversal symmetric. From this, we identify the explicit correspondence between the electric-type Love tensor and the Wilson tensor for a compact body of size $\mathcal{R}$,
\be\label{eq:WTesnorConsResp}
	k_{LL^{\prime}}^{\left(s\right)\text{Love}}\left(\omega\right) = \frac{2^{\ell-2}\Gamma\left(\ell+\frac{d-3}{2}\right)}{\pi^{\left(d-1\right)/2}}N^{\left(s\right)}_{\text{prop}}\frac{\left[C_{\text{el},LL^{\prime}}^{\left(s\right)}\left(\omega\right)\right]_{\text{TRS}}}{\mathcal{R}^{2\ell+d-3}} \,.
\ee
We see therefore that the Love tensor is defined from the \textit{conservative} response, i.e. the part of the response tensor invariant under the time-reversal transformations which corresponds to simultaneously flipping the sign of the frequency, $\omega\rightarrow-\omega$, and the exchange $L\leftrightarrow L^{\prime}$. This is implicit by the definition at the level of the action and the use of the in-out formalism since
\be\ba
	\sum_{\ell,\ell^{\prime}}\int\frac{d\omega}{2\pi}&\,\frac{\left[C_{\text{el},LL^{\prime}}^{\left(s\right)}\left(\omega\right)\right]_{\text{TRS}}}{2\ell!}\,\mathcal{E}^{\left(s\right)L}\left(-\omega\right)\mathcal{E}^{\left(s\right)L^{\prime}}\left(\omega\right) = \\
	&\sum_{\ell,\ell^{\prime}}\int\frac{d\omega}{2\pi}\,\frac{C_{\text{el},LL^{\prime}}^{\left(s\right)}\left(\omega\right)}{2\ell!}\,\mathcal{E}^{\left(s\right)L}\left(-\omega\right)\mathcal{E}^{\left(s\right)L^{\prime}}\left(\omega\right) \,.
\ea\ee

\subsubsection{Dissipation in EFT}
As we just saw, only $\left[C_{\text{el},LL^{\prime}}^{\left(s\right)}\left(\omega\right)\right]_{\text{TRS}}$ is relevant when computing $1$-point functions via the standard in-out formalism, i.e. local operators in the worldline EFT action capture only conservative effects. Dissipative effects are incorporated by introducing gapless internal degrees of freedom $X$. One then considers composite operators $Q_{L}^{\left(s\right)}\left(X\right)$ corresponding to the full multipole moments, including the dissipative multipole moments due to the internal degrees of freedom $X$, but whose exact dependence on $X$ is not known. These are then coupled to the field moments~\cite{Goldberger:2005cd,Goldberger:2019sya,Goldberger:2020fot,Goldberger:2020wbx},
\be
	S_{\text{diss}} = \sum_{s=0}^{2}\sum_{\ell=s}^{\infty}\frac{1}{\ell!}\int d\tau\,Q_{L}^{\left(s\right),\mathcal{E}}\left(X\right)\mathcal{E}^{\left(s\right)L}\left(x_{\text{cm}}\left(\tau\right)\right) + \left(\mathcal{E}\leftrightarrow\mathcal{B},\mathcal{T}\right) \,.
\ee
In order to account for dissipative effects at the level of the $1$-point function, one then employs the in-in (Schwinger-Keldysh) formalism~\cite{Schwinger:1960qe,Keldysh:1964ud,Goldberger:2005cd,Goldberger:2019sya,Goldberger:2020fot,Goldberger:2020wbx}. Within this framework~\cite{Ivanov:2022hlo},
\be\ba\label{eq:1ptInInResponse}
    {}&\vev{\delta\Phi^{\left(s\right)}\left(\omega,\mathbf{x}\right)} \supset
	\vcenter{\hbox{\begin{tikzpicture}
			\begin{feynman}
				\vertex[] (a0){};
				\vertex[left=0.2cm of a0] (a00);
				\vertex[above=0.4cm of a00] (a0t){};
				\vertex[below=0.4cm of a00] (a0b){};
				\vertex[left=0.4cm of a0] (p0);
				\vertex[above=0.4cm of p0] (p0t){$Q^{\left(s\right)}$};
				\vertex[below=0.4cm of p0] (p0b){$Q^{\left(s\right)}$};
				\vertex[below=1.5cm of a0] (p1);
				\vertex[above=1.5cm of a0] (p2);
				\vertex[below=0.6cm of a0] (p11);
				\vertex[above=0.6cm of a0] (p22);
				\vertex[below right=2cm of a0] (a1);
				\vertex[above right=2cm of a0] (a2);%{$\times$};
				\vertex[above right=2cm of a0] (a22){$\times$};
				\diagram*{
					(p1) -- [double,double distance=0.5ex] (p2),
					(a1) -- (a0b) , (a0t) -- (a2),
					(p11) -- [ghost](p22)
				};
			\end{feynman}
	\end{tikzpicture}}} \\
	&= \frac{\left(\ell-s\right)!}{\ell!}\sum_{\ell^{\prime}=s}^{\infty}\left[\frac{2^{\ell-2}\Gamma\left(\ell+\frac{d-3}{2}\right)}{\pi^{\left(d-1\right)/2}}\frac{\vev{Q_{L}^{\left(s\right),\mathcal{E}}Q_{L^{\prime}}^{\left(s\right),\mathcal{E}}}\left(-\omega\right)}{r^{2\ell+d-3}}\right]\bar{\mathcal{E}}^{\left(s\right)L^{\prime}}\left(\omega\right)x^{L}
	\,,
\ea\ee
and the full electric-type response tensors $k_{LL^{\prime}}^{\left(s\right)}\left(\omega\right)$ are matched onto the (retarded) $2$-point function~\cite{Ivanov:2022hlo},
\be\label{eq:WTesnorFullResp}
	k_{LL^{\prime}}^{\left(s\right)}\left(\omega\right) = \frac{2^{\ell-2}\Gamma\left(\ell+\frac{d-3}{2}\right)}{\pi^{\left(d-1\right)/2}}\frac{\vev{Q_{L}^{\left(s\right),\mathcal{E}}Q_{L^{\prime}}^{\left(s\right),\mathcal{E}}}\left(-\omega\right)}{\mathcal{R}^{2\ell+d-3}} \,.
\ee
Note that the real part of $k_{LL^{\prime}}^{\left(s\right)}$
is indistinguishable from $k_{LL^{\prime}}^{\left(s\right)\text{Love}}$
and therefore can be ignored. Its imaginary part 
though cannot be reproduced from a local 
world line action and thus it encodes 
non-conservative effects such as horizon absorption.

\subsection{Love numbers of 5-d rotating bodies}
So far we have formulated the response problem in terms of the response tensors $k_{LL^{\prime}}^{\left(s\right)}$. In practice, one is interested in the harmonic response coefficients arising after performing a harmonic expansion thanks to the $1$-to-$1$ correspondence between spatial STF tensors and spherical harmonics. From these, the Love numbers are identified as the conservative harmonic response coefficients~\cite{Charalambous:2021mea,Ivanov:2022hlo}. Isolating the conservative part of the harmonic response coefficients is in general non-trivial, but for some particular configurations, e.g. the remarkably integrable black hole perturbations, this decomposition allows one to identify the Love numbers as the real part of the harmonic response coefficients, while the imaginary part captures dissipative effects.

While this can be done in $d=1+3$ spacetime dimensions by performing an expansion into spherical harmonics over $\mathbb{S}^2$~\cite{LeTiec:2020bos}, it fails to work for a general rotating body in higher spacetime dimensions. For $d>4$, an expansion into spherical harmonics on $\mathbb{S}^{d-2}$ allows us to extract a simple isolating prescription of the conservative part of the response coefficients only for spherically symmetric and non-rotating bodies. For axisymmetric distributions, one should instead perform a modified harmonic expansion over $\left[\mathbb{S}^1\right]^{N}\subset \mathbb{S}^{d-2}$, with $N=\left\lfloor\frac{d-1}{2}\right\rfloor$ factors of $\mathbb{S}^1$, appropriate for the isometry subgroup $\left[U\left(1\right)\right]^{N}\subset SO\left(d-1\right)$ of such configurations. In $d=5$ spacetime dimensions, this is a modified harmonic expansion over the $\mathbb{S}^1\times\mathbb{S}^1$ part of $\mathbb{S}^3$ in accordance with the $U\left(1\right)\times U\left(1\right)$ azimuthal symmetries. This modified spherical harmonics basis for $d=5$ is introduced and analyzed in Appendix \ref{sec:ApSphericalHarmonics}.

To this end, we begin by expanding the $4$-dimensional spatial STF tensors $\bar{\mathcal{E}}_{L}^{\left(s\right)}$ into modified spherical harmonics of orbital number $\ell\in\mathbb{N}$,
\be
	\bar{\mathcal{E}}^{\left(s\right)L} = \sum_{m,j}\bar{\mathcal{E}}_{\ell m j}^{\left(s\right)}\mathcal{Y}_{\ell m j}^{L\ast} \,,
\ee
where the constant STF tensors $\mathcal{Y}_{\ell m j}^{L}$ are given by
\be
	\mathcal{Y}_{\ell m j}^{L} = \frac{\left(2\ell+2\right)!!}{4\pi^2\ell!}\oint_{\mathbb{S}^3}d\Omega_3\,\Omega^{\left\langle L \right\rangle}\tilde{Y}_{\ell m j}^{\ast}\left(\mathbf{\Omega}\right) \,,
\ee
with $\tilde{Y}_{\ell m j}\left(\mathbf{\Omega}\right)\equiv \tilde{Y}_{\ell m j}\left(\theta,\phi,\psi\right)$ the modified spherical harmonics on $\mathbb{S}^3$, $\Omega^{i}\equiv x^{i}/r$ and asterisks indicate complex conjugation. For future reference, the explicit limits of the sums over the azimuthal numbers $m$ and $j$ are
\be
	\sum_{m,j}\left(\dots\right)\equiv \sum_{m=-\ell}^{\ell}\left(\sum_{j=-\left(\ell-\left|m\right|\right),2}^{\ell-\left|m\right|}\left(\dots\right)\right) \,.
\ee
We note that the $j$-sum is being performed with a step $2$. This is merely a convention chosen such that the azimuthal number $m$ resembles the usual azimuthal number of scalar spherical harmonics on $\mathbb{S}^2$.

Then, the response coefficients $k_{\ell m j;\ell^{\prime} m^{\prime} j^{\prime}}^{\left(s\right)}\left(\omega\right)$ are related to the response tensor $k_{LL^{\prime}}^{\left(s\right)}\left(\omega\right)$ according to
\be\label{eq:RNsToRTs}
	k_{\ell m j;\ell^{\prime} m^{\prime} j^{\prime}}^{\left(s\right)}\left(\omega\right) = \frac{4\pi^2\ell!}{\left(2\ell+2\right)!!}k_{LL^{\prime}}^{\left(s\right)}\left(\omega\right)\mathcal{Y}_{\ell m j}^{L}\mathcal{Y}_{\ell^{\prime} m^{\prime} j^{\prime}}^{L^{\prime}\ast} \,.
\ee
Using the fact that the induced multipole moments $\delta Q_{L}^{\left(s\right)}\left(t\right)$ and source multipole moments $\mathcal{E}_{L}^{\left(s\right)}\left(t\right)$ are real in position space as well as the assumption that the response tensors $k_{LL^{\prime}}^{\left(s\right)}\left(\omega\right)$ are analytic in $\omega$, i.e. that
\be
	k_{LL^{\prime}}^{\left(s\right)}\left(\omega\right) = \sum_{n=0}^{\infty}k_{LL^{\prime};n}^{\left(s\right)}\left(i\omega\right)^{n} \,,
\ee
with real-valued modes $k_{LL^{\prime};n}^{\left(s\right)}$, we see that,
\be
	k_{LL^{\prime}}^{\left(s\right)\ast}\left(\omega\right)=k_{LL^{\prime}}^{\left(s\right)}\left(-\omega\right) \,.
\ee
From the complex conjugacy relation of the modified spherical harmonics, $\tilde{Y}_{\ell m j}^{\ast}=\tilde{Y}_{\ell,-m,-j}$, we then deduce the following complex conjugacy relation for the response coefficients
\be\label{eq:klmCC}
	k_{\ell m j;\ell^{\prime} m^{\prime} j^{\prime}}^{\left(s\right)\ast}\left(\omega\right) = k_{\ell,-m,-j;\ell^{\prime},-m^{\prime},-j^{\prime}}^{\left(s\right)}\left(-\omega\right) \,.
\ee

We can now translate the conservative/dissipative decomposition of the response tensor (see \eqref{eq:WTesnorConsResp},\eqref{eq:WTesnorFullResp}),
\be\ba
	k_{LL^{\prime}}^{\left(s\right)\text{Love}}\left(\omega\right) &= \frac{1}{2}\left(k_{LL^{\prime}}^{\left(s\right)}\left(\omega\right) + k_{L^{\prime}L}^{\left(s\right)}\left(-\omega\right)\right) \,, \\
	k_{LL^{\prime}}^{\left(s\right)\text{diss}}\left(\omega\right) &= \frac{1}{2}\left(k_{LL^{\prime}}^{\left(s\right)}\left(\omega\right) - k_{L^{\prime}L}^{\left(s\right)}\left(-\omega\right)\right) \,,
\ea\ee
at the level of the response coefficients $k_{\ell mj;\ell^{\prime}m^{\prime}j^{\prime}}^{\left(s\right)}\left(\omega\right)$. The definition \eqref{eq:RNsToRTs} and the complex conjugacy relation \eqref{eq:klmCC} immediately imply
\be\ba\label{eq:RCsConsDissGen}
	k_{\ell mj;\ell^{\prime}m^{\prime}j^{\prime}}^{\left(s\right)\text{Love}}\left(\omega\right) &= \frac{1}{2}\left(k_{\ell mj;\ell^{\prime}m^{\prime}j^{\prime}}^{\left(s\right)}\left(\omega\right)+k_{\ell^{\prime}m^{\prime}j^{\prime};\ell mj}^{\left(s\right)\ast}\left(\omega\right)\right) \,, \\
	k_{\ell mj;\ell^{\prime}m^{\prime}j^{\prime}}^{\left(s\right)\text{diss}}\left(\omega\right) &= \frac{1}{2i}\left(k_{\ell mj;\ell^{\prime}m^{\prime}j^{\prime}}^{\left(s\right)}\left(\omega\right)-k_{\ell^{\prime}m^{\prime}j^{\prime};\ell mj}^{\left(s\right)\ast}\left(\omega\right)\right) \,,
\ea\ee
such that
\be
	k_{\ell mj;\ell^{\prime}m^{\prime}j^{\prime}}^{\left(s\right)}\left(\omega\right) = k_{\ell mj;\ell^{\prime}m^{\prime}j^{\prime}}^{\left(s\right)\text{Love}}\left(\omega\right) + ik_{\ell mj;\ell^{\prime}m^{\prime}j^{\prime}}^{\left(s\right)\text{diss}}\left(\omega\right) \,.
\ee
We note, however, that $k_{\ell mj;\ell^{\prime}m^{\prime}j^{\prime}}^{\left(s\right)\text{Love}}\left(\omega\right)$ and $k_{\ell mj;\ell^{\prime}m^{\prime}j^{\prime}}^{\left(s\right)\text{diss}}\left(\omega\right)$ are in general complex numbers.

We now focus to axisymmetric configurations. The axisymmetry of the background implies the decoupling of $m$-modes and $j$-modes, while we further specialize here to the particular case where there is no $\ell$-mode mixing either, a case relevant for Myers-Perry black holes. Then,
\be
	k_{\ell m j;\ell^{\prime} m^{\prime} j^{\prime}}^{\left(s\right)}\left(\omega\right) = k_{\ell m j}^{\left(s\right)}\left(\omega\right)\delta_{\ell\ell^{\prime}}\delta_{mm^{\prime}}\delta_{jj^{\prime}} \,,
\ee
and the frequency space potential perturbation harmonic modes in the Newtonian limit simplify to
\be
	\delta\Phi^{\left(s\right)}_{\ell m j}\left(\omega,r\right) = \frac{\left(\ell-s\right)!}{\ell!}\left[1 + k_{\ell m j}^{\left(s\right)}\left(\omega\right)\left(\frac{\mathcal{R}}{r}\right)^{2\ell+2}\right]r^{\ell}\bar{\mathcal{E}}_{\ell m j}^{\left(s\right)}\left(\omega\right) \,.
\ee

These response coefficients $k_{\ell m j}^{\left(s\right)}\left(\omega\right)$ will in general be analytic functions in the angular momenta of the rotating body as well as the frequency $\omega$ with respect to an inertial observer. The complex conjugacy relation, which now reads $k_{\ell m j}^{\left(s\right)\ast}\left(\omega\right) = k_{\ell,-m,-j}^{\left(s\right)}\left(-\omega\right)$, then allows to explicitly separate the $m$- and $j$-dependencies of the response coefficients as
\be\label{eq:TLNs5d_mjExpansion}
	k_{\ell m j}^{\left(s\right)}\left(\omega\right) = k_{\ell}^{\left(0\right)}\left(\omega\right) + \chi\sum_{n_{\phi}=1}^{\infty}\sum_{n_{\psi}=1}^{\infty}k_{\ell}^{(n_{\phi},n_{\psi})}\left(\omega,\chi_{\phi},\chi_{\psi}\right) \left(im\right)^{n_{\phi}}\left(ij\right)^{n_{\psi}} \,,
\ee
with $\chi_{\phi}$ and $\chi_{\psi}$ the dimensionless spin parameters associated with the $J_{\phi}$ and $J_{\psi}$ angular momenta and the overall formal $\chi$ is to separate the non-spinning part $k_{\ell}^{\left(0\right)}\left(\omega\right)$. All $k_{\ell}^{(n_{\phi},n_{\psi})}\left(\omega,\chi_{\phi},\chi_{\psi}\right)$ are smooth functions of $\chi_{\phi}$ and $\chi_{\psi}$, satisfying the complex conjugation relation $k_{\ell}^{(n_{\phi},n_{\psi})\ast}\left(\omega,\chi_{\phi},\chi_{\psi}\right)=k_{\ell}^{(n_{\phi},n_{\psi})}\left(-\omega,\chi_{\phi},\chi_{\psi}\right)$.

Let us now extract a necessary condition for such a decoupling to occur. This analysis is the $d=5$ version of~\cite{LeTiec:2020bos}. Starting from the physically relevant part of the response tensor,
\be
	k_{L\left\langle L^{\prime} \right\rangle}^{\left(s\right)}\left(\omega\right) = \frac{4\pi^2\ell!}{\left(2\ell+2\right)!!}\sum_{m,j}k_{\ell m j}^{\left(s\right)}\left(\omega\right)\mathcal{Y}^{\ell m j\ast}_{L}\mathcal{Y}^{\ell m j}_{L^{\prime}} \,,
\ee
with $\ell^{\prime}=\ell$ understood, the expansion \eqref{eq:TLNs5d_mjExpansion} implies
\be\ba
	{}&k_{L\left\langle L^{\prime} \right\rangle}^{\left(s\right)}\left(\omega\right) = k_{\ell}^{\left(0\right)}\left(\omega\right)\delta_{L,L^{\prime}} + \chi\sum_{n_{\phi}=1}^{\infty}\sum_{n_{\psi}=1}^{\infty} \left(-1\right)^{n_{\phi}+n_{\psi}} \\
	&\times\bigg[k_{\ell}^{(2n_{\phi},2n_{\psi})}\left(\omega,\chi_{\phi},\chi_{\psi}\right)R_{LL^{\prime}}^{(2n_{\phi},2n_{\psi})} - k_{\ell}^{(2n_{\phi}-1,2n_{\psi}-1)}\left(\omega,\chi_{\phi},\chi_{\psi}\right)R_{LL^{\prime}}^{(2n_{\phi}-1,2n_{\psi}-1)} \\
	&+ k_{\ell}^{(2n_{\phi}-1,2n_{\psi})}\left(\omega,\chi_{\phi},\chi_{\psi}\right)I_{LL^{\prime}}^{(2n_{\phi}-1,2n_{\psi})} + k_{\ell}^{(2n_{\phi},2n_{\psi}-1)}\left(\omega,\chi_{\phi},\chi_{\psi}\right)I_{LL^{\prime}}^{(2n_{\phi},2n_{\psi}-1)}\bigg] \,,
\ea\ee
and the tensorial structure of $k_{L\left\langle L^{\prime} \right\rangle}^{\left(s\right)}$ is completely determined by two real-valued symmetric and two real-valued antisymmetric STF tensors,
\be\ba
	R_{LL^{\prime}}^{(2n_{\phi},2n_{\psi})} &\equiv \frac{8\pi^2\ell!}{\left(2\ell+2\right)!!}\sum_{m=1}^{\ell}\left(\sum_{j=-\left(\ell-m\right),2}^{\ell-m}m^{2n_{\phi}}j^{2n_{\psi}}\text{Re}\left\{\mathcal{Y}^{\ell mj\ast}_{L}\mathcal{Y}^{\ell mj}_{L^{\prime}}\right\}\right) \,, \\
	R_{LL^{\prime}}^{(2n_{\phi}-1,2n_{\psi}-1)} &\equiv \frac{8\pi^2\ell!}{\left(2\ell+2\right)!!}\sum_{m=1}^{\ell}\left(\sum_{j=-\left(\ell-m\right),2}^{\ell-m}m^{2n_{\phi}-1}j^{2n_{\psi}-1}\text{Re}\left\{\mathcal{Y}^{\ell mj\ast}_{L}\mathcal{Y}^{\ell mj}_{L^{\prime}}\right\}\right) \,, \\
	I_{LL^{\prime}}^{(2n_{\phi}-1,2n_{\psi})} &\equiv \frac{8\pi^2\ell!}{\left(2\ell+2\right)!!}\sum_{m=1}^{\ell}\left(\sum_{j=-\left(\ell-m\right),2}^{\ell-m}m^{2n_{\phi}-1}j^{2n_{\psi}}\text{Im}\left\{\mathcal{Y}^{\ell mj\ast}_{L}\mathcal{Y}^{\ell mj}_{L^{\prime}}\right\}\right) \,, \\
	I_{LL^{\prime}}^{(2n_{\phi},2n_{\psi}-1)} &\equiv \frac{8\pi^2\ell!}{\left(2\ell+2\right)!!}\sum_{m=1}^{\ell}\left(\sum_{j=-\left(\ell-m\right),2}^{\ell-m}m^{2n_{\phi}}j^{2n_{\psi}-1}\text{Im}\left\{\mathcal{Y}^{\ell mj\ast}_{L}\mathcal{Y}^{\ell mj}_{L^{\prime}}\right\}\right) \,,
\ea\ee
\be\ba
	R_{LL^{\prime}}^{(2n_{\phi},2n_{\psi})} = +R_{L^{\prime}L}^{(2n_{\phi},2n_{\psi})} \,,\quad R_{LL^{\prime}}^{(2n_{\phi}-1,2n_{\psi}-1)} = +R_{L^{\prime}L}^{(2n_{\phi}-1,2n_{\psi}-1)} \,, \\
	I_{LL^{\prime}}^{(2n_{\phi}-1,2n_{\psi})} = -I_{L^{\prime}L}^{(2n_{\phi}-1,2n_{\psi})} \,, \quad I_{LL^{\prime}}^{(2n_{\phi},2n_{\psi}-1)} = -I_{L^{\prime}L}^{(2n_{\phi},2n_{\psi}-1)} \,.
\ea\ee

Finally, let us write the conservative/dissipative decomposition of the response coefficients \eqref{eq:RCsConsDissGen} for the current special configuration, which is also the main result of interest of this analysis,
\be\ba\label{eq:RCsConsDiss}
	k_{\ell mj}^{\left(s\right)\text{Love}}\left(\omega\right) &= \text{Re}\left\{ k_{\ell mj}^{\left(s\right)}\left(\omega\right) \right\} \,, \\
	k_{\ell mj}^{\left(s\right)\text{diss}}\left(\omega\right) &= \text{Im}\left\{ k_{\ell mj}^{\left(s\right)}\left(\omega\right) \right\} \,.
\ea\ee
The Love numbers are therefore just the real part of the response coefficients, while the imaginary part encodes all the dissipative effects. We remark here that dissipative effects can survive even in the static limit due to frame dragging~\cite{Chia:2020yla,Charalambous:2021mea,Ivanov:2022hlo}.

\section{Scalar Love numbers of 5-d Myers-Perry black hole}
\label{sec:TLNsComputation}

We will now apply the tools presented in the previous section to compute the static scalar susceptibilities of $5$-dimensional Myers-Perry black holes and identify the corresponding static scalar ($s=0$) Love numbers. We will begin with a description of the equations of motion to be solved. These will be supplemented with boundary conditions as dictated by the $1$-body worldline EFT which motivates the use of the near-zone approximation. Then, this microscopic computation will be matched onto the worldline EFT $1$-point function definition to extract the conservative and dissipative contributions of the static scalar susceptibilities. We will finally analyze various cases of interest associated with the vanishing or the non-vanishing/RG flow of the static scalar Love numbers.

\subsection{Equations of motion and boundary conditions}
The $5$-dimensional Myers-Perry black hole geometry is presented in Appendix \ref{sec:Ap5dMPGeometry}. The massless Klein-Gordon equation in this background is separable,
\be
	\nabla^2 \Phi = \frac{4}{\Sigma}\left[\mathbb{O}_{\text{full}} - \mathbb{P}_{\text{full}}\right]\Phi = 0 \,,
\ee
with the radial and angular operators given by, after introducing the variable $\rho=r^2$,
\be\ba\label{eq:FullEOM}
	\mathbb{O}_{\text{full}} &\equiv \partial_{\rho}\,\Delta\,\partial_{\rho} - \frac{\left(a^2-b^2\right)}{4}\left(\frac{1}{\rho+a^2}\,\partial_{\phi}^2 - \frac{1}{\rho+b^2}\,\partial_{\psi}^2\right)  - \frac{1}{4}\left(\rho+a^2+b^2\right)\partial_{t}^2	\\
	&\quad- \frac{\left(\rho+a^2\right)\left(\rho+b^2\right)r_{s}^2}{4\Delta}\left(\partial_{t} + \frac{a}{\rho+a^2}\,\partial_{\phi} + \frac{b}{\rho+b^2}\,\partial_{\psi}\right)^2 \,, \\
	\mathbb{P}_{\text{full}} &\equiv -\frac{1}{4}\left[ \triangle_{\mathbb{S}^3}^{\left(0\right)} + \left(a^2\sin^2\theta+b^2\cos^2\theta\right)\partial_{t}^2 \right] \,.
\ea\ee
In the above expressions, the discriminant $\Delta$ is a quadratic polynomial in $\rho$,
\be
	\Delta = \left(\rho-\rho_{+}\right)\left(\rho-\rho_{-}\right) \,,
\ee
with $\rho_{\pm}$ the locations of the outer (``$+$'') and inner (``$-$'') horizons and the warp factor $\Sigma$ for the $5$-d Myers-Perry black hole is given by
\be
	\Sigma = r^2 + a^2\cos^2\theta + b^2\sin^2\theta \,.
\ee
We have also identified the scalar ($s=0$) Laplace-Beltrami operator on $\mathbb{S}^3$ in the current direction cosine angular coordinates,
\be
	\triangle_{\mathbb{S}^3}^{\left(0\right)} \equiv \frac{1}{\sin\theta\cos\theta}\partial_{\theta}\left(\sin\theta\cos\theta\,\partial_{\theta}\right) + \frac{1}{\sin^2\theta}\,\partial_{\phi}^2 + \frac{1}{\cos^2\theta}\,\partial_{\psi}^2 \,.
\ee
which is used to define a modified spherical harmonics expansion in the static case from which the Love numbers are more naturally extracted. This expansion is illustrated in detail in Appendix \ref{sec:ApSphericalHarmonics}.

After separating the variables,
\be\label{eq:PhiSepVar}
	\Phi_{\omega\ell m j}\left(t,\rho,\theta,\phi,\psi\right) = e^{-i\omega t}e^{im\phi}e^{ij\psi}R_{\omega\ell m j}\left(\rho\right) S_{\omega\ell m j}\left(\theta\right) \,,
\ee
this $5$-d scalar Teukolsky equation decomposes to
\be\ba
	\mathbb{O}_{\text{full}}\Phi_{\omega\ell m j} &= \hat{\ell}(\hat{\ell}+1)\Phi_{\omega\ell m j} \,, \\
	\mathbb{P}_{\text{full}}\Phi_{\omega\ell m j} &= \hat{\ell}(\hat{\ell}+1)\Phi_{\omega\ell m j} \,,
\ea\ee
with
\be
	\hat{\ell} \equiv \frac{\ell}{d-3} = \frac{\ell}{2} \,,
\ee
and $\ell$ an effective orbital number which is in general non-integer for $\omega\ne0$.

\subsubsection{Near-zone approximation}
In order to match observables onto the $1$-body worldline EFT according to \eqref{eq:1ptNewtonianMatching}, we should solve the massless Klein-Gordon equation in the appropriate regime. The physical setup consists of a binary system of compact bodies during the early stages of their inspiraling phase where a Post-Newtonian expansion is accurate. Centering the body of interest at the origin, the companion sources perturbations with frequency equal to the orbital frequency of the system $\omega=\omega_{\text{orb}}$. The system loses energy by emitting radiation with frequency $\omega_{\text{rad}}\propto \omega_{\text{orb}}$ which is then detected by an observer located at infinity through, for example, an interferometer.

The worldline EFT arises after integrating out the short scale internal degrees of freedom of the centered compact body, i.e. it is valid for low frequency perturbations. Furthermore, the $1$-body worldline EFT ignores the dynamics of the companion body sourcing the perturbations and a second condition for its validity is that the frequency of the perturbations is low with respect to the inverse separation of the two bodies. This combination of conditions defines the \textit{near-zone region}. For the current configuration of a black hole of outer horizon $r_{+}$, the near-zone approximation consists of working in the regime~\cite{Chia:2020yla,Charalambous:2021kcz,Starobinsky:1973aij,Starobinskil:1974nkd,Castro:2010fd,Maldacena:1997ih}
\be\label{eq:NZapprox}
	\omega \left(r-r_{+}\right) \ll 1 \,,\quad \omega r_{+} \ll 1\,.
\ee
In the near-zone region, one imposes the asymptotic boundary condition
\be\label{eq:Asympbc}
	R_{\omega\ell m j} \xrightarrow{r\rightarrow\infty} \bar{\mathcal{E}}_{\ell m j}\left(\omega\right)\,r^{\ell} = \bar{\mathcal{E}}_{\ell m j}\left(\omega\right)\,\rho^{\hat{\ell}} \,,
\ee
indicating the presence of a source at large distances with multipole moments $\bar{\mathcal{E}}_{\ell m j}\left(\omega\right)$, along with the ingoing boundary condition at the event horizon. Ingoing boundary condition at the future/past event horizon is imposed by requiring ingoing waves at the horizon in advanced ($+$)/retarded ($-$) coordinates,
\be
	\Phi_{\omega\ell m j} \xrightarrow{r\rightarrow r_{+}} T_{\ell m j}^{\left(\pm\right)}\left(\omega\right)\,e^{-i\omega t_{\pm}}e^{im\varphi_{\pm}}e^{ijy_{\pm}}S_{\omega\ell m j}\left(\theta\right) \,,
\ee
with $T_{\ell m j}^{\left(\pm\right)}\left(\omega\right)$ the transmission amplitudes. The relation between advanced/retarded coordinates $\left(t_{\pm},r,\theta,\varphi_{\pm},y_{\pm}\right)$ and Boyer-Lindquist coordinates  $\left(t,r,\theta,\phi,\psi\right)$ is given in Appendix \ref{sec:Ap5dMPGeometry} and implies
\be\label{eq:NHbc}
	\begin{gathered}
		R_{\omega\ell m j} \sim \left(\rho-\rho_{+}\right)^{\pm iZ_{+}\left(\omega\right)}\left(\rho-\rho_{-}\right)^{\pm iZ_{-}\left(\omega\right)}\,\,\,\,\,\,\, \text{as $\rho\rightarrow \rho_{+}$} \,, \\
		Z_{\pm}\left(\omega\right) \equiv \pm\frac{r_{\pm}}{2}\frac{\rho_{s}}{\rho_{+}-\rho_{-}}\left(m\Omega_{\phi}^{\left(\pm\right)}+j\Omega_{\psi}^{\left(\pm\right)}-\omega\right) \,,
	\end{gathered}
\ee
where $\Omega_{\phi}^{\left(\pm\right)}=\frac{a}{\rho_{\pm}+a^2}$ and $\Omega_{\psi}^{\left(\pm\right)}=\frac{b}{\rho_{\pm}+b^2}$. Physically, of course, we are only interested in regularity at the future event horizon and $\Omega_{\phi}^{\left(+\right)}\equiv\Omega_{\phi}$ and $\Omega_{\psi}^{\left(+\right)}\equiv\Omega_{\psi}$ are identified as the black hole angular velocities with respect to the two rotation planes.

An important remark here is that the near-zone approximation extends beyond the near-horizon or the low frequency regimes. In particular, not only does it preserve the near-horizon dynamics in the radial operator for any frequency $\omega$, but it also overlaps with the asymptotically flat far-zone region $r\gg r_{+}$ where outgoing boundary conditions are imposed. The overlapping intermediate region $r_{+}\ll r \ll \omega^{-1}$ then serves as a matching region that probes the response of the centered body in the outgoing waves that are detected at infinity.

It should be noted that the near-zone approximation is not unique as there are infinitely many ways to truncate the equations of motion as long as they differ by subleading terms. In practice, the truncation is done such that the equations of motion are exactly solvable in terms of elementary functions. There are two particular near-zone truncations of interest in the current work controlled by a sign $\sigma=\pm $. We split the radial operator as
\be\label{eq:NZRadial1}
	\mathbb{O}_{\text{full}} = \partial_{\rho}\,\Delta\,\partial_{\rho} + V_0^{\left(\sigma\right)} + \epsilon V_1^{\left(\sigma\right)} \,,
\ee
with $\epsilon$ a formal parameter which is equal to unity for the full equations of motion and equal to zero for the near-zone approximation and
\be\ba\label{eq:NZRadial2}
	V_0^{\left(\sigma\right)} = &- \frac{\rho_{s}^2\rho_{+}}{4\Delta}\left(\partial_{t}+\Omega_{\phi}\,\partial_{\phi}+\Omega_{\psi}\,\partial_{\psi}\right)^2 - \frac{a^2-b^2}{4\left(\rho-\rho_{-}\right)}\left(\partial_{\phi}^2-\partial_{\psi}^2\right) \\
	&- \frac{\rho_{s}^2\rho_{+}}{2\left(\rho_{+}-\rho_{-}\right)\left(\rho-\rho_{-}\right)} \left(\Omega_{\phi}-\sigma\,\Omega_{\psi}\right)\partial_{t}\left(\partial_{\phi}-\sigma\,\partial_{\psi}\right) \,,
\ea\ee
\be\ba\label{eq:NZRadial3}
	V_1^{\left(\sigma\right)} &= \frac{\rho_{s}\left(a-\sigma b\right)\left[\rho_{s}-\left(a+\sigma b\right)^2\right]}{4\left(\rho_{+}-\rho_{-}\right)\left(\rho-\rho_{-}\right)}\partial_{t}\left(\partial_{\phi}-\sigma\,\partial_{\psi}\right) \\
	&- \frac{\rho_{s}}{4\left(\rho-\rho_{-}\right)}\left[\rho_{s}\,\partial_{t}+\left(a+\sigma b\right)\left(\partial_{\phi}+\sigma\,\partial_{\psi}\right)\right]\partial_{t} - \frac{1}{4}\left(\rho+a^2+b^2+\rho_{s}\right)\partial_{t}^2 \,.
\ea\ee

For the angular operator, we use the splitting
\be\label{eq:NZAngular}
	\mathbb{P}_{\text{full}} = -\frac{1}{4}\left[ \triangle_{\mathbb{S}^3}^{\left(0\right)} + \epsilon\left(a^2\sin^2\theta+b^2\cos^2\theta\right)\partial_{t}^2 \right] \,,
\ee
that is, we are near-zone approximating it with the static angular operator. In the static limit, the angular problem is solved by $S_{\ell m j}\left(\theta\right)$, given in Appendix \ref{sec:ApSphericalHarmonics}, from which $\ell$ is set to be an orbital number ranging in the set of whole numbers, $\ell\in\mathbb{N}$, $m$ is identified as a spherical harmonics integer azimuthal number $\left|m\right|\le\ell$ and $j$ is a second integer azimuthal number ranging from $-\left(\ell-\left|m\right|\right)$ up to $\ell-\left|m\right|$, but with step $2$.

\subsection{Near-zone solution and scalar Love numbers}
After separating the variables as in \eqref{eq:PhiSepVar} and introducing
\be
	x \equiv \frac{r^2-r_{+}^2}{r_{+}^2-r_{-}^2} = \frac{\rho-\rho_{+}}{\rho_{+}-\rho_{-}} \,,
\ee
the near-zone equation of motion for the radial wavefunction can be massaged into
\be\label{eq:NZRadialX}
	\left[ \frac{d}{dx}\,x\left(1+x\right)\frac{d}{dx} + \frac{Z_{+}^2\left(\omega\right)}{x}-\frac{\tilde{Z}_{-}^{\left(\sigma\right)2}\left(\omega\right)}{1+x} \right] R_{\omega\ell m j} = \hat{\ell}(\hat{\ell}+1)R_{\omega\ell m j} \,,
\ee
with $Z_{+}\left(\omega\right)$ given in \eqref{eq:NHbc} and dictating the near-horizon behavior of the solution and
\be\label{eq:NZZetaMinus}
	\tilde{Z}_{-}^{\left(\sigma\right)}\left(\omega\right) \equiv -\frac{r_{+}}{2}\frac{\rho_{s}}{\rho_{+}-\rho_{-}}\left(m\Omega_{\psi}+j\Omega_{\phi}-\sigma\omega\right) \,.
\ee
We note in particular that $\tilde{Z}_{-}^{\left(\sigma\right)}\left(\omega=0\right)=Z_{-}\left(\omega=0\right)$ in \eqref{eq:NHbc} reflecting how the near-zone approximation becomes exact in the static limit. The above differential equation has three regular singular points at $x=0$, $x=-1$ and $x\rightarrow\infty$ with the characteristic exponents given in Table~\ref{tbl:NZRadialExponents}.

\begin{table}[!t]
	\centering
	\begin{tabular}{|c||c|c|c|}
        \hline
		Singular point & $x=0$ & $x=-1$ & $x\rightarrow\infty$ \\
		\hline
		Characteristic exponent 1 & $+iZ_{+}\left(\omega\right)$ & $+i\tilde{Z}_{-}^{\left(\sigma\right)}\left(\omega\right)$ & $\hat{\ell}$ \\
		\hline
		Characteristic exponent 2 & $-iZ_{+}\left(\omega\right)$ & $-i\tilde{Z}_{-}^{\left(\sigma\right)}\left(\omega\right)$ & $-\hat{\ell}-1$ \\
        \hline
	\end{tabular}
    \caption{Characteristic exponents of the near-zone-truncated radial equation of motion, Eq.~\eqref{eq:NZRadialX}, for scalar perturbations of the $5$-d Myers-Perry black hole. These are expressed in terms of the rescaled orbital number $\hat{\ell}=\ell/2$ and the parameters $Z_{+}\left(\omega\right)$ and $\tilde{Z}_{-}^{\left(\sigma\right)}\left(\omega\right)$, given in Eq.~\eqref{eq:NHbc} and Eq.~\eqref{eq:NZZetaMinus} respectively.}
    \label{tbl:NZRadialExponents}
\end{table}

The differential equation can be solved analytically in terms of Euler's hypergeometric functions. For future convenience, we introduce the parameters
\be\ba\label{eq:GammaPlusMinus}
	\Gamma^{\left(\sigma\right)}_{\pm\sigma}\left(\omega\right) &\equiv Z_{+}\left(\omega\right) \mp \sigma\, \tilde{Z}_{-}^{\left(\sigma\right)}\left(\omega\right) \\
	&= \frac{r_{+}}{2}\frac{\rho_{s}}{\rho_{+}-\rho_{-}}\left[\left(m\pm \sigma j\right)\left(\Omega_{\phi}\pm\sigma\Omega_{\psi}\right)-\omega\left(1\pm1\right)\right] \,.
\ea\ee
The solution that is regular at the future event horizon then reads
\be\ba\label{eq:NZRadialSolution}
	{}&R_{\omega\ell m j} = \bar{R}_{\ell m j}\left(\omega\right)\,\left(\frac{x}{1+x}\right)^{iZ_{+}\left(\omega\right)} \\
	&\left(1+x\right)^{i\Gamma_{\pm\sigma}^{\left(\sigma\right)}\left(\omega\right)}\,{}_2F_1\left(\hat{\ell}+1+i\Gamma^{\left(\sigma\right)}_{\pm\sigma}\left(\omega\right),-\hat{\ell}+i\Gamma^{\left(\sigma\right)}_{\pm\sigma}\left(\omega\right);1+2iZ_{+}\left(\omega\right);-x\right) \,,
\ea\ee
where $\bar{R}_{\ell m j}\left(\omega\right)$ is fixed from the asymptotic boundary condition \eqref{eq:Asympbc} to be proportional to the source moments harmonic modes $\bar{\mathcal{E}}_{\ell mj}\left(\omega\right)$ according to (see Appendix~\ref{sec:ApSchwarzschildLimit} for more details)
\be
	\bar{R}_{\ell mj}\left(\omega\right) = \bar{\mathcal{E}}_{\ell mj}\left(\omega\right) \left(\rho_{+}-\rho_{-}\right)^{\hat{\ell}}\frac{\Gamma\left(\hat{\ell}+1+i\Gamma^{\left(\sigma\right)}_{+\sigma}\left(\omega\right)\right)\Gamma\left(\hat{\ell}+1+i\Gamma^{\left(\sigma\right)}_{-\sigma}\right)}{\Gamma\left(2\hat{\ell}+1\right)\Gamma\left(1+2iZ_{+}\left(\omega\right)\right)} \,.
\ee

Up to this point in solving the static problem, we only assumed that the orbital number $\ell$ is a generic real number. 
This prescription is known
as ``analytic continuation.''
It is often used for the
Newtonian matching within the worldline EFT from which the Love numbers are defined~\cite{Kol:2011vg,Chia:2020yla,Charalambous:2021mea,Ivanov:2022hlo}, 
see also~\cite{Starobinsky:1973aij,Starobinskil:1974nkd} for the first practical 
use of this approach. Namely, in order to extract the response coefficients from the above microscopic computation, one should first analytically continue $\ell$ to be a real number, then expand around large $r$ to read the coefficient in front of the $r^{-\ell-2}$ term and then send $\ell$ to its physical integer values at the end. Doing this, we find, before sending $\ell$ to take its physical values,
\be\label{eq:NZSLNs}
	k_{\ell m j}\left(\omega\right) = \frac{\Gamma\left(-2\hat{\ell}-1\right)\Gamma\left(\hat{\ell}+1+i\Gamma^{\left(\sigma\right)}_{+\sigma}\left(\omega\right)\right)\Gamma\left(\hat{\ell}+1+i\Gamma^{\left(\sigma\right)}_{-\sigma}\right)}{\Gamma\left(2\hat{\ell}+1\right)\Gamma\left(-\hat{\ell}+i\Gamma^{\left(\sigma\right)}_{+\sigma}\left(\omega\right)\right)\Gamma\left(-\hat{\ell}+i\Gamma^{\left(\sigma\right)}_{-\sigma}\right)}\left(\frac{\rho_{+}-\rho_{-}}{\rho_{s}}\right)^{2\hat{\ell}+1} \,,
\ee
which can be massaged using the mirror formula for the $\Gamma$-functions into the more transparent result
\be\ba
	k_{\ell m j}&\left(\omega\right) = A_{\ell m j}\left(\omega\right)\times\bigg\{- i\sinh2\pi Z_{+}\left(\omega\right) \\	&+\tan\pi\hat{\ell}\cosh\pi\Gamma^{\left(\sigma\right)}_{+\sigma}\left(\omega\right)\cosh\pi\Gamma^{\left(\sigma\right)}_{-\sigma} - \cot\pi\hat{\ell}\sinh\pi\Gamma^{\left(\sigma\right)}_{+\sigma}\left(\omega\right)\sinh\pi\Gamma^{\left(\sigma\right)}_{-\sigma} \bigg\} \,,
\ea\ee
where $A_{\ell m j}\left(\omega\right)$ is a real constant given by
\be
	A_{\ell m j}\left(\omega\right) \equiv \frac{\left|\Gamma\left(\hat{\ell}+1+i\Gamma^{\left(\sigma\right)}_{+\sigma}\left(\omega\right)\right)\right|^2\left|\Gamma\left(\hat{\ell}+1+i\Gamma^{\left(\sigma\right)}_{-\sigma}\right)\right|^2}{2\pi\,\Gamma\left(2\hat{\ell}+1\right)\Gamma\left(2\hat{\ell}+2\right)}\left(\frac{\rho_{+}-\rho_{-}}{\rho_{s}}\right)^{2\hat{\ell}+1} \,.
\ee
The conservative/dissipative decomposition \eqref{eq:RCsConsDiss} then implies
\be\label{eq:5dMPRCsDiss}
	k_{\ell mj}^{\text{diss}}\left(\omega\right) = -A_{\ell mj}\left(\omega\right)\sinh2\pi Z_{+}\left(\omega\right) \,,
\ee
\be\ba\label{eq:5dMPLove}
	k_{\ell mj}^{\text{Love}}\left(\omega\right) = A_{\ell mj}\left(\omega\right)\times&\bigg\{\tan\pi\hat{\ell}\cosh\pi\Gamma^{\left(\sigma\right)}_{+\sigma}\left(\omega\right)\cosh\pi\Gamma^{\left(\sigma\right)}_{-\sigma} \\
	&- \cot\pi\hat{\ell}\sinh\pi\Gamma^{\left(\sigma\right)}_{+\sigma}\left(\omega\right)\sinh\pi\Gamma^{\left(\sigma\right)}_{-\sigma} \bigg\} \,.
\ea\ee
At this point, we would like stress out that the above results should be trusted only for small values of $\omega$. Indeed, the near-zone approximation is accurate only for low frequencies. In particular, in the near-zone split \eqref{eq:NZRadial1}-\eqref{eq:NZAngular}, we are already approximating at order $\mathcal{O}\left(\omega a,\omega b,\omega^2\right)$~\cite{Charalambous:2022rre}. Nevertheless, the above result is accurate in the static limit which is also the case of interest. In the rest of this work, we will be using the above $\omega$-dependent expressions but it should always be kept in mind that they are accurate only in the static limit for non-zero spin parameters.
The Love numbers behaviors following from Eq.~\eqref{eq:5dMPRCsDiss}
are listed in table~\ref{tbl:StaticSLNs}
and in table~\ref{tbl:WSLNs}
for static and time-dependent cases, 
respectively. Let us discuss the 
Love number phenomenology in detail.

\begin{table}[!t]
	\centering
	\begin{tabular}{|l|l||l|}
		\hline
		\multicolumn{2}{|c||}{Range of parameters} & Behavior of $k^{\text{Love}}_{\ell mj}\left(\omega=0\right)$ \\
		\hline\hline
		\multicolumn{2}{|c||}{$\ell\in2\mathbb{N}+1$} & \multicolumn{1}{c|}{Running} \\
		\hline
		\multirow{2}{*}{$\ell\in2\mathbb{N}$} & \multicolumn{1}{c||}{$\left|a\right|=\left|b\right|$ OR $\left|m\right|=\left|j\right|$} & \multicolumn{1}{c|}{Vanishing} \\
		\cline{2-3}
		& \multicolumn{1}{c||}{$\left|a\right|\ne\left|b\right|$ AND $\left|m\right|\ne\left|j\right|$} & \multicolumn{1}{c|}{Running} \\
		\hline
	\end{tabular}
	\caption{Behavior of static scalar Love numbers as a function of the $5$-d Myers-Perry black hole angular momenta $a$ and $b$ and the scalar field perturbation orbital number $\ell$ and azimuthal numbers $m$ and $j$. For generic angular momenta and azimuthal and orbital numbers, the static scalar Love numbers exhibit a classical RG flow. The only exception is when the orbital number is even ($\hat{\ell}$ is integer) and the angular momenta or the azimuthal numbers are equal in magnitude in which case the static Love numbers turn out to vanish.
	}
	\label{tbl:StaticSLNs}
\end{table}

\begin{table}[!t]
	\centering
	\begin{tabular}{|l|l||l|}
		\hline
		\multicolumn{2}{|c||}{Range of parameters} & Behavior of $k^{\text{Love}}_{\ell mj}\left(\omega\ne0\right)$ \\
		\hline\hline
		\multicolumn{2}{|c||}{\Gape[5pt]{$i\Gamma^{\left(\sigma\right)}_{\pm}\left(\omega\right)-\hat{\ell} \notin \mathbb{Z}$}} & \multicolumn{1}{c|}{Running} \\
		\hline
		\multirow{2}{*}{$\begin{matrix} i\Gamma^{\left(\sigma\right)}_{+}\left(\omega\right)-\hat{\ell} = k\in\mathbb{Z} \\ \text{OR} \\ i\Gamma^{\left(\sigma\right)}_{-}\left(\omega\right)-\hat{\ell} = k\in\mathbb{Z} \end{matrix}$} & \multicolumn{1}{c||}{\Gape[9pt]{$-\ell \le k \le 0$}} & \multicolumn{1}{c|}{Vanishing} \\
		\cline{2-3}
		& \multicolumn{1}{c||}{\Gape[10pt]{$k>0$ OR $k<-\ell$}} & \multicolumn{1}{c|}{Non-running and non-vanishing} \\
		\hline
	\end{tabular}
	\caption{Behavior of $\omega$-dependent near-zone scalar Love numbers as a function of the parameters $\Gamma^{\left(\sigma\right)}_{\pm}\left(\omega\right)$, given in \eqref{eq:GammaPlusMinus}, and the generalized orbital number $\hat{\ell}=\frac{\ell}{2}$ of the perturbation. For generic angular momenta and azimuthal and orbital numbers, the static scalar Love numbers exhibit a classical RG flow. However, there is a discrete series of imaginary-valued $\Gamma^{\left(\sigma\right)}_{\pm}$'s for which the near-zone Love numbers do not run. For the $\Gamma^{\left(\sigma\right)}_{-\sigma}$ branch, these are unphysical, accompanied by conical singularities in the scalar field profile. For the $\Gamma^{\left(\sigma\right)}_{+\sigma}\left(\omega\right)$ branch, the $-\ell\le k<0$ modes acquire the interpretation of Total Transmission Modes.
 }
	\label{tbl:WSLNs}
\end{table}

\subsection{Running Love}
We begin analyzing our above result for the scalar Love numbers by first addressing the divergent behavior
associated with RG running. 
For general non-zero spin parameters $a$ and $b$ and general azimuthal numbers $m$ and $j$ and frequency $\omega$, i.e. for general non-zero $\Gamma^{\left(\sigma\right)}_{\pm\sigma}\left(\omega\right)$, the scalar Love numbers \textit{always} diverge, either as $\cot\pi\hat{\ell}$ for integer $\hat{\ell}$ (even $\ell$) or as $\tan\pi\hat{\ell}$ for half-integer $\hat{\ell}$ (odd $\ell$). More specifically, in the limit $\varepsilon\rightarrow0$ where $2\hat{\ell}=n-\varepsilon$ approaches a whole number $n\in\mathbb{N}$, the response coefficients \eqref{eq:NZSLNs} develop a simple pole due to the diverging $\Gamma(-2\hat{\ell}-1)$. From the residue of the $\Gamma$-function near negative integers, $\Gamma\left(-n+\varepsilon\right) = \frac{\left(-1\right)^{n}}{n!\,\varepsilon} + \mathcal{O}\left(\varepsilon^0\right)$, the developed pole can be worked out to be
\be\label{eq:kRes}
	k_{\ell m j}\left(\omega\right) = -\frac{\left(-1\right)^{n}}{n!\,\varepsilon}\frac{\Gamma\left(\frac{n}{2}+1+i\Gamma^{\left(\sigma\right)}_{+\sigma}\left(\omega\right)\right)\Gamma\left(\frac{n}{2}+1+i\Gamma^{\left(\sigma\right)}_{-\sigma}\right)}{\Gamma\left(-\frac{n}{2}+i\Gamma^{\left(\sigma\right)}_{+\sigma}\left(\omega\right)\right)\Gamma\left(-\frac{n}{2}+i\Gamma^{\left(\sigma\right)}_{-\sigma}\right)\left(n+1\right)!}\left(\frac{\rho_{+}-\rho_{-}}{\rho_{s}}\right)^{n+1} + \mathcal{O}\left(\varepsilon^{0}\right) \,.
\ee

The full solution, however, is regular due to a compensating divergence in the ``source'' part of the scalar field profile. More specifically, as is illustrated in detail in Appendix \ref{sec:ApSchwarzschildLimit}, the source/response split is performed prior to sending the orbital number to take its physical values with the end result
\be\ba\label{eq:SourceReponseSplit}
	R_{\omega\ell m j}\left(\rho\right) &= \bar{\mathcal{E}}_{\ell m j}\left(\omega\right)\,\rho^{\hat{\ell}}\left[Z_{\omega\ell m j}^{\text{source}}\left(\rho\right) + k_{\ell m j}\left(\omega\right)\,\left(\frac{\rho_{s}}{\rho}\right)^{2\hat{\ell}+1} Z_{\omega\ell m j}^{\text{response}}\left(\rho\right)\right] \,, \\
	Z_{\omega\ell m j}^{\text{source}}\left(\rho\right) &= \left(1-\frac{\rho_{+}}{\rho}\right)^{\hat{\ell}} \left(\frac{\rho-\rho_{-}}{\rho-\rho_{+}}\right)^{\mp i\sigma\tilde{Z}_{-}^{\left(\sigma\right)}\left(\omega\right)} \\
	&\times\,{}_2F_1\left(-\hat{\ell}+i\Gamma^{\left(\sigma\right)}_{\mp\sigma}\left(\omega\right),-\hat{\ell}-i\Gamma^{\left(\sigma\right)}_{\pm\sigma}\left(\omega\right);-2\hat{\ell};\frac{\rho_{+}-\rho_{-}}{\rho_{+}-\rho}\right) \,, \\
	Z_{\omega\ell m j}^{\text{response}}\left(\rho\right) &= \left(1-\frac{\rho_{+}}{\rho}\right)^{-\hat{\ell}-1} \left(\frac{\rho-\rho_{-}}{\rho-\rho_{+}}\right)^{\mp i\sigma\tilde{Z}_{-}^{\left(\sigma\right)}\left(\omega\right)} \\
	&\times\,{}_2F_1\left(\hat{\ell}+1+i\Gamma^{\left(\sigma\right)}_{\mp\sigma}\left(\omega\right),\hat{\ell}+1-i\Gamma^{\left(\sigma\right)}_{\pm\sigma}\left(\omega\right);2\hat{\ell}+2;\frac{\rho_{+}-\rho_{-}}{\rho_{+}-\rho}\right) \,.
\ea\ee
In the limit $\varepsilon\rightarrow0$ where $2\hat{\ell}=n-\varepsilon$ approaches a whole number $n\in\mathbb{N}$, $\rho^{\hat{\ell}}Z_{\omega\ell m j}^{\text{source}}$ also develops a simple pole. The diverging component of the ``source'' part of the solution can be obtained from the following residue formula for the hypergeometric function
\be\ba
	{}_2F_1\left(a,b;-n+\varepsilon;z\right) &= \Gamma\left(-n+\varepsilon\right)\frac{\Gamma\left(a+n+1\right)\Gamma\left(b+n+1\right)}{\Gamma\left(a\right)\Gamma\left(b\right)\left(n+1\right)!} \\
	&\times z^{n+1}{}_2F_1\left(a+n+1,b+n+1;n+2;z\right) + \mathcal{O}\left(\varepsilon^0\right) \,,
\ea\ee
and, consequently,
\be\ba
	Z_{\omega\ell m j}^{\text{source}} &= -\frac{1}{n!\,\varepsilon}\frac{\Gamma\left(\frac{n}{2}+1+i\Gamma^{\left(\sigma\right)}_{+\sigma}\left(\omega\right)\right)\Gamma\left(\frac{n}{2}+1-i\Gamma^{\left(\sigma\right)}_{-\sigma}\right)}{\Gamma\left(-\frac{n}{2}+i\Gamma^{\left(\sigma\right)}_{+\sigma}\left(\omega\right)\right)\Gamma\left(-\frac{n}{2}-i\Gamma^{\left(\sigma\right)}_{-\sigma}\right)\left(n+1\right)!}\left(\frac{\rho_{+}-\rho_{-}}{\rho_{s}}\right)^{n+1} \\
	&\times \,\left(\frac{\rho_{s}}{\rho}\right)^{n+1}Z_{\omega\ell m j}^{\text{response}} + \mathcal{O}\left(\varepsilon^0\right) \,,
\ea\ee
which exactly cancels with the divergence in the scalar Love numbers whenever $n$ is an integer.

From the EFT point of view, this diverging behavior of the scalar Love numbers is interpreted as a classical RG flow. More specifically, power counting arguments reveal that whenever $2\hat{\ell}\in\mathbb{N}$, the Wilson coefficients defining the Love numbers get renormalized from an overlapping with the ``source'' part of the $1$-point function, namely, from the following type of diagrams~\cite{Kol:2011vg,Charalambous:2022rre,Ivanov:2022hlo}
\be
	\Phi \supset \vcenter{\hbox{\begin{tikzpicture}
			\begin{feynman}
				\vertex[dot] (a0);
				\vertex[below=1cm of a0] (p1);
				\vertex[above=1cm of a0] (p2);
				\vertex[right=0.4cm of a0, blob] (gblob){};							
				\vertex[right=1.5cm of p1] (b1);
				\vertex[right=1.5cm of p2] (b2);
				\vertex[right=1.19cm of p2] (b22){$\times$};
				\vertex[above=0.7cm of a0] (g1);
				\vertex[above=0.4cm of a0] (g2);
				\vertex[right=0.05cm of a0] (gdtos){$\vdots$};
				\vertex[below=0.7cm of a0] (gN);
				\vertex[left=0.2cm of gN] (gN1);
				\vertex[left=0.2cm of g1] (g11);
				\diagram*{
					(p1) -- [double,double distance=0.5ex] (p2),
					(g1) -- [photon] (gblob),
					(g2) -- [photon] (gblob),
					(gN) -- [photon] (gblob),
					(b1) -- (gblob) -- (b2),
				};
				\draw [decoration={brace}, decorate] (gN1.south west) -- (g11.north west)
				node [pos=0.55, left] {\(2\hat{\ell}+1\)};
			\end{feynman}
	\end{tikzpicture}}} \,.
\ee
From the theory of differential equations point of view, whenever $2\hat{\ell}\in\mathbb{N}$, the series solutions of the radial differential equation fall into the degenerate case where the characteristic exponents near $x\rightarrow\infty$ differ by an integer number and, thus, according to Fuchs' theorem, only one of the independent solutions can be written as a Frobenius series around there, while the second independent solution will unavoidably contain logarithms. More explicitly, the solution regular at the future event horizon is still given by \eqref{eq:NZRadialSolution}, but its analytic continuation at large distances must be taken as a limiting case with the end result being
\be\ba
	{}&R_{\omega\ell m j} = \bar{\mathcal{E}}_{\ell m j}\left(\omega\right) \rho_{s}^{\hat{\ell}}\left(\frac{\rho-\rho_{-}}{\rho-\rho_{+}}\right)^{i\tilde{Z}_{-}^{\left(\sigma\right)}\left(\omega\right)} \\
	&\times\bigg\{ \left(\frac{\rho-\rho_{+}}{\rho_{s}}\right)^{\hat{\ell}}\sum_{k=0}^{\ell}\frac{\left(-\hat{\ell}+i\Gamma^{\left(\sigma\right)}_{+\sigma}\left(\omega\right)\right)_{k}}{\left(-k+\hat{\ell}+1+i\Gamma^{\left(\sigma\right)}_{-\sigma}\right)_{k}}\frac{\left(\ell-k\right)!}{\ell!\,k!}\left(-x\right)^{-k} \\
	&\,\,\,+\underset{\varepsilon\rightarrow0}{\text{Res}}\left\{k_{\ell-\varepsilon, m j}\left(\omega\right)\right\} \left(\frac{\rho-\rho_{+}}{\rho_{s}}\right)^{-\hat{\ell}-1}\sum_{k=0}^{\infty}\frac{\left(\hat{\ell}+1+i\Gamma^{\left(\sigma\right)}_{+\sigma}\left(\omega\right)\right)_{k}}{\left(-k-\hat{\ell}+i\Gamma^{\left(\sigma\right)}_{-\sigma}\right)_{k}}\frac{\left(\ell+1\right)!}{\left(k+\ell+1\right)!\,k!}\left(+x\right)^{-k} \\
	{}&\,\,\,\times\bigg[\log x + \psi\left(k+1\right) + \psi\left(k+\ell+2\right) - \psi\left(k+\hat{\ell}+1+i\Gamma^{\left(\sigma\right)}_{+\sigma}\left(\omega\right)\right) - \psi\left(-k-\hat{\ell}+i\Gamma^{\left(\sigma\right)}_{-\sigma}\right)\bigg] \bigg\} \,,
\ea\ee
where we have identified the coefficient multiplying the second term as the residue \eqref{eq:kRes} of the response coefficients as $\ell$ approaches a whole number. From the EFT point of view, this residue is precisely the $\beta$-function dictating the classical RG flow of the Love numbers~\cite{Kol:2011vg,Ivanov:2022hlo,Charalambous:2022rre},
\be
	L\frac{d k_{\ell mj}}{dL} = -\frac{\left(-1\right)^{\ell}}{\ell!}\frac{\Gamma\left(\frac{\ell}{2}+1+i\Gamma^{\left(\sigma\right)}_{+\sigma}\left(\omega\right)\right)\Gamma\left(\frac{\ell}{2}+1+i\Gamma^{\left(\sigma\right)}_{-\sigma}\right)}{\Gamma\left(-\frac{\ell}{2}+i\Gamma^{\left(\sigma\right)}_{+\sigma}\left(\omega\right)\right)\Gamma\left(-\frac{\ell}{2}+i\Gamma^{\left(\sigma\right)}_{-\sigma}\right)\left(\ell+1\right)!}\left(\frac{\rho_{+}-\rho_{-}}{\rho_{s}}\right)^{\ell+1} \,.
\ee
This $\beta$-function is evidently real and is therefore entirely associated with the running of the Love numbers, while the dissipative response exhibits no RG flow.

\subsection{Vanishing static Love}
We now turn to the possible resonant conditions for which the $\beta$-function associated with the near-zone Love numbers is zero. Let us start with the case of static scalar Love numbers, which is after all the only regime within which the near-zone approximation is accurate for generic spin parameters. The static scalar Love numbers vanish only if $\hat{\ell}\in\mathbb{N}$ ($\ell$ is an even integer) and
\be\label{eq:VanishingLove_IntL_Static}
	\begin{gathered}
		\Gamma^{\left(\sigma\right)}_{+\sigma}\left(\omega=0\right) = 0 \Rightarrow \frac{j+\sigma m}{2}\left(\Omega_{\psi}+\sigma\Omega_{\phi}\right) = 0 \,, \\
		\text{OR} \\
		\Gamma^{\left(\sigma\right)}_{-\sigma} = 0 \Rightarrow \frac{j-\sigma m}{2}\left(\Omega_{\psi}-\sigma\Omega_{\phi}\right) = 0 \,.
	\end{gathered}
\ee
In other words, these conditions 
are satisfied if either $\left|a\right|=\left|b\right|$ or $\left|m\right|=\left|j\right|$. The first case describes an equi-rotating Myers-Perry black hole in $5$ spacetime dimensions, which has the property of being equipped with an enhanced isometry group $U\left(1\right) \times U\left(1\right) \rightarrow U\left(2\right)$, while the second case can be regarded as a higher-dimensional generalization of ``axisymmetric'' perturbations~\cite{Pani:2015hfa,Gurlebeck:2015xpa,LeTiec:2020bos} that actually includes
here non-axisymmetric cases ($m,j\ne0$). This situation is also similar to the higher-dimensional Schwarzschild black holes~\cite{Kol:2011vg}: for integer $\hat{\ell}$, the static scalar Love numbers vanish, while for half-integer $\hat{\ell}$ they are non-zero and exhibit a classical RG flow discussed above. The corresponding static scalar field profile for $\hat{\ell}\in\mathbb{N}$ is given by
\be\ba\label{eq:StaticRadialVanishingLove}
	R_{\omega=0,\ell m j}\bigg|_{\Gamma^{\left(\sigma\right)}_{\pm\sigma}=0} &= \bar{R}_{\ell m j}\left(\omega=0\right)\,\left(\frac{x}{1+x}\right)^{i\Gamma^{\left(\sigma\right)}_{\mp\sigma}/2} \,{}_2F_1\left(\hat{\ell}+1,-\hat{\ell};1+i\Gamma^{\left(\sigma\right)}_{\mp\sigma};-x\right) \\
	&= \bar{R}_{\ell m j}\left(\omega=0\right)\,\left(\frac{x}{1+x}\right)^{i\Gamma^{\left(\sigma\right)}_{\mp\sigma}/2} \sum_{n=0}^{\hat{\ell}} \frac{\left(\hat{\ell}+n\right)!}{\left(\hat{\ell}-n\right)!} \frac{1}{\left(1+i\Gamma^{\left(\sigma\right)}_{\mp\sigma}\right)_{n}}\frac{x^{n}}{n!} \,,
\ea\ee
where we have used the polynomial form of the hypergeometric function whose one of first two parameters is a negative integer and $\left(a\right)_{n}$ is the Pochhammer symbol.

The vanishing of static Love numbers raises naturalness concerns from the point of view of the worldline EFT~\cite{tHooft:1979rat,Porto:2016zng}. In the absence of any selection rules imposed by an enhanced symmetry structure, the Love numbers are expected to be $\mathcal{O}\left(1\right)$ numbers and exhibit running. In contrast, we find here situations where the static Love numbers vanish at all scales and call upon a symmetry explanation to be presented in the next section. Related to this, for generic real values of the orbital number, the ``source'' and ``response'' parts are given by two infinite series in inverse powers of $\rho$, see Appendix \ref{sec:ApSchwarzschildLimit}. These two series overlap when $\ell$ takes its physical integer values. When the static Love numbers vanish, the final result after summing these two series is the quasi-polynomial form shown above. However, the ``source'' and ``response'' parts are still given by two infinite series which now conspire to give the quasi-polynomial radial solution. In particular, it is these infinite cancellations resulting in the quasi-polynomial form that need to be addressed by a symmetry argument.

\subsection{Non-running Love}
It is also interesting to investigate other situations where $k_{\ell mj}^{\text{Love}}\left(\omega\right)$ 
are fine tuned. 
For even $\ell$ (integer $\hat{\ell}$), we find that the scalar Love numbers \eqref{eq:NZSLNs} exhibit no RG flow under the conditions
\be\label{eq:VanishingLove_IntL}
	\begin{gathered}
		i\Gamma^{\left(\sigma\right)}_{+\sigma}\left(\omega\right) \in \mathbb{Z} \Rightarrow \omega_{k} = \frac{j+\sigma m}{2}\left(\Omega_{\psi}+\sigma\Omega_{\phi}\right) + \frac{i}{\beta}k \,, \\
		\text{OR} \\
		i\Gamma^{\left(\sigma\right)}_{-\sigma} \in \mathbb{Z} \Rightarrow \frac{j-\sigma m}{2}\left(\Omega_{\psi}-\sigma\Omega_{\phi}\right) = -\frac{i}{\beta}k \,,
	\end{gathered}
\ee
where $k\in\mathbb{Z}$ and we have introduced the inverse Hawking temperature of the $5$-dimensional Myers-Perry black hole,
\be\label{eq:betaMP}
	\beta = \frac{1}{2\pi T_{H}} = \frac{\rho_{s}r_{+}}{\rho_{+}-\rho_{-}} \,.
\ee
In particular, for $-\hat{\ell}\le k\le\hat{\ell}$ the scalar Love numbers vanish identically, while for $k\ge\hat{\ell}+1$ or $k\le -\hat{\ell}-1$, they are non-zero but still exhibit no running.

Similarly, for odd $\ell$ (half-integer $\hat{\ell}$), the conditions read
\be\label{eq:VanishingLove_HIntL}
	\begin{gathered}
		i\Gamma^{\left(\sigma\right)}_{+\sigma}\left(\omega\right) \in \mathbb{Z} + \frac{1}{2} \Rightarrow \omega_{k} = \frac{j+\sigma m}{2}\left(\Omega_{\psi}+\sigma\Omega_{\phi}\right) + \frac{i}{\beta}\left(k+\frac{1}{2}\right) \,, \\
		\text{OR} \\
		i\Gamma^{\left(\sigma\right)}_{-\sigma} \in \mathbb{Z} + \frac{1}{2} \Rightarrow \frac{j-\sigma m}{2}\left(\Omega_{\psi}-\sigma\Omega_{\phi}\right) = -\frac{i}{\beta}\left(k+\frac{1}{2}\right) \,,
	\end{gathered}
\ee
with vanishing Love numbers whenever $-\hat{\ell}\le k+\frac{1}{2}\le \hat{\ell}$.

Regarding the conditions on $\Gamma^{\left(\sigma\right)}_{-\sigma}$, these are in general accompanied by conical deficits in the scalar field profile because they imply imaginary azimuthal numbers which break the periodicity of the scalar field with respect to azimuthal rotations. The only situation where this does not happen is when $\Gamma^{\left(\sigma\right)}_{-\sigma}=0$ for $\hat{\ell}\in\mathbb{N}$.

Despite their unphysical
nature in certain cases, the vanishing/non-running of scalar Love numbers beyond the static limit and for all the situations in \eqref{eq:VanishingLove_IntL}-\eqref{eq:VanishingLove_HIntL} is still an interesting result. The corresponding near-zone radial wavefunction for integer $\hat{\ell}$ takes the form
\be\ba\label{eq:NZRadialVanishingLove}
	R_{\omega\ell m j}\bigg|_{i\Gamma^{\left(\sigma\right)}_{\pm\sigma}\left(\omega\right)=k} &= \bar{R}_{\ell m j}\left(\omega\right)\,\left(\frac{x}{1+x}\right)^{i\Gamma^{\left(\sigma\right)}_{\mp\sigma}\left(\omega\right)} \\
	&\left[x\left(1+x\right)\right]^{k/2}\,{}_2F_1\left(\hat{\ell}+1+k,-\hat{\ell}+k;1+k+i\Gamma^{\left(\sigma\right)}_{\mp\sigma}\left(\omega\right);-x\right) \,.
\ea\ee
For $k\le\hat{\ell}$, this again takes a particular quasi-polynomial form to be addressed in the next section via symmetry arguments as well. As we will see, it is the highest-weight property that dictates this quasi-polynomial form. Nevertheless, for $k<-\hat{\ell}$ the polynomial starts developing $\rho^{-\hat{\ell}-1}$ terms. The absence of logarithms indicates that this is not due to an overlapping with PN corrections to the Newtonian source and therefore these are interpreted as non-vanishing and non-running Love numbers.

As we will see in the next section, the highest-weight representation relevant for the properties of the near-zone Love numbers is actually an indecomposable $\SL$ representation of type ``$\circ[\circ[\circ$''. The states ``sandwiched'' between the two highest-weight modes are the ones for which the near-zone Love numbers vanish, while the states spanning the irreducible (lower) highest-weight representation have non-vanishing and non-running near-zone Love numbers. As for the vectors spanning the upper ladder above the reducible (higher) highest-weight representation, these will be spanned by the states characterized by the resonant conditions with $k\ge\hat{\ell}+1$. For half-integer $\hat{\ell}$ for which $i\Gamma^{\left(\sigma\right)}_{+\sigma}\left(\omega\right)$ or $i\Gamma^{\left(\sigma\right)}_{-\sigma}$ is a half-integer, the same conclusions are drawn after replacing $k\rightarrow k+\frac{1}{2}$.

\section{Love symmetries}
\label{sec:SL2R}

In~\cite{Charalambous:2021kcz,Charalambous:2022rre}, we have demonstrated the emergence of an enhanced $\SL$ symmetry, dubbed ``Love symmetry'', of the near-zone equations of motion for black hole perturbations. 
In the context of this symmetry, the vanishing of Love numbers appears  
as a constraint imposed 
by 
the highest-weight 
Love symmetry representation 
structure, to which the relevant 
 black hole perturbation solutions belong.
In particular, the highest-weight property dictates a (quasi-)polynomial form of the regular radial wavefunctions which is the behavior indicative of the vanishing of the Love numbers. The intricate structure of the scalar Love numbers for $5$-d Myers-Perry black holes extracted in the previous section sets a good example of examining this hypothesis. In this section, we will demonstrate the existence of $\SL$ structures of the near-zone Klein-Gordon equation for all values of the spin parameters and azimuthal numbers. As we will see, it is only when the scalar Love numbers vanish that the corresponding regular scalar field solution belongs to a highest-weight representation of the corresponding $\SL$ and the vanishing of the scalar Love numbers will be immediately implied through the highest-weight property.

To demonstrate the enhanced symmetry structure of the near-zone equations of motion, it is convenient to introduce the following sum/difference azimuthal angles,
\be
	\psi_{\pm}=\psi\pm\phi \,. 
\ee
The corresponding angular velocities and azimuthal numbers with respect to these two directions are then
\be
	\Omega_{\pm} = \Omega_{\psi} \pm \Omega_{\phi} \,,\quad m_{\pm}= \frac{j\pm m}{2} \,.
\ee

The two near-zone splits \eqref{eq:NZRadial1}-\eqref{eq:NZRadial2} have the property of each being equipped with an $\SL$ structure. Indeed, for each sign $\sigma=\pm$, we can find a set of three vector fields satisfying the $\SL$ algebra,
\be
	\left[L_{m}^{\left(\sigma\right)},L_{n}^{\left(\sigma\right)}\right] = \left(m-n\right)L_{m+n}^{\left(\sigma\right)} \,,\quad m,n=0,\pm1 \,.
\ee
These generators are given by the vector fields
\be\label{eq:LoveGen}
	\begin{gathered}
		L_0^{\left(\sigma\right)} = -\beta\left(\partial_{t} + \Omega_{\sigma}\,\partial_{\sigma}\right) \,, \\
		L_{\pm1}^{\left(\sigma\right)} = e^{\pm t/\beta}\left[\mp\sqrt{\Delta}\,\partial_{\rho} + \partial_{\rho}\left(\sqrt{\Delta}\right)\,\beta\left(\partial_{t} + \Omega_{\sigma}\,\partial_{\sigma}\right) + \frac{\rho_{+}-\rho_{-}}{2\sqrt{\Delta}} \beta\Omega_{-\sigma}\,\partial_{-\sigma}\right] \,,
	\end{gathered}
\ee
with $\partial_{\pm}\equiv\partial_{\psi_{\pm}}=\left(\partial_{\psi}\pm\partial_{\phi}\right)/2$ and $\beta$ the inverse Hawking temperature \eqref{eq:betaMP} of the $5$-d Myers-Perry black hole. The corresponding Casimirs,
\be\ba\label{eq:LoveCasimir}
	{}&\mathcal{C}_2^{\left(\sigma\right)} = L_0^{\left(\sigma\right)2} - \frac{1}{2}\left(L_{+1}^{\left(\sigma\right)}L_{-1}^{\left(\sigma\right)}+L_{-1}^{\left(\sigma\right)}L_{+1}^{\left(\sigma\right)}\right) \\
	&= \partial_{\rho}\,\Delta\,\partial_{\rho} - \frac{\rho_{s}^2\rho_{+}}{4\Delta}\left(\partial_{t}+\Omega_{+}\,\partial_{+}+\Omega_{-}\,\partial_{-}\right)^2 - \frac{\rho_{+}-\rho_{-}}{\rho-\rho_{-}}\beta^2\,\left(\partial_{t}+\Omega_{\sigma}\,\partial_{\sigma}\right)\Omega_{-\sigma}\,\,\partial_{-\sigma} \,,
\ea\ee
are precisely equal to the two near-zone truncations of the radial operator \eqref{eq:NZRadial1}-\eqref{eq:NZRadial2} when working in the initial Boyer-Lindquist azimuthal coordinates $\left(\phi,\psi\right)$. 
We will denote the individual algebras generated for each sign $\sigma$ as $\SL_{\left(\sigma\right)}$. A crucial property of these generators is that they are regular at both the future and the past event horizons as can be seen by switching to the advanced and retarded null coordinates respectively (see Appendix \ref{sec:ApSL2RGenerators}).

Solutions of the near-zone equations of motion then form representations of the corresponding $\SL_{\left(\sigma\right)}$ symmetry, labeled by their Casimir eigenvalues, which are equal to the angular eigenvalues $\hat{\ell}(\hat{\ell}+1)$, and the $L_0$-weights~\cite{Miller1968,Howe1992},
\be
	\mathcal{C}_2^{\left(\sigma\right)}\Phi_{\omega\ell mj} = \hat{\ell}(\hat{\ell}+1)\Phi_{\omega\ell mj} \,,\quad L_0^{\left(\sigma\right)}\Phi_{\omega\ell mj} = h^{\left(\sigma\right)}\Phi_{\omega\ell mj} \,.
\ee
We remark, in particular, that
\be
	h^{\left(\sigma\right)}=i\beta\left(\omega-m_{\sigma}\Omega_{\sigma}\right) = -i\Gamma^{\left(\sigma\right)}_{+\sigma}\left(\omega\right) \,.
\ee

\subsection{Highest-weight banishes static Love}
We saw in the previous section that the static scalar Love numbers vanish whenever $m_{\sigma}\Omega_{\sigma}=0$ for $\sigma=+$ or $\sigma=-$ and only for integer $\hat{\ell}$ (see Eq. \eqref{eq:VanishingLove_IntL_Static}). We will show here that the vanishing of the static Love numbers follows from the fact that the relevant solution of the near-zone equations of motion belongs to a highest-weight representation of $\SL_{\left(\sigma\right)}$.

Let us construct the highest-weight representation of $\SL_{\left(\sigma\right)}$ with highest-weight $h^{\left(\sigma\right)}_{-\hat{\ell},0}=-\hat{\ell}$~\cite{Miller1968,Howe1992,Charalambous:2021kcz,Charalambous:2022rre}. The primary state $\upsilon_{-\hat{\ell},0}^{\left(\sigma\right)}$ satisfies,
\be
	L_0^{\left(\sigma\right)}\upsilon_{-\hat{\ell},0}^{\left(\sigma\right)}=-\hat{\ell}\,\upsilon_{-\hat{\ell},0}^{\left(\sigma\right)} \,,\quad L_{+1}^{\left(\sigma\right)}\upsilon_{-\hat{\ell},0}^{\left(\sigma\right)} = 0 \,.
\ee
Supplementing with the condition of definite azimuthal numbers,
\be
	J_0^{\left(\pm\right)}\upsilon_{-\hat{\ell},0}^{\left(\sigma\right)} = m_{\pm }\upsilon_{-\hat{\ell},0}^{\left(\sigma\right)} \,,
\ee
where $J_0^{\left(\pm\right)}=-i\partial_{\pm}$ are the two $\mathfrak{so}\left(3\right)$ $J_0$-generators of the rotation group algebra $\mathfrak{so}\left(4\right)\simeq \mathfrak{so}\left(3\right)\oplus\mathfrak{so}\left(3\right)$ (see Appendix \ref{sec:ApSphericalHarmonics}), we find, up to an overall normalization constant,
\be\label{eq:SL2RHighestL}
	\upsilon_{-\hat{\ell},0}^{\left(\sigma\right)} = \mathcal{F}_{\sigma}\left(\rho\right) e^{im_{\sigma}\left(\psi_{\sigma}-\Omega_{\sigma}t\right)}e^{im_{-\sigma}\psi_{-\sigma}} \left[e^{t/\beta}\sqrt{\Delta}\right]^{\hat{\ell}} \,,
\ee
with the form factor given by
\be
	\mathcal{F}_{\sigma}\left(\rho\right) \equiv \left(\frac{\rho-\rho_{+}}{\rho-\rho_{-}}\right)^{im_{-\sigma}\beta\Omega_{-\sigma}/2} = \left(\frac{\rho-\rho_{+}}{\rho-\rho_{-}}\right)^{i\Gamma^{\left(\sigma\right)}_{-\sigma}/2} \,.
\ee
This highest-weight vector is regular at the future event horizon but singular at the past event horizon. From the regularity of the generators, all the descendants will also be regular at the future event horizon and singular at the past one. For generic parameters, the highest-weight representation is an infinite-dimensional Verma module and is spanned by the vectors~\cite{Howe1992,Miller1968,Miller1970}
\be
	\upsilon_{-\hat{\ell},n}^{\left(\sigma\right)} = \left[L_{-1}^{\left(\sigma\right)}\right]^{n}\upsilon_{-\hat{\ell},0}^{\left(\sigma\right)} \,,\quad n\ge0\,,
\ee
whose charge under $L_0^{\left(\sigma\right)}$ is
\be
	h_{-\hat{\ell},n}^{\left(\sigma\right)} = n-\hat{\ell} \,.
\ee

Let us compare with the properties of the regular static solution with $m_{\sigma}\Omega_{\sigma}=0$ and $\hat{\ell}\in\mathbb{N}$ which corresponds to vanishing static Love numbers. This is a null state of $L_0^{\left(\sigma\right)}$ regular at the future event horizon and is therefore identified as the $n=\hat{\ell}$ descendant in the highest-weight multiplet,
\be
	\Phi_{\omega=0,\ell mj}\bigg|_{m_{\sigma}\Omega_{\sigma}=0} \propto \upsilon_{-\hat{\ell},\hat{\ell}}^{\left(\sigma\right)} = \left[L_{-1}^{\left(\sigma\right)}\right]^{\hat{\ell}}\upsilon_{-\hat{\ell},0}^{\left(\sigma\right)} \,.
\ee
Noticing that, for generic $\upsilon\left(\rho\right)$,
\be\ba\label{eq:Lp1n}
	\left[L_{+1}^{\left(\sigma\right)}\right]^{n}&\left(\mathcal{F}_{\sigma}\left(\rho\right)e^{im_{\sigma}\left(\psi_{\sigma}-\Omega_{\sigma}t\right)}e^{im_{-\sigma}\psi_{-\sigma}}\upsilon\left(\rho\right)\right) = \\
	&\mathcal{F}_{\sigma}\left(\rho\right)e^{im_{\sigma}\left(\psi_{\sigma}-\Omega_{\sigma}t\right)}e^{im_{-\sigma}\psi_{-\sigma}}\left[-e^{t/\beta}\sqrt{\Delta}\right]^{n}\frac{d^{n}}{d\rho^{n}}\upsilon\left(\rho\right) \,,
\ea\ee
we see that the highest-property,
\be
	\left[L_{+1}^{\left(\sigma\right)}\right]^{\hat{\ell}+1}\Phi_{\omega=0,\ell mj}\bigg|_{m_{\sigma}\Omega_{\sigma}=0} = 0 \,,
\ee
dictates a quasi-polynomial form of the near-zone radial wavefunction,
\be
	R_{\omega=0,\ell mj}\bigg|_{m_{\sigma}\Omega_{\sigma}=0} = \mathcal{F}_{\sigma}\left(\rho\right) \sum_{n=0}^{\hat{\ell}}c_{n}\rho^{n} \,.
\ee
This is precisely the quasi-polynomial form \eqref{eq:StaticRadialVanishingLove} that we wanted to address. In conclusion, we have seen how the vanishing of static scalar Love numbers of the $5$-dimensional Myers-Perry black hole is automatically outputted as a selection rule following from the fact that corresponding solution belongs to a highest-weight representation of the near-zone $\SL_{\left(\sigma\right)}$. On the opposite route, the conditions for the regular at the future event horizon static solution to belong to a highest-weight representation are that $\hat{\ell}\in\mathbb{N}$ and that $m_{\sigma}\Omega_{\sigma}=0$, which are precisely the conditions of vanishing static scalar Love numbers.

Let us briefly comment on the structure of the lowest-weight representation of $\SL_{\left(\sigma\right)}$, spanned by ascendants $\bar{\upsilon}_{+\hat{\ell},n}^{\left(\sigma\right)}$ of a lowest-weight state $\bar{\upsilon}_{+\hat{\ell},0}^{\left(\sigma\right)}$,
\be
	\bar{\upsilon}_{+\hat{\ell},n}^{\left(\sigma\right)} = \left[-L_{+1}^{\left(\sigma\right)}\right]^{n}\bar{\upsilon}_{+\hat{\ell},0}^{\left(\sigma\right)} \,,
\ee
satisfying
\be
	L_0^{\left(\sigma\right)}\bar{\upsilon}_{+\hat{\ell},0}^{\left(\sigma\right)}=+\hat{\ell}\,\bar{\upsilon}_{+\hat{\ell},0}^{\left(\sigma\right)} \,,\quad L_{-1}^{\left(\sigma\right)}\bar{\upsilon}_{+\hat{\ell},0}^{\left(\sigma\right)} = 0 \,,
\ee
and having definite azimuthal numbers. We find
\be\label{eq:SL2RLowestL}
	\bar{\upsilon}_{+\hat{\ell},0}^{\left(\sigma\right)} = \bar{\mathcal{F}}_{\sigma}\left(\rho\right) e^{im_{\sigma}\left(\psi_{\sigma}-\Omega_{\sigma}t\right)}e^{im_{-\sigma}\psi_{-\sigma}} \left[e^{-t/\beta}\sqrt{\Delta}\right]^{\hat{\ell}} \,,\quad \bar{\mathcal{F}}_{\sigma}\left(\rho\right) \equiv \left(\frac{\rho-\rho_{+}}{\rho-\rho_{-}}\right)^{-i\Gamma^{\left(\sigma\right)}_{+\sigma}/2} \,.
\ee
This state is a solution of the near-zone equations of motion that is regular at the past event horizon, but singular at the future event horizon. As a result, from the regularity of the generators, all the ascendants will also be regular at the past event horizon and singular at the future one, with $L_0$-charges $\bar{h}^{\left(\sigma\right)}_{\hat{\ell},n}=\hat{\ell}-n$. In particular, when $\hat{\ell}$ is an integer, the $n=\hat{\ell}$ ascendant $\bar{\upsilon}_{+\hat{\ell},\hat{\ell}}^{\left(\sigma\right)}$ will be the static solution of the Klein-Gordon equation with $m_{\sigma}\Omega_{\sigma}=0$ that is singular at the future event horizon. We have just demonstrated that the static solutions regular and singular at the future event horizon belong to different, locally distinguishable, representations of $\SL_{\left(\sigma\right)}$; the highest-weight representation and lowest-weight representation respectively. This is the algebraic manifestation of the absence of RG flow for the static scalar Love numbers in the particular case of $\hat{\ell}\in\mathbb{N}$ and $m_{\sigma}\Omega_{\sigma}=0$. The construction of these highest-weight and lowest-weight representations is demonstrated graphically in Figure \ref{fig:SL2R_HW_LW}.

\begin{figure}
	\centering
	\begin{subfigure}[b]{0.49\textwidth}
		\centering
		\begin{tikzpicture}
			\node at (0,1) (uml3) {$\upsilon^{\left(\sigma\right)}_{-\hat{\ell},\hat{\ell}}$};
			\node at (0,2) (uml2) {$\upsilon^{\left(\sigma\right)}_{-\hat{\ell},2}$};
			\node at (0,3) (uml1) {$\upsilon^{\left(\sigma\right)}_{-\hat{\ell},1}$};
			\node at (0,4) (uml0) {$\upsilon^{\left(\sigma\right)}_{-\hat{\ell},0}$};
			
			\draw (1,1) -- (5,1);
			\node at (3,1.5) (up) {$\vdots$};
			\node at (3,0.5) (um) {$\vdots$};
			\draw (1,2) -- (5,2);
			\draw (1,3) -- (5,3);
			\draw (1,4) -- (5,4);
			\draw [snake=zigzag] (1,4.1) -- (5,4.1);
			
			\draw[red] [<-] (2.5,2) -- node[left] {$L_{-1}^{\left(\sigma\right)}$} (2.5,3);
			\draw[red] [<-] (2,3) -- node[left] {$L_{-1}^{\left(\sigma\right)}$} (2,4);
			\draw[blue] [->] (4,3) -- node[right] {$L_{+1}^{\left(\sigma\right)}$} (4,4);
			\draw[blue] [->] (3.5,2) -- node[right] {$L_{+1}^{\left(\sigma\right)}$} (3.5,3);
		\end{tikzpicture}
		\caption{The highest-weight representation spanned by states $\{\upsilon^{\left(\sigma\right)}_{-\hat{\ell},n}|n\in\mathbb{N}\}$ which are regular (singular) at the future (past) event horizon and have weights $h^{\left(\sigma\right)}_{-\hat{\ell},n}=n-\hat{\ell}$.}
	\end{subfigure}
	\hfill
	\begin{subfigure}[b]{0.49\textwidth}
		\centering
		\begin{tikzpicture}
			\node at (0,3) (upll) {$\bar{\upsilon}^{\left(\sigma\right)}_{+\hat{\ell},\hat{\ell}}$};
			\node at (0,2) (upl2) {$\bar{\upsilon}^{\left(\sigma\right)}_{+\hat{\ell},2}$};
			\node at (0,1) (upl1) {$\bar{\upsilon}^{\left(\sigma\right)}_{+\hat{\ell},1}$};
			\node at (0,0) (upl0) {$\bar{\upsilon}^{\left(\sigma\right)}_{+\hat{\ell},0}$};
			
			\draw [snake=zigzag] (1,-0.1) -- (5,-0.1);
			\draw (1,0) -- (5,0);
			\draw (1,1) -- (5,1);
			\draw (1,2) -- (5,2);
			\node at (3,2.5) (up) {$\vdots$};
			\draw (1,3) -- (5,3);
			\node at (3,3.5) (um) {$\vdots$};
			
			\draw[blue] [->] (2.5,0) -- node[left] {$L_{+1}^{\left(\sigma\right)}$} (2.5,1);
			\draw[blue] [->] (2,1) -- node[left] {$L_{+1}^{\left(\sigma\right)}$} (2,2);
			\draw[red] [<-] (4,1) -- node[right] {$L_{-1}^{\left(\sigma\right)}$} (4,2);
			\draw[red] [<-] (3.5,0) -- node[right] {$L_{-1}^{\left(\sigma\right)}$} (3.5,1);
		\end{tikzpicture}
		\caption{The lowest-weight representation spanned by states $\{\bar{\upsilon}^{\left(\sigma\right)}_{+\hat{\ell},n}|n\in\mathbb{N}\}$ which are singular (regular) at the future (past) event horizon and have weights $\bar{h}^{\left(\sigma\right)}_{+\hat{\ell},n}=\hat{\ell}-n$.}
	\end{subfigure}
	\caption{Infinite-dimensional highest-weight and lowest-weight representations of $\SL_{\left(\sigma\right)}$ containing near-zone solutions for a massless scalar field in the $5$-d Myers-Perry black hole background with multipolar index $\ell$. Whenever $\hat{\ell}=\frac{\ell}{2}\in\mathbb{N}$ and $m_{\sigma}\Omega_{\sigma}=0$, the solution regular (singular) at the future event horizon is the $\hat{\ell}$'th descendant (ascendant) in the highest-weight (lowest-weight) representation.}
	\label{fig:SL2R_HW_LW}
\end{figure}

\subsection{Highest-weight banishes near-zone Love}
Interestingly, the near-zone symmetries can address all the situations in Eq.~\eqref{eq:VanishingLove_IntL} and Eq.~\eqref{eq:VanishingLove_HIntL}, even the unphysical ones.
%Recall that in the near-zone truncations \eqref{eq:NZRadial1}-\eqref{eq:NZRadial2}, scalar Love numbers may vanish whenever $i\Gamma^{\left(\sigma\right)}_{+\sigma}\left(\omega\right)\in\mathbb{Z}$ or $i\Gamma^{\left(\sigma\right)}_{-\sigma}\in\mathbb{Z}$ if $\hat{\ell}$ is an integer, or whenever $i\Gamma^{\left(\sigma\right)}_{+\sigma}\left(\omega\right)\in\mathbb{Z}+\frac{1}{2}$ or $i\Gamma^{\left(\sigma\right)}_{-\sigma}\in\mathbb{Z}+\frac{1}{2}$ if $\hat{\ell}$ is a half-integer (see \eqref{eq:VanishingLove_IntL}-\eqref{eq:VanishingLove_HIntL}).

If $\hat{\ell}$ is an integer and $i\Gamma^{\left(\sigma\right)}_{+\sigma}\left(\omega\right)=k\in\mathbb{Z}$, then the corresponding near-zone solution has the form
\be
	\Phi_{\omega\ell mj}\bigg|_{i\Gamma^{\left(\sigma\right)}_{+\sigma}\left(\omega\right)=k} \propto e^{kt/\beta}e^{im_{\sigma}\left(\psi_{\sigma}-\Omega_{\sigma}t\right)}e^{im_{-\sigma}\psi_{-\sigma}}R_{\omega\ell mj}\left(\rho\right)\bigg|_{i\Gamma^{\left(\sigma\right)}_{+\sigma}\left(\omega\right)=k} \,,
\ee
where we are suppressing the $\theta$-dependence, and therefore satisfies
\be
	L_0^{\left(\sigma\right)}\Phi_{\omega\ell mj}\bigg|_{i\Gamma^{\left(\sigma\right)}_{+\sigma}\left(\omega\right)=k} = -k\,\Phi_{\omega\ell mj}\bigg|_{i\Gamma^{\left(\sigma\right)}_{+\sigma}\left(\omega\right)=k} \,.
\ee
If $k\le\hat{\ell}$, the regular at the future event horizon solution is then recognized to be the $n=\hat{\ell}-k$ descendant,
\be
	\Phi_{\omega\ell mj}\bigg|_{i\Gamma^{\left(\sigma\right)}_{+\sigma}\left(\omega\right)=k\le\hat{\ell}} \propto \upsilon_{-\hat{\ell},\hat{\ell}-k}^{\left(\sigma\right)} = \left[L_{-1}^{\left(\sigma\right)}\right]^{\hat{\ell}-k}\upsilon_{-\hat{\ell},0}^{\left(\sigma\right)} \,.
\ee
The highest-property,
\be
	\left[L_{+1}^{\left(\sigma\right)}\right]^{\hat{\ell}-k+1}\Phi_{\omega\ell mj}\bigg|_{i\Gamma^{\left(\sigma\right)}_{+\sigma}\left(\omega\right)=k\le\hat{\ell}} = 0 \,,
\ee
does not directly imply any useful quasi-polynomial form itself. Rather, it is the fact that
\be
	\upsilon_{-\hat{\ell},\hat{\ell}-k}^{\left(\sigma\right)} =
	\begin{cases}
	    \left(-1\right)^{k}\left[L_{+1}^{\left(\sigma\right)}\right]^{k}\upsilon_{-\hat{\ell},\hat{\ell}}^{\left(\sigma\right)} \quad \text{for }0\le k \le \hat{\ell} \\
	    \left[L_{-1}^{\left(\sigma\right)}\right]^{-k}\upsilon_{-\hat{\ell},\hat{\ell}}^{\left(\sigma\right)} \quad\quad\quad\, \text{for }k<0
    \end{cases}
\ee
and the quasi-polynomial form of the static element $\upsilon_{-\hat{\ell},\hat{\ell}}^{\left(\sigma\right)}$ that give a useful expression for the relevant near-zone solution. From \eqref{eq:Lp1n}, and an analogous relation for the action of $L_{-1}^{\left(\sigma\right)}$ we conclude that
\be
	R_{\omega\ell mj}\left(\rho\right)\bigg|_{i\Gamma^{\left(\sigma\right)}_{+\sigma}\left(\omega\right)=k\le\hat{\ell}} = \mathcal{F}_{\sigma}\left(\rho\right) \Delta^{k/2} \sum_{n=0}^{\hat{\ell}-k}c_{n}\rho^{n} \,,
\ee
which has the exact quasi-polynomial form of \eqref{eq:NZRadialVanishingLove} with $k\le\hat{\ell}$.

Another remark here is there are in general two possible highest-weight representations of $\SL$; one has $h^{\left(\sigma\right)}=-\hat{\ell}$, which is the one we saw, and the other has $h^{\left(\sigma\right)}=+\hat{\ell}+1$. Both highest-weight representations contain solutions that are regular at the future event horizon. Since the weights differ by the integer amount $2\hat{\ell}+1$, we see that the primary state with $h^{\left(\sigma\right)}=+\hat{\ell}+1$ is actually a descendant of the primary state with $h^{\left(\sigma\right)}=-\hat{\ell}$,
\be
    \upsilon_{+\hat{\ell}+1,0} = \left[L_{-1}^{\left(\sigma\right)}\right]^{2\hat{\ell}+1}\upsilon_{-\hat{\ell},0} \,.
\ee
Therefore, the highest-weight Verma module of $\SL_{\left(\sigma\right)}$ we have been working with so far is actually of type ``$[\circ[\circ$''. With this fact in hand, we see an interesting algebraic interpretation of the elements of this representation: The irreducible (lower) highest-weight is spanned by states with non-vanishing but also non-running near-zone scalar Love numbers, while the quotient representation sandwiched between the two primary states is spanned by all the possible near-zone solutions regular at the future event horizon that have vanishing scalar Love numbers.

Let us now supplement with the case where $k\ge\hat{\ell}+1$ for which the scalar Love numbers are not zero but still exhibit no running. For the sake of this, we need to look into states $\{\upsilon^{\left(\sigma\right)}_{-\hat{\ell},-\left(n+1\right)}|n\in\mathbb{N}\}$ that span the ladder above this highest-weight representation. These states are ascendants of the state $\upsilon^{\left(\sigma\right)}_{-\hat{\ell},-1}$,
\be
	\upsilon^{\left(\sigma\right)}_{-\hat{\ell},-\left(n+1\right)} = \frac{1}{n!(2\hat{\ell}+2)_{n}}\left[L_{+1}^{\left(\sigma\right)}\right]^{n}\upsilon^{\left(\sigma\right)}_{-\hat{\ell},-1} \,,
\ee
which satisfies
\be
	L_0^{\left(\sigma\right)}\upsilon^{\left(\sigma\right)}_{-\hat{\ell},-1} = -(\hat{\ell}+1)\,\upsilon^{\left(\sigma\right)}_{-\hat{\ell},-1} \,,\quad L_{-1}^{\left(\sigma\right)}\upsilon^{\left(\sigma\right)}_{-\hat{\ell},-1} = \upsilon^{\left(\sigma\right)}_{-\hat{\ell},0} \,.
\ee
The resulting first-order inhomogeneous differential equation can be solved to get
\be\ba
	\upsilon^{\left(\sigma\right)}_{-\hat{\ell},-1} &= \frac{\mathcal{F}_{\sigma}\left(\rho\right)e^{im_{\sigma}\left(\psi_{\sigma}-\Omega_{\sigma}t\right)}e^{im_{-\sigma}\Omega_{-\sigma}}}{\left(\rho_{+}-\rho_{-}\right)\left(\hat{\ell}+1+i\Gamma^{\left(\sigma\right)}_{-\sigma}\right)}\left(e^{t/\beta}\sqrt{\Delta}\right)^{\hat{\ell}+1} \\
	&\quad\quad\times{}_2F_1\left(1,2\hat{\ell}+2;2+\hat{\ell}+i\Gamma^{\left(\sigma\right)}_{-\sigma};-\frac{\rho-\rho_{+}}{\rho_{+}-\rho_{-}}\right) \,,
\ea\ee
which is regular at the future event horizon and singular at the past one. Consequently, all the ascendants $\upsilon^{\left(\sigma\right)}_{-\hat{\ell},-\left(n+1\right)}$, with $n\in\mathbb{N}$ will also be regular at the future event horizon near-zone solutions, with $L_0$-eigenvalues $h^{\left(\sigma\right)}_{-\hat{\ell},-\left(n+1\right)}=-(n+\hat{\ell}+1)$. From this, we therefore identify
\be
	\Phi_{\omega\ell mj}\bigg|_{i\Gamma^{\left(\sigma\right)}_{+\sigma}\left(\omega\right)=k\ge\hat{\ell}+1} \propto \upsilon_{-\hat{\ell},-\left(k-\hat{\ell}\right)}^{\left(\sigma\right)} \propto \left[L_{+1}^{\left(\sigma\right)}\right]^{k-\hat{\ell}-1}\upsilon_{-\hat{\ell},-1}^{\left(\sigma\right)} \,.
\ee
As already discussed around \eqref{eq:NZRadialVanishingLove}, we do not expect to find any conspiring quasi-polynomial solution for $k\ge\hat{\ell}+1$. However, we have supplemented with the algebraic property of these remaining states to span the \textit{entire} type ``$\circ[\circ[\circ$'' representation of $\SL_{\left(\sigma\right)}$ for which the highest-weight state has weight $h^{\left(\sigma\right)}_{-\hat{\ell},0}=-\hat{\ell}$ (see~\cite{Howe1992} for our notation).

As for the near-zone solutions singular at the future event horizon, these can be similarly worked out to span the entire type ``$\circ]\circ]\circ$'' representation of $\SL_{\left(\sigma\right)}$ for which the lowest-weight state has weight $\bar{h}^{\left(\sigma\right)}_{+\hat{\ell},0}=+\hat{\ell}$, thus providing us with an algebraic argument of the absence of RG flow of the Love numbers. These constructions are shown graphically in Figure \ref{fig:SL2R_oHW_oLW}.

\begin{figure}
	\centering
	\begin{subfigure}[b]{0.49\textwidth}
		\centering
		\begin{tikzpicture}
		    \node at (0,-1) (uml4) {$\upsilon^{\left(\sigma\right)}_{-\hat{\ell},2\hat{\ell}+2}$};
		    \node at (0,0) (uml3) {$\upsilon^{\left(\sigma\right)}_{-\hat{\ell},2\hat{\ell}+1}$};
			\node at (0,1) (uml2) {$\upsilon^{\left(\sigma\right)}_{-\hat{\ell},2\hat{\ell}}$};
			\node at (0,2) (uml1) {$\upsilon^{\left(\sigma\right)}_{-\hat{\ell},1}$};
			\node at (0,3) (uml0) {$\upsilon^{\left(\sigma\right)}_{-\hat{\ell},0}$};
			\node at (0,4) (umlm1) {$\upsilon^{\left(\sigma\right)}_{-\hat{\ell},-1}$};
			\node at (0,5) (umlm2) {$\upsilon^{\left(\sigma\right)}_{-\hat{\ell},-2}$};
			
			\node at (3,-1.5) (um) {$\vdots$};
			\draw (1,-1) -- (5,-1);
			\draw [snake=zigzag] (1,0.1) -- (5,0.1);
			\draw (1,0) -- (5,0);
			\draw (1,1) -- (5,1);
			\node at (3,1.5) (up) {$\vdots$};
			\draw (1,2) -- (5,2);
			\draw (1,3) -- (5,3);
			\draw (1,4) -- (5,4);
			\draw [snake=zigzag] (1,3.1) -- (5,3.1);
			\draw (1,5) -- (5,5);
			\node at (3,5.5) (up) {$\vdots$};
			
			\draw[red] [<-] (3,3) -- node[left] {$L_{-1}^{\left(\sigma\right)}$} (3,4);
			\draw[red] [<-] (2,2) -- node[left] {$L_{-1}^{\left(\sigma\right)}$} (2,3);
			\draw[red] [<-] (3,0) -- node[left] {$L_{-1}^{\left(\sigma\right)}$} (3,1);
			\draw[red] [<-] (2,-1) -- node[left] {$L_{-1}^{\left(\sigma\right)}$} (2,0);
			\draw[blue] [->] (4,-1) -- node[right] {$L_{+1}^{\left(\sigma\right)}$} (4,0);
			\draw[blue] [->] (4,2) -- node[right] {$L_{+1}^{\left(\sigma\right)}$} (4,3);
			\draw[red] [<-] (2,4) -- node[left] {$L_{-1}^{\left(\sigma\right)}$} (2,5);
			\draw[blue] [->] (4,4) -- node[right] {$L_{+1}^{\left(\sigma\right)}$} (4,5);
		\end{tikzpicture}
		\caption{The type ``$\circ[\circ[\circ$'' representation spanned by states $\{\upsilon^{\left(\sigma\right)}_{-\hat{\ell},j}|j\in\mathbb{Z}\}$ which are regular (singular) at the future (past) event horizon and have weights $h^{\left(\sigma\right)}_{-\hat{\ell},j}=j-\hat{\ell}$.}
	\end{subfigure}
	\hfill
	\begin{subfigure}[b]{0.49\textwidth}
		\centering
		\begin{tikzpicture}
		    \node at (0,4) (upl4) {$\bar{\upsilon}^{\left(\sigma\right)}_{+\hat{\ell},2\hat{\ell}+2}$};
		    \node at (0,3) (upl3) {$\bar{\upsilon}^{\left(\sigma\right)}_{+\hat{\ell},2\hat{\ell}+1}$};
			\node at (0,2) (upl2) {$\bar{\upsilon}^{\left(\sigma\right)}_{+\hat{\ell},2\hat{\ell}}$};
			\node at (0,1) (upl1) {$\bar{\upsilon}^{\left(\sigma\right)}_{+\hat{\ell},1}$};
			\node at (0,0) (upl0) {$\bar{\upsilon}^{\left(\sigma\right)}_{+\hat{\ell},0}$};
			\node at (0,-1) (uplm1) {$\bar{\upsilon}^{\left(\sigma\right)}_{+\hat{\ell},-1}$};
			\node at (0,-2) (uplm2) {$\bar{\upsilon}^{\left(\sigma\right)}_{+\hat{\ell},-2}$};
			
			\draw [snake=zigzag] (1,-0.1) -- (5,-0.1);
			\draw (1,0) -- (5,0);
			\draw (1,1) -- (5,1);
			\node at (3,1.5) (up) {$\vdots$};
			\draw (1,2) -- (5,2);
			\node at (3,4.5) (um) {$\vdots$};
			\draw [snake=zigzag] (1,2.9) -- (5,2.9);
			\draw (1,3) -- (5,3);
			\draw (1,4) -- (5,4);
			\draw (1,-1) -- (5,-1);
			\draw (1,-2) -- (5,-2);
			\node at (3,-2.5) (up) {$\vdots$};
			
			\draw[blue] [->] (3,-1) -- node[left] {$L_{+1}^{\left(\sigma\right)}$} (3,0);
			\draw[blue] [->] (2,0) -- node[left] {$L_{+1}^{\left(\sigma\right)}$} (2,1);
			\draw[blue] [->] (3,2) -- node[left] {$L_{+1}^{\left(\sigma\right)}$} (3,3);
			\draw[blue] [->] (2,3) -- node[left] {$L_{+1}^{\left(\sigma\right)}$} (2,4);
			\draw[red] [<-] (4,3) -- node[right] {$L_{-1}^{\left(\sigma\right)}$} (4,4);
			\draw[red] [<-] (4,0) -- node[right] {$L_{-1}^{\left(\sigma\right)}$} (4,1);
			\draw[blue] [->] (2,-1) -- node[left] {$L_{+1}^{\left(\sigma\right)}$} (2,-2);
			\draw[red] [<-] (4,-1) -- node[right] {$L_{-1}^{\left(\sigma\right)}$} (4,-2);
		\end{tikzpicture}
		\caption{The type ``$\circ]\circ]\circ$'' representation spanned by states $\{\bar{\upsilon}^{\left(\sigma\right)}_{+\hat{\ell},j}|j\in\mathbb{Z}\}$ which are singular (regular) at the future (past) event horizon and have weights $\bar{h}^{\left(\sigma\right)}_{+\hat{\ell},j}=\hat{\ell}-j$.}
	\end{subfigure}
	\caption{The type ``$\circ[\circ[\circ$'' and type ``$\circ]\circ]\circ$'' representations of $\SL_{\left(\sigma\right)}$ containing all the near-zone solutions for a massless scalar field in the $5$-d Myers-Perry black hole background that have vanishing/non-running Love numbers regarding the conditions on $\Gamma^{\left(\sigma\right)}_{+\sigma}$ (see \eqref{eq:VanishingLove_IntL}-\eqref{eq:VanishingLove_HIntL}). For integer $\hat{\ell}$, the relevant condition is $i\Gamma^{\left(\sigma\right)}_{+\sigma}=k\in\mathbb{Z}$ and is depicted above. For half-integer $\hat{\ell}$, the condition becomes $i\Gamma^{\left(\sigma\right)}_{+\sigma}=k+\frac{1}{2}$, $k\in\mathbb{Z}$, and the structure of the representations is as in the above figures after replacing $k\rightarrow k+\frac{1}{2}$.}
	\label{fig:SL2R_oHW_oLW}
\end{figure}

Last, if $\hat{\ell}$ is a half-integer and $i\Gamma^{\left(\sigma\right)}_{+\sigma}\left(\omega\right)=k+\frac{1}{2}$, with $k\in\mathbb{Z}$, then we are looking at near-zone solutions of the form
\be
	\Phi_{\omega\ell mj}\bigg|_{i\Gamma^{\left(\sigma\right)}_{+\sigma}\left(\omega\right)=k+\frac{1}{2}} \propto e^{\left(k+1/2\right)t/\beta}e^{im_{\sigma}\left(\psi_{\sigma}-\Omega_{\sigma}t\right)}e^{im_{-\sigma}\psi_{-\sigma}}R_{\omega\ell mj}\left(\rho\right)\bigg|_{i\Gamma^{\left(\sigma\right)}_{+\sigma}\left(\omega\right)=k+\frac{1}{2}} \,,
\ee
which satisfy
\be
	L_0^{\left(\sigma\right)}\Phi_{\omega\ell mj}\bigg|_{i\Gamma^{\left(\sigma\right)}_{+\sigma}\left(\omega\right)=k+\frac{1}{2}} = -\left(k+\frac{1}{2}\right)\Phi_{\omega\ell mj}\bigg|_{i\Gamma^{\left(\sigma\right)}_{+\sigma}\left(\omega\right)=k+\frac{1}{2}} \,.
\ee
The above analysis is then carried away identically, after replacing $k\rightarrow k+\frac{1}{2}$.

\subsection{Local near-zone $\SL\times\SL$}
Somewhat surprisingly, we can address the remaining situations where $i\Gamma^{\left(\sigma\right)}_{-\sigma}\in\mathbb{Z}$ for integer $\hat{\ell}$ or $i\Gamma^{\left(\sigma\right)}_{-\sigma}\in\mathbb{Z}+\frac{1}{2}$ for half-integer $\hat{\ell}$, even for $\omega\ne0$. This is ought to the observation that the particular near-zone truncations are equipped with a larger $\SL_{\left(\sigma\right),\text{L}}\times\SL_{\left(\sigma\right),\text{R}}$ structure. The first $\SL$ factor is the Love symmetry $\SL$,
\be
	\SL_{\left(\sigma\right),\text{L}}=\SL_{\left(\sigma\right)} \,,
\ee
generated by the globally defined vector fields \eqref{eq:LoveGen}, $L_{m}^{\left(\sigma\right),\text{L}}=L_{m}^{\left(\sigma\right)}$, $m=0,\pm1$. The second, $\SL_{\left(\sigma\right),\text{R}}$, factor is generated by the following vector fields
\be\label{eq:LoveGenSpBroken}
	\begin{gathered}
		L_0^{\left(\sigma\right),\text{R}} = -\beta\Omega_{-\sigma}\,\partial_{-\sigma} \,, \\
		L_{\pm1}^{\left(\sigma\right),\text{R}} = e^{\pm\psi_{-\sigma}/\left(\beta\Omega_{-\sigma}\right)}\left[\mp\sqrt{\Delta}\partial_{\rho}+\partial_{\rho}\left(\sqrt{\Delta}\right)\beta\Omega_{-\sigma}\,\partial_{-\sigma}+\frac{\rho_{+}-\rho_{-}}{2\sqrt{\Delta}}\beta\left(\partial_{t}+\Omega_{\sigma}\,\partial_{\sigma}\right)\right] \,.
	\end{gathered}
\ee
The Casimirs of the two commuting $\SL$'s are exactly the same,
\be
	\mathcal{C}_2^{\left(\sigma\right),\text{R}} = \mathcal{C}_2^{\left(\sigma\right),\text{L}} = \mathcal{C}_2^{\left(\sigma\right)} \,.
\ee
In addition, the sets of vector fields generating the two $\SL$'s are regular at both the future and the past event horizons with respect to the radial variable, i.e. they do not develop poles as $\rho\rightarrow \rho_{+}$. However, $\SL_{\left(\sigma\right),\text{R}}$ is spontaneously broken down to $U\left(1\right)_{\left(\sigma\right),\text{R}}$ by the periodic identification of the azimuthal angles, $\psi_{\pm}\sim\psi_{\pm}+2\pi$, under which $L_{\pm1}^{\left(\sigma\right),\text{R}}$ develop conical deficits.

Despite this breaking of $\SL_{\left(\sigma\right),\text{R}}$ by the periodic identification of the azimuthal angles, it can still be used to explain the vanishing/non-running corresponding to the situations where $i\Gamma^{\left(\sigma\right)}_{-\sigma}\in\mathbb{Z}$ for integer $\hat{\ell}$ or $i\Gamma^{\left(\sigma\right)}_{-\sigma}\in\mathbb{Z}+\frac{1}{2}$ for half-integer $\hat{\ell}$ in a similar fashion as in the previous subsection. Solutions of the near-zone equations of motion will furnish representations labeled by the Casimir and $L_0$-eigenvalues,
\be
	\mathcal{C}_2^{\left(\sigma\right),\text{R}}\Phi_{\omega\ell mj}=\hat{\ell}(\hat{\ell}+1)\Phi_{\omega\ell mj} \,,\quad L_0^{\left(\sigma\right),\text{R}}\Phi_{\omega\ell mj} = h^{\left(\sigma\right),\text{R}}\Phi_{\omega\ell mj} \,,
\ee
with
\be
	h^{\left(\sigma\right),\text{R}} = -i\beta m_{-\sigma}\Omega_{-\sigma} = -i\Gamma^{\left(\sigma\right)}_{-\sigma} \,.
\ee
Let us construct the analogous highest-weight representation of $\SL_{\left(\sigma\right),\text{R}}$. The primary state with highest-weight $h_{-\hat{\ell},0}^{\left(\sigma\right),\text{R}}=-\hat{\ell}$, satisfying
\be
	L_0^{\left(\sigma\right),\text{R}}\upsilon_{-\hat{\ell},0}^{\left(\sigma\right),\text{R}} = -\hat{\ell}\,\upsilon_{-\hat{\ell},0}^{\left(\sigma\right),\text{R}} \,,\quad 	L_{+1}^{\left(\sigma\right),\text{R}}\upsilon_{-\hat{\ell},0}^{\left(\sigma\right),\text{R}} = 0 \,,
\ee
and having definite azimuthal numbers and frequency is given by, up to an overall normalization constant,
\be
	\upsilon_{-\hat{\ell},0}^{\left(\sigma\right),\text{R}} = \mathcal{F}_{\sigma}^{\text{R}}\left(\rho\right)e^{-i\omega t}e^{im_{\sigma}\psi_{\sigma}}e^{\hat{\ell}\psi_{-\sigma}/\left(\beta\Omega_{-\sigma}\right)}\Delta^{\hat{\ell}/2} \,, \quad \mathcal{F}_{\sigma}^{\text{R}}\left(\rho\right) \equiv \left(\frac{\rho-\rho_{+}}{\rho-\rho_{-}}\right)^{i\Gamma^{\left(\sigma\right)}_{+\sigma}\left(\omega\right)/2} \,.
\ee
This state is singular at the past event horizon and develops conical deficits as we go around the azimuthal circles, but develops no pole at the future event horizon with respect to the radial variable. The descendants,
\be
	\upsilon_{-\hat{\ell},n}^{\left(\sigma\right),\text{R}} = \left[L_{+1}^{\left(\sigma\right),\text{R}}\right]^{n}\upsilon_{-\hat{\ell},0}^{\left(\sigma\right),\text{R}} \,,
\ee
share the same boundary conditions, with the conical deficit measured by their charge under $L_0^{\left(\sigma\right),\text{R}}$,
\be
	h_{-\hat{\ell},n}^{\left(\sigma\right),\text{R}} = n-\hat{\ell} \,.
\ee

If $\hat{\ell}\in\mathbb{N}$, there exists a particular descendant that develops no conical deficit and is therefore truly regular at the future event horizon. This is the $n=\hat{\ell}$ descendant which is a null state under $L_0^{\left(\sigma\right),\text{R}}$ and corresponds to the regular near-zone solution with $\Gamma^{\left(\sigma\right)}_{+\sigma}=0$. Noticing that
\be\ba
	\left[L_{+1}^{\left(\sigma\right)}\right]^{n}&\left(\mathcal{F}_{\sigma}^{\text{R}}\left(\rho\right)e^{i\omega t}e^{im_{\sigma}\psi_{\sigma}}\upsilon\left(\rho\right)\right) = \\
	&\mathcal{F}_{\sigma}^{\text{R}}\left(\rho\right)e^{i\omega t}e^{im_{\sigma}\psi_{\sigma}}\left[-e^{\psi_{-\sigma}/\left(\beta\Omega_{-\sigma}\right)}\sqrt{\Delta}\right]^{n}\frac{d^{n}}{d\rho^{n}}\upsilon\left(\rho\right) \,,
\ea\ee
for generic $\upsilon\left(\rho\right)$, we see that the highest-weight property,
\be
	\left[L_{+1}^{\left(\sigma\right),\text{R}}\right]^{\hat{\ell}+1}\Phi_{\omega\ell mj}\bigg|_{\Gamma^{\left(\sigma\right)}_{-\sigma}=0} = 0 \,,
\ee
implies the following quasi-polynomial form for the radial wavefunction
\be
	R_{\omega\ell mj}\bigg|_{\Gamma^{\left(\sigma\right)}_{-\sigma}=0} = \mathcal{F}_{\sigma}^{\text{R}}\left(\rho\right)\sum_{n=0}^{\hat{\ell}}c_{n}\rho^{n} \,,
\ee
which exactly matches the one from \eqref{eq:StaticRadialVanishingLove} we wanted to address. The absence of RG flow is also encoded in the representation theory analysis of $\SL_{\left(\sigma\right),\text{R}}$, with the solution singular at the future event horizon (and regular at the past event horizon) being the $n=\hat{\ell}$ ascendant of the locally distinguishable lowest-weight representation of $\SL_{\left(\sigma\right),\text{R}}$ with lowest-weight $\bar{h}_{+\hat{\ell},0}^{\left(\sigma\right),\text{R}}=+\hat{\ell}$.

For the other situations of vanishing/non-running scalar Love numbers that regard the conditions on $\Gamma^{\left(\sigma\right)}_{-\sigma}$ for which the relevant near-zone solutions develop conical deficits, we can follow the same procedure as in the previous subsection and show that all the relevant regular (singular) at the future event horizon near-zone solutions span the entire representation of type ``$\circ[\circ[\circ$'' (type ``$\circ]\circ]\circ$'').
\section{Properties}
\label{sec:Properties}

In this section we will present a number of interesting properties of the Love symmetries. First, we will show how the near-zone $\SL$ symmetries acquire a geometric interpretation as isometries of effective geometries within the framework of subtracted geometries~\cite{Cvetic:2011dn,Cvetic:2011hp}. Then, we will demonstrate how both near-zone symmetries can be realized as subalgebras of a larger, infinite-dimensional $\SL\ltimes \hat{U}\left(1\right)^2_{\mathcal{V}}$ extension, which is interpreted as the $5$-d version of the $\SL\ltimes \hat{U}\left(1\right)_{\mathcal{V}}$ infinite extension of the Love symmetry for Kerr-Newman black holes~\cite{Charalambous:2021kcz,Charalambous:2022rre}. We will also present another interesting generalization of the near-zone symmetries which will exhaust all the possible near-zone truncations of the equations of motion that are equipped with an enhanced $\SL$ symmetry and acquire a subtracted geometry interpretation. We will close this section by proposing a physical interpretation of the states in the highest-weight Love multiplet we saw in the previous section. In particular, by employing a partial wave analysis of the black hole scattering problem, we will argue that states with vanishing Love numbers can be interpreted as total transmission modes~\cite{Hod:2013fea,Cook:2016fge,Cook:2016ngj,Ivanov:2022qqt}.

\subsection{Near-zone symmetries as isometries of subtracted geometries}

To introduce the notion of subtracted geometries, we begin by writing the geometry of the $5$-d Myers-Perry black hole as a fibration over a $4$-d base space~\cite{Chong:2006zx,Cvetic:2011hp},
\be
	\begin{gathered}
		ds^2 = -\Delta_0^{-2/3}G\left(dt+\mathcal{A}\right)^2 + \Delta_0^{1/3}ds_4^2 \,, \\
		ds_4^2 = \frac{d\rho^2}{4X} + \,d\theta^2 + \frac{1}{4}\sum_{i,j=1}^{2}\gamma_{ij}d\phi^{i}d\phi^{j} \,,
	\end{gathered}
\ee
where we use small Latin indices from the middle of alphabet to label the azimuthal angles, with $\phi^1\equiv\phi$ and $\phi^2\equiv\psi$. In the notation more conventionally used to write down the line element of the geometry (see Appendix \ref{sec:Ap5dMPGeometry}),
\be
	\begin{gathered}
		X = \Delta \,,\quad \Delta_0=\Sigma^3 \,,\quad G=\Sigma\left(\Sigma-\rho_{s}\right) \,, \\
		\mathcal{A} = \frac{\rho_{s}\Sigma}{G}\left(a\sin^2\theta\,d\phi + b\cos^2\theta\,d\psi\right) \,, \\
		\frac{1}{4}\sum_{i,j}\gamma_{ij}d\phi^{i}d\phi^{j} = \frac{\rho_{s}}{G}\left(a\sin^2\theta\,d\phi+b\cos^2\theta\,d\psi\right)^2 + \frac{\rho+a^2}{\Sigma}d\phi^2+\frac{\rho+b^2}{\Sigma}d\psi^2 \,.
	\end{gathered}
\ee

Let us forget for a moment what the explicit expressions for $\Delta_0$, $G$, $\mathcal{A}$, $X$ and $\gamma_{ij}$ are. The generic effective geometry then describes a stationary, axisymmetric black hole whose event horizon is the larger root of the function $X$. The function $G$ captures the static limit at the surface $G=0$, setting the boundaries of the ergosphere. The thermodynamic properties of the black hole are completely independent of the warp-factor $\Delta_0$, while they only depend on the near-horizon behavior of the function $G$, the angular potential $\mathcal{A}$ and the induced metric $\gamma_{ij}$~\cite{Cvetic:2011hp}. The warp-factor $\Delta_0$ can therefore be interpreted as to encode information about the environment surrounding the black hole, rather than its internal structure. A subtracted geometry is then a geometry obtained by a modification of the warp-factor, while keeping all the other metric functions fixed~\cite{Cvetic:2011hp,Cvetic:2011dn}.

To reveal a connection to the Love symmetries, it is more convenient to look at the inverse metric,
\be
	\begin{gathered}
		g^{\mu\nu}\partial_{\mu}\partial_{\nu} = \frac{4}{\Delta_0^{1/3}}\left\{X\partial_{\rho}^2 + \frac{1}{4}\,\partial_{\theta}^2 - \frac{\Delta_0}{4G}\partial_{t}^2 + \sum_{i,j}\gamma^{ij}D_{i}D_{j}\right\} \,,\quad D_{i}\equiv \partial_{i} - \mathcal{A}_{i}\partial_{t} \,,
	\end{gathered}
\ee
with $\gamma^{ij}$ the components of the inverse of the induced azimuthal metric $\gamma_{ij}$. The Love symmetry $\SL_{\left(\sigma\right)}$ can then be realized as an isometry of the effective geometry with inverse metric
\be
	\tilde{g}^{\mu\nu}_{\left(\sigma\right)}\partial_{\mu}\partial_{\nu} = \frac{4}{\tilde{\Delta}_{0}^{\left(\sigma\right)1/3}}\left\{X\partial_{\rho}^2 + \frac{1}{4}\,\partial_{\theta}^2 - \frac{\tilde{\Delta}_{0}^{\left(\sigma\right)}}{4G}\partial_{t}^2 + \sum_{i,j}\gamma^{ij}\tilde{D}_{i}^{\left(\sigma\right)}\tilde{D}_{j}^{\left(\sigma\right)}\right\} \,,\quad \tilde{D}_{i}^{\left(\sigma\right)} = \partial_{i} - \tilde{\mathcal{A}}_{i}^{\left(\sigma\right)}\partial_{t} \,,
\ee
where
\be
	\begin{gathered}
		\tilde{\Delta}_{0}^{\left(\sigma\right)} = 16\rho_{+}\rho_{s}^2\left[1+\beta^2\left(\Omega_{\phi}-\sigma\Omega_{\psi}\right)^2\right] \,, \\
		\tilde{\mathcal{A}}^{\left(\sigma\right)} = -\frac{\rho_{+}\rho_{s}^2}{4\Delta}\left(\Omega_{\phi}\,\partial_{\phi}+\Omega_{\psi}\,\partial_{\psi}\right) - \frac{\rho_{+}-\rho_{-}}{\rho-\rho_{-}}\frac{\beta^2}{4}\left(\Omega_{\phi}-\sigma\Omega_{\psi}\right)\left(\partial_{\phi}-\sigma\partial_{\psi}\right) \,.
	\end{gathered}
\ee
More importantly, this effective geometry has exactly the same $4$-d base space $ds_4^2$ and preserves the entire form of the function $G$ which captures information about the static limit. The only difference relevant to the original definition of subtracted geometries~\cite{Cvetic:2011hp,Cvetic:2011dn} is that the angular potential is itself modified, but in such a way that the thermodynamic properties of the black hole remain unaltered.

As a side note, we remark here that it might be possible to realize subtracted geometries by a scaling limit of the full geometry, see e.g.~\cite{Cvetic:2012tr}. It would be interesting to investigate whether the effective geometries associated with the Love symmetries can be manifested as similar scaling limits of the full $5$-d Myers-Perry black hole geometry. We leave such an analysis for future work.

\subsection{Infinite-dimensional extension}
In the previous section, we presented how two different near-zone truncations of the massless Klein-Gordon equation admit two different $\SL$ symmetries, $\SL_{\left(\sigma\right)}$, $\sigma=\pm$. Both of these $\SL$ algebras can be realized as subalgebras of the semi-direct product $\SL_{\left[0,0\right]}\ltimes U\left(1\right)^2_{\mathcal{V}}$, where $\SL_{\left[0,0\right]}$ is generated by the following vector fields
\be
	\begin{gathered}
		L_0^{\left[0,0\right]} = -\beta\,\partial_{t} \,, \\
		L_{\pm1}^{\left[0,0\right]} = e^{\pm t/\beta}\left[\mp\sqrt{\Delta}\,\partial_{\rho} + \partial_{\rho}\left(\sqrt{\Delta}\right)\beta\,\partial_{t} + \frac{\rho_{+}-\rho_{-}}{2\sqrt{\Delta}}\beta\left(\Omega_{+}\,\partial_{+}+\Omega_{-}\,\partial_{-}\right)\right] \,.
	\end{gathered}
\ee
Each of the $U\left(1\right)_{\mathcal{V}}$ factors is generated by vector fields of the form $\upsilon\,\beta\Omega_{\sigma}\,\partial_{\sigma}$ with $\upsilon$ belonging to a representation $\mathcal{V}$ of $\SL_{\left[0,0\right]}$, for each sign $\sigma=+,-$ respectively.

Let us construct one such representation $\mathcal{V}=\left\{\upsilon_{0,k},k\in\mathbb{Z}\right\}$. We first specify $\upsilon_{0,0}=-1$, which belongs to the singleton representation of $\SL_{\left[0,0\right]}$, satisfying $L_{\pm1}^{\left[0,0\right]}\upsilon_{0,0}=0$, and has vanishing azimuthal numbers, therefore further satisfying $L_0^{\left[0,0\right]}\upsilon_{0,0}=0$. We then built the $\upsilon_{0,\pm1}$ states, under the conditions that they can reach the singleton state via the action of $L_{\mp1}^{\left[0,0\right]}$ and that they have weights $h=\mp1$,
\be
	L_0^{\left[0,0\right]}\upsilon_{0,\pm1} = \mp \upsilon_{0,\pm1} \,,\quad L_{\mp1}^{\left[0,0\right]}\upsilon_{0,\pm1} = \mp\upsilon_{0,0} \,.
\ee
Solving these, we arrive at the following basic states of the representation $\mathcal{V}$
\be
	\upsilon_{0,0}=-1 \,,\quad \upsilon_{0,\pm1}=e^{\pm t/\beta}\sqrt{\frac{\rho-\rho_{+}}{\rho-\rho_{-}}} \,,
\ee
which are automatically regular at both the future and the past event horizons. The rest of the representation $\mathcal{V}$ can then be constructed by climbing up or down the ladder,
\be
	\upsilon_{0,\pm n} = \left[L_{\pm1}^{\left[0,0\right]}\right]^{n-1}\upsilon_{0,\pm1} = \left(\pm1\right)^{n-1}\left(n-1\right)!\,e^{\pm nt/\beta}\left(\frac{\rho-\rho_{+}}{\rho-\rho_{-}}\right)^{n/2} \,,
\ee
with integer $n\ge1$. This construction is depicted in Figure \ref{fig:VSL2R}.

\begin{figure}
	\centering
	\begin{tikzpicture}
		\node at (0,0) (um3) {$\upsilon_{0,-3}$};
		\node at (0,1) (um2) {$\upsilon_{0,-2}$};
		\node at (0,2) (um1) {$\upsilon_{0,-1}$};
		\node at (0,3) (u0) {$\upsilon_{0,0}$};
		\node at (0,4) (up1) {$\upsilon_{0,+1}$};
		\node at (0,5) (up2) {$\upsilon_{0,+2}$};
		\node at (0,6) (up3) {$\upsilon_{0,+3}$};
		
		\node at (3,-0.4) (um) {$\vdots$};
		\draw (1,0) -- (5,0);
		\draw (1,1) -- (5,1);
		\draw (1,2) -- (5,2);
		\draw [snake=zigzag] (1,2.9) -- (5,2.9);
		\draw (1,3) -- (5,3);
		\draw [snake=zigzag] (5,3.1) -- (1,3.1);
		%		\draw [snake=zigzag] (1,3.1) -- (5,3.1);
		\draw (1,4) -- (5,4);
		\draw (1,5) -- (5,5);
		\draw (1,6) -- (5,6);
		\node at (3,6.4) (up) {$\vdots$};
		
		\draw[blue] [->] (2,0) -- node[left] {$L_{+1}^{\left[0,0\right]}$} (2,1);
		\draw[blue] [->] (2.5,1) -- node[left] {$L_{+1}^{\left[0,0\right]}$} (2.5,2);
		\draw[blue] [->] (3,2) -- node[left] {$L_{+1}^{\left[0,0\right]}$} (3,3);
		\draw[blue] [->] (2.5,4) -- node[left] {$L_{+1}^{\left[0,0\right]}$} (2.5,5);
		\draw[blue] [->] (2,5) -- node[left] {$L_{+1}^{\left[0,0\right]}$} (2,6);
		\draw[red] [<-] (4,0) -- node[right] {$L_{-1}^{\left[0,0\right]}$} (4,1);
		\draw[red] [<-] (3.5,1) -- node[right] {$L_{-1}^{\left[0,0\right]}$} (3.5,2);
		\draw[red] [<-] (3,3) -- node[right] {$L_{-1}^{\left[0,0\right]}$} (3,4);
		\draw[red] [<-] (3.5,4) -- node[right] {$L_{-1}^{\left[0,0\right]}$} (3.5,5);
		\draw[red] [<-] (4,5) -- node[right] {$L_{-1}^{\left[0,0\right]}$} (4,6);
	\end{tikzpicture}
	\caption{A representation $\mathcal{V}$ of $\SL_{\left[0,0\right]}$ used to construct the $\SL_{\left[0,0\right]}\ltimes\hat{U}\left(1\right)^2_{\mathcal{V}}$ extension.}
	\label{fig:VSL2R}
\end{figure}

Consequently, we can extend $\SL_{\left[0,0\right]}$ into $\SL_{\left[0,0\right]}\ltimes\hat{U}\left(1\right)^2_{\mathcal{V}}$, via the $U\left(1\right)^2_{\mathcal{V}}$ elements
\be
	\upsilon = \sum_{k\in\mathbb{Z}}\upsilon_{0,k}\left(\alpha_{k,+}\,\beta\Omega_{+}\,\partial_{+} + \alpha_{k,-}\,\beta\Omega_{-}\,\partial_{-}\right) \,.
\ee

Within this infinite extension lies a particular $2$-parameter family of $\SL$ subalgebras,
\be
	\SL_{\left[\alpha_{+},\alpha_{-}\right]}\subset \SL_{\left[0,0\right]}\ltimes U\left(1\right)^2_{\mathcal{V}} \,,
\ee
generated by the vector fields
\be
	L_{m}^{\left[\alpha_{+},\alpha_{-}\right]} = L_{m}^{\left[0,0\right]} + \upsilon_{0,m}\left(\alpha_{+}\,\beta\Omega_{+}\,\partial_{+} + \alpha_{-}\,\beta\Omega_{-}\,\partial_{-}\right) \,,\quad m=0,\pm1 \,.
\ee
The corresponding Casimir operator is given by
\be\ba
	{}&\mathcal{C}_2^{\left[\alpha_{+},\alpha_{-}\right]} = \partial_{\rho}\,\Delta\,\partial_{\rho}- \frac{\rho_{s}^2\rho_{+}}{4\Delta}\left(\partial_{t} + \Omega_{+}\,\partial_{+} + \Omega_{-}\,\partial_{-}\right)^2 \\
	&+ \frac{\rho_{+}-\rho_{-}}{\rho-\rho_{-}}\beta^2\left(\partial_{t} + \alpha_{+}\,\Omega_{+}\,\partial_{+} + \alpha_{-}\,\Omega_{-}\,\partial_{-}\right)\left[\left(\alpha_{+}-1\right)\Omega_{+}\,\partial_{+} + \left(\alpha_{-}-1\right)\Omega_{-}\,\partial_{-}\right] \,.
\ea\ee
Note that for an arbitrary pair $\left(\alpha_{+},\alpha_{-}\right)$, these Casimirs do not correspond to any
consistent physical near-zone truncation of the Klein-Gordon equation in the background of the $5$-d Myers-Perry black hole, except in two cases;
\be
	\alpha^{\text{NZ}}_{\sigma} = 1 \,\quad \text{AND} \quad\, \alpha^{\text{NZ}}_{-\sigma} = 0 \,,
\ee
for $\sigma=+$ or $\sigma=-$. These are precisely our two $\SL$ symmetries of the near-zone truncations \eqref{eq:NZRadial1}-\eqref{eq:NZRadial2} which are now realized as subalgebras of the same larger structure. In the current notation,
\be
	\SL_{\left(+\right)} = \SL_{\left[1,0\right]} \,,\quad \SL_{\left(-\right)} = \SL_{\left[0,1\right]} \,.
\ee
However, the Casimirs with generic $\alpha_{+}$ and $\alpha_{-}$ have another remarkable property: they are the most general globally defined and time-reversal symmetric truncations of the equations of motion which preserve the characteristic exponents in the vicinity of the event horizon (see Appendix \ref{sec:ApSL2RGenerators}).
%\textcolor{blue}{\textbf{In fact, all of these $\SL$ subalgebras can be realized as isometries of subtracted geometries. The corresponding subtracted geometry for the pair $\left(\alpha_{+},\alpha_{-}\right)$ is given by,}}
%\todo[inline]{Write down metric of subtracted geometry}

\subsection{Infinite zones of Love from local time translations}

Beyond the infinite extension described above involving subtracted geometry truncations of the radial Klein-Gordon operator equipped with an enhanced $\SL$ symmetry, there is another, different type of generalization that gives rise to all the possible near-zone truncations of the equations of motion such that an $\SL$ symmetry emerges.
The corresponding generators make up two towers of near-zone $\SL$'s and are given by
\be
	\begin{gathered}
		L_0^{\left(\sigma\right)}\left[g\left(\rho\right)\right] = L_0^{\left(\sigma\right)}\left[0\right] \,, \\
		L_{\pm1}^{\left(\sigma\right)}\left[g\left(\rho\right)\right] = e^{\pm g\left(\rho\right)/\beta}L_{\pm1}^{\left(\sigma\right)}\left[0\right] \pm e^{\pm \left(t+g\left(\rho\right)\right)/\beta}\sqrt{\Delta}\left(\partial_{\rho}g\left(\rho\right)\right)\partial_{t} \,,
	\end{gathered}
\ee
where $L_{m}^{\left(\sigma\right)}\left[0\right]$ are the Love symmetries generators \eqref{eq:LoveGen} and $g\left(\rho\right)$ is an arbitrary radial function which is regular and non-vanishing at the event horizon. All of these $\SL$ algebras, however, can be realized as cousins of the Love symmetries, corresponding to local $\rho$-dependent temporal translations,
\be
	\tilde{t} = t + g\left(\rho\right) \,.
\ee
Indeed, when writing the generators using this time coordinate, they acquire the same form as the Love symmetries generators,
\be
	\begin{gathered}
		L_0^{\left(\sigma\right)}\left[g\left(\rho\right)\right] = -\beta\left(\partial_{\tilde{t}} + \Omega_{\sigma}\,\partial_{\sigma}\right) \,, \\
		L_{\pm1}^{\left(\sigma\right)}\left[g\left(\rho\right)\right] = e^{\pm \tilde{t}/\beta}\left[\mp\sqrt{\Delta}\,\partial_{\rho} + \partial_{\rho}\left(\sqrt{\Delta}\right)\,\beta\left(\partial_{\tilde{t}} + \Omega_{\sigma}\,\partial_{\sigma}\right) + \frac{\rho_{+}-\rho_{-}}{2\sqrt{\Delta}} \beta\Omega_{-\sigma}\,\partial_{-\sigma}\right] \,.
	\end{gathered}
\ee
The associated Casimir is then given by \eqref{eq:LoveCasimir} with $t$ replaced $\tilde{t}$. Furthermore, the argument of vanishing static Love numbers remains unaltered even when using these generalized Love symmetries. Namely, when the conditions for vanishing static Love numbers are satisfied, the corresponding regular at the future event horizon static solution is a descendant in the highest-weight representation of these generalized near-zone $\SL$'s and the highest-weight property dictates the (quasi-)polynomial form of the solution.

\subsection{Vanishing Love numbers as Total Transmission Modes}
In Section~\ref{sec:SL2R} we have demonstrated that highest-weight multiplets are spanned by near-zone solutions with vanishing Love numbers. We will argue here that another interpretation of these states is that they are total transmission modes of the effective near-zone geometry. To show this, we consider the full radial operator in \eqref{eq:FullEOM} and treat it as a scattering problem. The field redefinition
\be
	\Phi = A\left(r\right)\, \Psi \,,\quad A\left(r\right)\equiv \sqrt{\frac{r}{\left(r^2+a^2\right)\left(r^2+b^2\right)}}
\ee
brings the radial problem to its canonical form,
\be
	\left[\partial_{r_{\ast}}^2 - \mathcal{K}^2 - \frac{r^2\Delta}{\left(r^2+a^2\right)^2\left(r^2+b^2\right)^2}\mathcal{V}\right]\Psi = 0 \,,
\ee
where $\mathcal{K}=\partial_{t} + \frac{a}{r^2+a^2}\partial_{\phi} + \frac{b}{r^2+b^2}\partial_{\psi}$ and with the reduced scalar potential given by
\be
	\mathcal{V} = 4\mathbb{P}_{\text{full}} - \frac{a^2b^2}{r^2}\left(\partial_{t}+\frac{1}{a}\,\partial_{\phi}+\frac{1}{b}\,\partial_{\psi}\right)^2 - 2\left(a\,\partial_{\phi}+b\,\partial_{\psi}\right)\partial_{t} - \frac{1}{rA\left(r\right)}\partial_{r}\left(\frac{\Delta}{r}A^{\prime}\left(r\right)\right) \,.
\ee

The setup then consists of an incident scalar wave of frequency $\omega$ coming from infinity and scattered by the black hole. After separating the variables
\be
	\Psi_{\omega\ell mj} = e^{-i\omega t}e^{im\phi}e^{ij\psi}P_{\omega\ell mj}\left(r\right)S_{\omega\ell mj}\left(\theta\right) \,,
\ee
the asymptotic radial wavefunction $P^{\infty}_{\omega\ell mj}$ satisfies the following differential equation in the far-zone region $r\gg r_{+}$,
\be
	\left[\frac{d^2}{dr^2} + \omega^2 - \frac{\ell\left(\ell+2\right)-\frac{3}{4}}{r^2}\right]P^{\infty}_{\omega\ell mj} = 0 \,,
\ee
with $\ell$ an effective orbital number in terms of which the angular eigenvalues are $\ell\left(\ell+2\right)$, but which is non-integer for $\omega\ne0$. The asymptotic solution can then be found in terms of Hankel functions to be
\be\label{eq:FZRadialSolution}
	P^{\infty}_{\omega\ell mj} = \sqrt{\frac{\pi\omega r}{2}}e^{-i\frac{\pi}{2}\left(\ell+\frac{3}{2}\right)}\mathcal{I}_{\ell mj} \left[ H_{\ell+1}^{\left(2\right)}\left(\omega r\right) + \mathcal{R}_{\ell mj}\left(\omega\right)e^{i\pi\left(\ell+\frac{3}{2}\right)}H_{\ell+1}^{\left(1\right)}\left(\omega r\right) \right] \,.
\ee
The integration constants were fixed such that
\be
	P^{\infty}_{\omega\ell mj} \xrightarrow{r\rightarrow\infty} \mathcal{I}_{\ell mj}\left[e^{-i\omega r} + \mathcal{R}_{\ell mj}\left(\omega\right) e^{+i\omega r}\right] \,,
\ee
that is, $\left|\mathcal{I}_{\ell mj}\right|^2$ is the incoming flux and $\mathcal{R}_{\ell mj}\left(\omega\right)$ is the reflection amplitude.

Let us now look at what happens in the near-zone region, $\omega r\ll 1$. For the asymptotic wavefunction solution \eqref{eq:FZRadialSolution} to be supported in the near-zone we must of course also have $\omega r_{s}\ll1$ such that the intermediate region $r_{+}\ll r \ll \omega^{-1}$ is non-empty. From the asymptotic behaviors of the Hankel functions for small arguments, we see that we get an expression of the form
\be
	P^{\infty}_{\omega\ell mj} \xrightarrow{\omega r\ll1} \tilde{\mathcal{I}}_{\ell mj}\left(\omega\right) r^{\ell+\frac{3}{2}}\left[1+k_{\ell mj}\left(\omega\right) \left(\frac{r_{s}}{r}\right)^{2\ell+2}\right] \,,
\ee
from which we can match the response coefficients onto the reflection amplitude. This is in turn related to the real-valued (conservative) phase-shifts $\delta_{\ell}\left(\omega\right)$ and (dissipative) transmission factors $\eta_{\ell}\left(\omega\right)$ that enter a partial wave analysis of the scattering problem~\cite{Matzner1978ApJS,Futterman:1988ni,Dolan:2008kf}. Namely, to linear order in the response coefficients, we have~\cite{Ivanov:2022qqt}
\be
	\eta_{\ell}\left(\omega\right)e^{2i\delta_{\ell}\left(\omega\right)} = e^{i\pi\left(\ell+\frac{3}{2}\right)}\mathcal{R}_{\ell00}\left(\omega\right) = 1 +i\frac{2\pi\sin^2\pi\left(\ell+1\right)}{\Gamma\left(\ell+1\right)\Gamma\left(\ell+2\right)}\left(\frac{\omega r_{s}}{2}\right)^{2\ell+2} k_{\ell 00}\left(\omega\right)
\ee
and, therefore, we can extract a matching condition between the real and imaginary parts of the response coefficients and the phase-shifts and transmission factors respectively~\cite{Ivanov:2022qqt},
\be\ba
	\eta_{\ell}\left(\omega\right) &= 1 -\frac{2\pi\sin^2\pi\left(\ell+1\right)}{\Gamma\left(\ell+1\right)\Gamma\left(\ell+2\right)}\left(\frac{\omega r_{s}}{2}\right)^{2\ell+2} \text{Im}\left\{k_{\ell00}\left(\omega\right)\right\} \,, \\
	\delta_{\ell}\left(\omega\right) &= \frac{\pi\sin^2\pi\left(\ell+1\right)}{\Gamma\left(\ell+1\right)\Gamma\left(\ell+2\right)}\left(\frac{\omega r_{s}}{2}\right)^{2\ell+2} \text{Re}\left\{k_{\ell00}\left(\omega\right)\right\} \,,
\ea\ee
which is an alternative way to see that the Love numbers enter only in the conservative dynamics. From these expressions, one can then compute elastic and absorption cross-sections. In $d=1+4$ spacetime dimensions a partial wave analysis of the scattering problem results in
\be\ba
	\sigma_{\text{elastic}} &= \frac{16\pi}{\omega^3}\sum_{\ell=0}^{\infty}\left(\ell+1\right)^2\left(\ell+2\right)\sin^2\delta_{\ell}\left(\omega\right) \,, \\
	\sigma_{\text{absorption}} &= \frac{4\pi}{\omega^3}\sum_{\ell=0}^{\infty}\left(\ell+1\right)^2\left(\ell+2\right)\left[1-\eta_{\ell}^2\left(\omega\right)\right] \,.
\ea\ee
Consequently, vanishing (dynamical) Love numbers
for certain frequencies 
imply a vanishing partial elastic cross-section,
\be
	\sigma_{\text{elastic},\ell} = 0 \quad \text{ if } \quad k_{\ell00}^{\text{Love}}\left(\omega\right) = 0 \,,
\ee
while the corresponding partial absorption cross-section is maximized. Vanishing Love numbers are, thus, interpreted as reflectionless, total transmission modes~\cite{Hod:2013fea,Cook:2016fge,Cook:2016ngj,Ivanov:2022qqt}. We note, however, that strictly speaking, 
the connection of the Love symmetries highest-weight multiplet states and total transmission modes is not $1$-to-$1$. For instance, the analysis of~\cite{Cook:2016fge} indicates that, for the $d=4$ Kerr black hole, the corresponding Love symmetry highest-weight multiplet~\cite{Charalambous:2021kcz,Charalambous:2022rre} does appear to capture these algebraically special modes but is also over-counting them, in the sense that it also contains states that are not true algebraically special quasinormal modes of the full geometry that 
includes the asymptotically flat region. 
This feature is quite surprising because the Love symmetries manifest themselves only in the near-zone region and, therefore, their highest-weight multiplet states are expected to be accurate only in the regime of low perturbation frequencies. We leave a better understanding of this connection for future work.
\section{Relation to NHE isometries}
\label{sec:NHE}
We have seen that the near-zone $\SL$ symmetries can be realized as particular subalgebras of the infinite extension $\SL\ltimes U\left(1\right)^2_{\mathcal{V}}$ and that all of these $\SL$ subalgebras of the infinite extension are approximate symmetries of the $5$-d Myers-Perry black hole, in the sense that they are isometries of geometries that preserve the internal structure of the black hole. Here, we will relate these approximate symmetries with the exact isometries of the near-horizon region of the extremal $5$-d Myers-Perry black hole.

We will start with a brief review of how an enhanced $\SL$ symmetry for the extremal $5$-d Myers-Perry black hole arises in the near-horizon region~\cite{Bardeen:1999px,Kunduri:2007vf,Galajinsky:2012vh,Galajinsky:2013mla,Hakobyan:2017qee,Figueras:2008qh}. Then, we will demonstrate how to take an appropriate extremal limit of the approximated $\SL$ symmetries which will precisely recover the $\SL$ Killing vectors of the near-horizon extremal (NHE) geometry.
 
\subsection{Enhanced symmetries for NHE Myers-Perry black hole}
Consider the extremal $5$-d Myers-Perry black hole geometry. The extremality condition reads,
\be
	\left|a\right|+\left|b\right| = r_{s} \,,
\ee
with the degenerate horizon located at,
\be
	\rho_{+} = \rho_{-} = \frac{\rho_{s}-a^2-b^2}{2} = \left|ab\right| \,.
\ee
%
%The metric in $\left(t,\rho,\phi,\psi\right)$ coordinates is given by,
%\be
%	ds^2 = -dt^2 + \frac{\rho_{s}}{\Sigma}\left(dt-a\sin^2\theta\,d\phi-b\cos^2\theta\,d\psi\right)^2 + \Sigma\left(\frac{d\rho^2}{4\Delta} + d\theta^2\right) + \left(\rho+a^2\right)\sin^2\theta\,d\phi^2 + \left(\rho+b^2\right)\cos^2\theta\,d\psi^2 \,.
%\ee
To obtain the near-horizon geometry, we perform the following change into co-rotating coordinates~\cite{Bardeen:1999px,Galajinsky:2012vh,Galajinsky:2013mla}
\be
	\tilde{\rho} = \frac{\rho-\rho_{+}}{\lambda} \,,\quad \tau=\lambda t \,,\quad \tilde{\phi} = \phi - \Omega_{\phi}\,t \,,\quad  \tilde{\psi} = \psi - \Omega_{\psi}\,t \,,
\ee
and take the scaling limit $\lambda\rightarrow 0$. The resulting NHE geometry is then given by~\cite{Galajinsky:2012vh,Galajinsky:2013mla,Hakobyan:2017qee,Figueras:2008qh}
\be
	ds^2_{\text{NHE}} = \frac{\Sigma_{+}}{\rho_0}\left[-\left(\frac{\tilde{\rho}}{\rho_0}\right)^2d\tau^2 + \left(\frac{\rho_0}{2\tilde{\rho}}\right)^2\frac{d\tilde{\rho}^2}{\rho_0}+\rho_0\,d\theta^2\right] + \sum_{i,j=1}^{2}\tilde{\gamma}_{ij}D\tilde{\phi}^{i}D\tilde{\phi}^{j} \,.
\ee
In the above expression, $\rho_0^3=\rho_{+}\rho_{s}^2$, $\Sigma_{+} = \rho_{+}+a^2\cos^2\theta+b^2\sin^2\theta$, small Latin indices label the azimuthal angles with $\tilde{\phi}^1\equiv\tilde{\phi}$ and $\tilde{\phi}^2\equiv\tilde{\psi}$, $\tilde{\gamma}_{ij}$ is the induced metric at the horizon along the azimuthal directions,
\be\ba
	\sum_{i,j=1}^{2}\tilde{\gamma}_{ij}d\tilde{\phi}^{i}d\tilde{\phi}^{j} &= \frac{\rho_{s}}{\Sigma_{+}}\left(a\sin^2\theta\,d\tilde{\phi} + b\cos^2\theta\,d\tilde{\psi}\right)^2 \\
	&+ \left(\rho_{+}+a^2\right)\sin^2\theta\,d\tilde{\phi}^2 + \left(\rho_{+}+b^2\right)\cos^2\theta\,d\tilde{\psi}^2 \,,
\ea\ee
and $D\tilde{\phi}^{i} = d\tilde{\phi}^{i}+k^{i}\tilde{\rho}\,d\tau$ with
\be
	k^{\tilde{\phi}} = \frac{\Omega_{\phi}}{\Sigma_{+}}\left[\cos^2\theta+\frac{b^2}{\rho_{+}}\sin^2\theta\right] \,,\quad k^{\tilde{\psi}} = \frac{\Omega_{\psi}}{\Sigma_{+}}\left[\sin^2\theta+\frac{a^2}{\rho_{+}}\cos^2\theta\right] \,.
\ee
This form of the NHE metric makes the enhanced isometry of the geometry manifest. More explicitly, the NHE geometry of the $5$-d Myers-Perry black hole has an $\SL\times U\left(1\right)^2$ isometry, with $U\left(1\right)^2$ generated by the azimuthal Killing vectors and the $\SL$ Killing vectors given by
%\be
%	\begin{gathered}
%		\xi_0 = \tau\,\partial_{\tau} - \tilde{\rho}\,\partial_{\tilde{\rho}} \,\,\,,\,\,\, \xi_{+1} = \partial_{\tau} \,, \\
%		\xi_{-1} = \left(\frac{\rho_{+}\rho_{s}^2}{4\tilde{\rho}^2} + \tau^2\right)\,\partial_{\tau} - 2\tau\tilde{\rho}\,\partial_{\tilde{\rho}} - \frac{\rho_{+}\rho_{s}^2}{2\tilde{\rho}}\sum_{i=1}^{2}\frac{1}{\rho_{+}+a_{i}^2}\Omega_{\phi^{i}}\,\partial_{\tilde{\phi}^{i}} \,,
%	\end{gathered}
%\ee
%where $a_1=a$ and $a_2=b$ are understood. In the initial $\left(t,\rho,\phi,\psi\right)$ coordinates, the $\SL$ Killing vectors read,
\be\label{eq:SL2RKV_BH}
	\begin{gathered}
		\xi_0 = \tau\,\partial_{\tau} - \tilde{\rho}\,\partial_{\tilde{\rho}} \,\,\,,\,\,\, \xi_{+1} = \partial_{\tau} \,, \\
		\xi_{-1} = \left(\frac{\rho_{+}\rho_{s}^2}{4\tilde{\rho}^2} + \tau^2\right)\,\partial_{\tau} - 2\tau\tilde{\rho}\,\partial_{\tilde{\rho}} - \frac{\rho_{+}\rho_{s}}{2\tilde{\rho}}\left(\frac{1}{a}\,\partial_{\tilde{\phi}}+\frac{1}{b}\,\partial_{\tilde{\psi}}\right) \,.
	\end{gathered}
\ee
In the initial $\left(t,\rho,\phi,\psi\right)$ coordinates, the $\SL$ Killing vectors read
%\be\ba
%	\xi_0 &= t\left(\partial_{t}+\Omega_{\phi}\,\partial_{\phi}+\Omega_{\psi}\,\partial_{\psi}\right) - \left(\rho-\rho_{+}\right)\partial_{\rho} \,, \\
%	\xi_{+1} &= \lambda\left(\partial_{t}+\Omega_{\phi}\,\partial_{\phi}+\Omega_{\psi}\,\partial_{\psi}\right) \,, \\
%	\xi_{-1} &= \lambda\bigg[ \left(\frac{\rho_{+}\rho_{s}^2}{4\left(\rho-\rho_{+}\right)^2} + t^2 \right)\left(\partial_{t} + \Omega_{\phi}\,\partial_{\phi} + \Omega_{\psi}\partial_{\psi}\right) \\
%	&\quad\quad-2t\left(\rho-\rho_{+}\right)\partial_{\rho} - \frac{\rho_{+}\rho_{s}}{2\left(\rho-\rho_{+}\right)}\left(\frac{1}{a}\,\partial_{\phi} + \frac{1}{b}\,\partial_{\psi}\right) \bigg] \,.
%\ea\ee
\be\label{eq:SL2RKV_BL}
	\begin{gathered}
		\xi_0 = t\,K - \left(\rho-\rho_{+}\right)\partial_{\rho} \,,\quad \xi_{+1} = \lambda^{-1}\,K \,, \\
		\xi_{-1} = \lambda\bigg[ \left(\frac{\rho_{+}\rho_{s}^2}{4\left(\rho-\rho_{+}\right)^2} + t^2 \right)K -2t\left(\rho-\rho_{+}\right)\partial_{\rho} - \frac{\rho_{+}\rho_{s}}{2\left(\rho-\rho_{+}\right)}\left(\frac{1}{a}\,\partial_{\phi} + \frac{1}{b}\,\partial_{\psi}\right) \bigg] \,,
	\end{gathered}
\ee
where $K$ is the Killing vector that becomes null at the event horizon,
\be
	K = \partial_{t} + \Omega_{\phi}\,\partial_{\phi} + \Omega_{\psi}\,\partial_{\psi} = \lambda\,\partial_{\tau} \,.
\ee

The Casimir associated with this $\SL$ algebra is given by
%\be\ba
%	\mathcal{C}_2^{\SL} &= \partial_{\tilde{\rho}}\,\tilde{\rho}^2\,\partial_{\tilde{\rho}} - \frac{\rho_{+}\rho_{s}^2}{4\tilde{\rho}^2}\,\partial_{\tau}^2 + \frac{\rho_{+}\rho_{s}^2}{2\tilde{\rho}}\sum_{i=1}^{2}\frac{1}{\rho_{+}+a_{i}^2}\,\Omega_{\phi^{i}}\,\partial_{\tau}\partial_{\tilde{\phi}^{i}} \\
%	&= \,.
%\ea\ee
\be\ba\label{eq:SL2RKV_Casimir}
%	\mathcal{C}_2^{\SL} &= \partial_{\tilde{\rho}}\,\tilde{\rho}^2\,\partial_{\tilde{\rho}} - \frac{\rho_{+}\rho_{s}^2}{4\tilde{\rho}^2}\,\partial_{\tau}^2 + \frac{\rho_{+}\rho_{s}}{2\tilde{\rho}}\left(\frac{1}{a}\,\partial_{\tilde{\phi}}+\frac{1}{b}\,\partial_{\tilde{\psi}}\right)\partial_{\tau} \\
%	&= \partial_{\rho}\left(\rho-\rho_{+}\right)^2\partial_{\rho} -\frac{\rho_{+}\rho_{s}^2}{4\left(\rho-\rho_{+}\right)^2}\left(\partial_{t}+\Omega_{\phi}\,\partial_{\phi}+\Omega_{\psi}\,\partial_{\psi}\right)^2 \\
%	&\quad+ \frac{\rho_{+}\rho_{s}}{2\left(\rho-\rho_{+}\right)}\left(\frac{1}{a}\,\partial_{\phi}+\frac{1}{b}\,\partial_{\psi}\right)\left(\partial_{t}+\Omega_{\phi}\,\partial_{\phi}+\Omega_{\psi}\,\partial_{\psi}\right) \,,
	\mathcal{C}_2^{\SL} &= \partial_{\tilde{\rho}}\,\tilde{\rho}^2\,\partial_{\tilde{\rho}} - \frac{\rho_{+}\rho_{s}^2}{4\tilde{\rho}^2}\,\partial_{\tau}^2 + \frac{\rho_{+}\rho_{s}}{2\tilde{\rho}}\left(\frac{1}{a}\,\partial_{\tilde{\phi}}+\frac{1}{b}\,\partial_{\tilde{\psi}}\right)\partial_{\tau} \\
	&= \partial_{\rho}\left(\rho-\rho_{+}\right)^2\partial_{\rho} -\frac{\rho_{+}\rho_{s}^2}{4\left(\rho-\rho_{+}\right)^2}K^2 + \frac{\rho_{+}\rho_{s}}{2\left(\rho-\rho_{+}\right)}\left(\frac{1}{a}\,\partial_{\phi}+\frac{1}{b}\,\partial_{\psi}\right)K \,,
\ea\ee
and correctly reproduces the full radial operator for the Klein-Gordon equation in the NHE limit after supplementing with the $U\left(1\right)^2$ contributions,
%\be
%	\mathbb{O}_{\text{full}}^{\left(0\right)} = \mathcal{C}_2^{\SL}-\frac{1}{4}\left(2\rho_{s}-\rho_{+}\right)\left( \Omega_{\phi}\,\partial_{\tilde{\phi}}+\Omega_{\psi}\,\partial_{\tilde{\psi}} \right)^2 + \mathcal{O}\left(\lambda\right) \,.
%\ee
\be
	\mathbb{O}_{\text{full}} = \mathcal{C}_2^{\SL}-\frac{1}{4}\left(2\rho_{s}-\rho_{+}\right)\left( \Omega_{\phi}\,\partial_{\phi}+\Omega_{\psi}\,\partial_{\psi} \right)^2 + \mathcal{O}\left(\lambda\right) \,.
\ee

\subsection{NHE algebra from infinite extension}
Let us demonstrate now how the NHE $\SL$ Killing vectors can be recovered from the non-extremal $\SL\ltimes U\left(1\right)^2_{\mathcal{V}}$. We parameterize the extremal limit in terms of the Hawking temperature, $T_{H}\rightarrow0$. For example,
\be
	\left|a\right|+\left|b\right| = r_{s}\left(1-4\pi^2T_{H}^2r_{s}^2\right) + \mathcal{O}\left(T_{H}^3\right) \,.
\ee
Consider the $\SL$ subalgebra of $\SL\ltimes U\left(1\right)^2_{\mathcal{V}}$ corresponding to the choices\footnote{We note here that we can always consider a family of such $\SL$ subalgebras for which one adds arbitrary $\mathcal{O}\left(T_{H}^2\right)$ terms in $\alpha_{\pm}$. We also remark that $\alpha_{\sigma_{ab}} = 1 + \frac{r_{+}}{2}\pi T_{H} + \mathcal{O}\left(T_{H}^2\right)$ and $\alpha_{-\sigma_{ab}}=\mathcal{O}\left(T_{H}^2\right)$, where $\sigma_{ab}$ is the sign of the product of the two spin parameters of the black hole.}
\be
	\alpha_{\pm}^{\text{NHE}} = 1 + \frac{r_{+}}{\Omega_{\pm}}\left(\frac{1}{a}\pm\frac{1}{b}\right)\pi T_{H} \,.
\ee
The generators of this algebra are given by
\be
	\begin{gathered}
		L_0^{\text{NHE}} = -\frac{K}{2\pi T_{H}} - \frac{r_{+}}{2}\left(\frac{1}{a}\,\partial_{\phi} + \frac{1}{b}\,\partial_{\psi}\right) \,, \\
		L_{\pm1}^{\text{NHE}} = e^{\pm 2\pi T_{H}t}\left[\mp\sqrt{\Delta}\,\partial_{\rho} + \partial_{\rho}\left(\sqrt{\Delta}\right)\frac{K}{2\pi T_{H}} + \sqrt{\frac{\rho-\rho_{+}}{\rho-\rho_{-}}}\,\frac{r_{+}}{2}\left(\frac{1}{a}\,\partial_{\phi} + \frac{1}{b}\,\partial_{\psi}\right)\right] \,,
	\end{gathered}
\ee
and produce the following Casimir operator
\be\ba
	\mathcal{C}_2^{\text{NHE}} &= \partial_{\rho}\,\Delta\,\partial_{\rho} - \frac{\rho_{+}\rho_{s}^2}{4\Delta}\,K^2 + \frac{\rho_{+}\rho_{s}}{2\left(\rho-\rho_{-}\right)} \left(\frac{1}{a}\,\partial_{\phi} + \frac{1}{b}\,\partial_{\psi}\right)K \\
	&\quad +\frac{\rho_{+}-\rho_{-}}{\rho-\rho_{-}}\frac{\rho_{+}}{4}\left(\frac{1}{a}\,\partial_{\phi} + \frac{1}{b}\,\partial_{\psi}\right)^2 \,.
\ea\ee
Even though this Casimir does not give rise to a near-zone truncation in the non-extremal case, it has the special property of precisely reproducing the Casimir operator \eqref{eq:SL2RKV_Casimir} associated with the NHE $\SL$ Killing vectors when taking the extremal limit,
\be
	\mathcal{C}_2^{\text{NHE}} = \mathcal{C}_2^{\SL} + \mathcal{O}\left(T_{H}\right) \,.
\ee
In fact, by considering the following linear combinations in the extremal limit
\be\ba
	\xi_0 &= \lim_{T_{H}\rightarrow 0} \frac{L_{+1}^{\text{NHE}} - L_{-1}^{\text{NHE}}}{2} \,, \\
	\xi_{+1} &= \lambda^{-1}\lim_{T_{H}\rightarrow0}2\pi T_{H}\,L_0^{\text{NHE}} \,, \\
	\xi_{-1} &= \lambda\lim_{T_{H}\rightarrow0} \frac{L_{+1}^{\text{NHE}} + L_{-1}^{\text{NHE}} +2 L_0^{\text{NHE}}}{2\pi T_{H}} \,,
\ea\ee
we precisely recover the NHE $\SL$ Killing vectors \eqref{eq:SL2RKV_BL} after identifying $\lambda$ with the near-horizon scaling parameter. We see, therefore, that the infinite extension $\SL\ltimes U\left(1\right)^2_{\mathcal{V}}$ contains both the Love symmetries $\SL_{\left(\sigma\right)}$ associated with the non-extremal near-zone truncations as well as a family of $\SL$ subalgebras which in the extremal limit recover the exact $\SL$ Killing vectors of the NHE geometry.
\section{Summary and Discussion}
\label{sec:Discussion}

In this work, we have extended the proposal of the Love symmetry resolution of the seemingly unnatural values of the black hole Love numbers~\cite{Charalambous:2021kcz,Charalambous:2022rre} to higher-dimensional rotating black holes in General Relativity. Namely, we have explored in full the case of static scalar responses of the $5$-dimensional Myers-Perry black hole.

Compared to the examples of Kerr-Newman black holes in $d=4$ spacetime dimensions~\cite{Charalambous:2021mea} and Schwarzschild black holes in $d=5$ spacetime dimensions~\cite{Kol:2011vg}, we find some interesting exact results. To start with, static scalar Love numbers do not in general vanish in $d=5$ for generic spin parameters, not even when $\hat{\ell}\in\mathbb{N}$, in contrast to the Schwarzschild case~\cite{Kol:2011vg}. Beyond vanishing for ``axisymmetric'' perturbations~\cite{Landry:2015zfa,Pani:2015hfa,Gurlebeck:2015xpa}, we also find that the static Love numbers vanish
for equi-rotating black holes,
which does not have a counterpart in $d=4$. 
We remark here that the current results can be straightforwardly extended to include the case of $5$-d electrically charged Myers-Perry black holes, mainly due to the fact that the discriminant function remains a quadratic polynomial in $\rho$~\cite{Chong:2005hr,Chong:2006zx}. Scalar Love numbers for $5$-d charged Myers-Perry black holes were also considered in~\cite{Consoli:2022eey}, who however focused on their slowly-rotating limits thus missing the classical RG flow feature, which we study in detail here.

It appears that the vanishing of static Love numbers for rotating black holes in $d=4$ is an exception rather than the norm. Indeed, as we have demonstrated in this work, Love numbers for rotating black holes in $d=5$ are in general non-zero and exhibit running, in agreement with
Wilsonian 
naturalness arguments. Regardless, we were still able to find near-zone truncations acquiring $\SL$ Love symmetries just like in $d=4$ Kerr-Newman black holes and $d\ge4$ Reissner-Nordstr\"{o}m black holes~\cite{Bertini:2011ga,Kim:2012mh,Charalambous:2021kcz,Charalambous:2022rre}. In the special situations where Love numbers do vanish, however, it is the highest-property of the corresponding Love symmetry that outputs this vanishing as a selection rule. We see therefore that the existence of near-zone Love symmetries appears to be routed in black holes in General Relativity, rather than only with background geometries and perturbations with vanishing Love numbers.

At the same time, we have demonstrated here that the highest-weight representation of the near-zone $\SL$'s, along with its full extension into the representation of type ``$\circ[\circ[\circ$'', plays a special role in the scalar response problem: it is entirely spanned by near-zone solutions with vanishing/non-running Love numbers. These properties are in fact \'{a} posteriori seen to be shared with the Love symmetry presented in~\cite{Charalambous:2021kcz,Charalambous:2022rre} for the $d=4$ Kerr-Newman black hole. These two features, the existence of near-zone $\SL$ symmetries and the vanishing/non-running of Love numbers, appear therefore to be mutually inclusive, with the solutions of vanishing Love numbers furnishing a quotient representation of the highest-weight Verma module of the near-zone $\SL$. We remind here, though, that only the static results can be trusted within the near-zone regime.

On that account, it is interesting to further study this hypothesis. On the one hand, it is instructive to extend the analysis to other general-relativistic black holes. The obvious next candidate to analyze is the higher-dimensional Myers-Perry black holes whose scalar field perturbations are still separable~\cite{Frolov:2006pe,Frolov:2008jr}. A technical obstacle in this approach, however, is the fact that the angular eigenvalues in $d>5$ are not known in closed form, but can be obtained as an expansion in spin parameters ratios, see e.g.~\cite{Cho:2011yp}. It would be interesting, in particular, to analyze the fate of scalar Love numbers for equi-rotating Myers-Perry black holes in odd spacetime dimensions which have the enhanced isometry subgroup $U\left(1\right)^{N}\rightarrow U\left(N\right)$. Moreover, it only deems appropriate to extend to higher-spin fields, namely, electromagnetic and gravitational perturbations. At least for spin-$1$ perturbations, this should be very similar to the work done here thanks to the separability of electromagnetic perturbations in the background of Myers-Perry black holes~\cite{Lunin:2017drx}.

On the other hand, it is still an open question whether Love symmetry exists in theories of gravity beyond General Relativity. A preliminary analysis around this was done in~\cite{Charalambous:2022rre}, where a sufficient geometric condition was extracted for spherically symmetric black holes. It would be interesting to supplement that analysis with sufficient \text{and} necessary constraints, investigate what type of theories of gravity support such geometries and whether the corresponding Love symmetries live up to their names, i.e. whether they can address the potential vanishing of Love numbers. As a counterexample, it was shown in~\cite{Charalambous:2022rre} that Love symmetry does not exist for the case of Riemann-cubed modifications of general relativity, see also~\cite{Cai:2019npx,Cardoso:2018ptl,DeLuca:2022tkm}. This nicely fitted with the corresponding computation of static scalar Love numbers which were found to be non-zero and exhibit the expected RG flow.

The nature of the Love symmetry is still not fully understood. 
From arguments stated here and in~\cite{Charalambous:2022rre}, 
the approximate Love symmetry for non-extremal black holes can be 
interpreted to be a remnant of enhanced isometries of extremal black holes. 
This could be further supported by studying the perturbations of black objects with non-spherical horizons, namely, black $p$-branes~\cite{Duff:1993ye}. One would then attempt to identify near-zone truncations admitting an $SO\left(p+1,2\right)$ symmetry. We leave this for future work. If such an analysis turns out to yield affirmative results, then the Love symmetry may shine more light on the potential holographic descriptions of asymptotically flat general-relativistic black holes~\cite{Bardeen:1999px,Guica:2008mu,Castro:2010fd,Lu:2008jk}.

There are some unconventional features of the Love symmetries regarding their property to offer IR selection rules. In particular, they have the feature of mixing IR and UV modes as can be seen from the fact that representations of the near-zone $\SL$ have non-zero frequencies and, thus, they are not directly manifested at the level of the worldline EFT. The Love symmetry proposal is then somewhat different from the standard 't Hooft's notion of naturalness~\cite{tHooft:1979rat}. However, the Love symmetry does provide selection rules that dictate the vanishing of Love numbers and hence restore the naturalness in the broader sense. Furthermore, while there are non-zero corrections at next-to-leading order in the near-zone expansion (see, e.g., \cite{Charalambous:2021mea}), these are all suppressed within the near-zone regime and, more importantly, the near-zone approximation becomes exact for static perturbations. Regarding the spectrum of the Love symmetry highest-weight multiplets, the damping of these frequencies comes in the form $-\text{Im}\left\{\omega_{n}\right\} = n/\beta$ and, therefore, the closer to extremality the black hole is (large $\beta$), the more states in the multiplet will be within the near-zone regime, at least for the simplified case of non-rotating black holes. Moreover, in the extremal regime the Love symmetry reduces to the exact isometry of the near-horizon Myers-Perry black hole throat. As one can see, in this case the vanishing of static Love numbers also follows from the generator algebra and the highest-weight property of the relevant representation, see~\cite{Charalambous:2022rre} for the $d=4$ version of this. Away from extremity, our arguments are still valid, but the exact form of black hole perturbations requires corrections which formally become order-one for generic black hole spin values. Static perturbations, nevertheless, do not acquire any corrections and hence the results remain exact for them.

The frequencies of the non-static states furnishing representations of the Love symmetry are themselves also phenomenologically interesting. For the highest-weight representation, the corresponding frequencies have the same form as the ``near-horizon'' modes presented in~\cite{Zimmerman:2011dx}. More interestingly, the purely imaginary spacing is precisely equal to the universal QNMs level spacing as extracted from Padmanabhan's argument~\cite{Padmanabhan:2003fx}. Another possible connection to the QNM spectrum has been suggested in~\cite{Charalambous:2022rre}, where the complex frequencies of highest-weight elements were contrasted to highly damped QNMs and total transmission modes~\cite{Cook:2016fge,Cook:2016ngj}. We would like to stress that the near-zone expansion that makes the 
Love symmetry explicit, can capture only those properties of QNMs that are sensitive to the near-horizon part of black hole effective potential, and therefore may not be sufficient for the recovery of the QNM spectrum.
A more direct way to reveal such ``beyond-near-zone'' connections would be to identify symmetry-breaking parameters that depart from Love symmetric near-zone configurations and apply a spurion analysis to extract relations similar to the Gell-Mann-Okubo mass formulas~\cite{Gell-Mann:1961omu,Okubo:1961jc}.
A potential approach along these lines would be to identify the effective black hole geometries, for which Love symmetries are isometries, as scaling limits of the full asymptotically flat Myers-Perry black hole solution, see e.g.~\cite{Cvetic:2012tr}.

\paragraph{Acknowledgments}
We are grateful to Sergei Dubovsky for many 
insightful comments on the draft of this paper. 
We also thank Barak Kol 
and Zihan Zhou for useful discussions. PC is partially supported by the
2022-2023 Dean's Dissertation Fellowship.

\appendix
\section{Geometry of 5-d Myers-Perry black hole}
\label{sec:Ap5dMPGeometry}

The Myers-Perry geometry describes an electrically neutral, rotating black hole in higher dimensional spacetimes~\cite{Myers:1986un}. In the present work, we focus to the $5$-dimensional case with the geometry in Boyer-Lindquist coordinates given by
\be\ba
	ds^2 = -dt^2 &+ \frac{r_{s}^2}{\Sigma}\left(dt-a\sin^2\theta\,d\phi - b\cos^2\theta\,d\psi\right)^2 + \frac{r^2\Sigma}{\Delta}\,dr^2 + \Sigma\,d\theta^2 \\
	&+ \left(r^2+a^2\right)\sin^2\theta\,d\phi^2 + \left(r^2+b^2\right)\cos^2\theta\,d\psi^2 \,,
\ea\ee
where
\be
	\begin{gathered}
		\Sigma = r^2 + a^2\cos^2\theta + b^2\sin^2\theta \,, \\
		\Delta = \left(r^2+a^2\right)\left(r^2+b^2\right)-r_{s}^2r^2 \,,
	\end{gathered}
\ee
and $\theta\in\left[0,\frac{\pi}{2}\right]$ is a direction cosine angle, while $\phi\in\left[0,2\pi\right)$ and $\psi\in\left[0,2\pi\right)$ are periodically identified azimuthal angles. This geometry describes an asymptotically flat black hole with mass $M$ and angular momenta $J_{\phi}$ and $J_{\psi}$ along the two orthogonal planes of rotation, related to the parameters $r_{s}$, $a$ and $b$ appearing in the line element above according to
\be
	M = \frac{3\pi}{8 G}r_{s}^2\,\,\,,\,\,\,J_{\phi} = \frac{2}{3}Ma\,\,\,,\,\,\,J_{\psi} = \frac{2}{3}Mb \,.
\ee

The horizons correspond to the roots of the discriminant $\Delta$ and are located at radial distances $r=r_{\pm}$, with
\be
	r_{\pm}^2 = \frac{1}{2}\left[r_{s}^2-a^2-b^2\pm\sqrt{\left(r_{s}^2-a^2-b^2\right)^2-4a^2b^2}\right] \,.
\ee
The absence of a naked singularity imposes the inequality
\be
	\left|a\right|+\left|b\right| \le r_{s} \,,
\ee
with its saturation indicating the extremality condition. The event horizon $r_{\text{h}}=r_{+}$ and the Cauchy horizon $r_{\text{C}}=r_{-}$ are Killing horizons with respect to the Killing vectors
\be
	K^{\left(\pm\right)} = \partial_{t} + \Omega_{\phi}^{\left(\pm\right)}\,\partial_{\phi} + \Omega_{\psi}^{\left(\pm\right)}\,\partial_{\psi}
\ee
respectively, where $\Omega_{\phi}^{\left(\pm\right)} = \frac{a}{r_{\pm}^2+a^2}$ and $\Omega_{\psi}^{\left(\pm\right)} = \frac{b}{r_{\pm}^2+b^2}$. In particular, $\Omega_{\phi}^{\left(+\right)}\equiv\Omega_{\phi}$ and $\Omega_{\psi}^{\left(+\right)}\equiv\Omega_{\psi}$ are the black hole angular velocities as realized by a static observer at the exterior. The inverse metric components can be extracted to be
\be\ba
	%g^{\mu\nu}\partial_{\mu}\partial_{\nu} = \frac{1}{\Sigma} &\bigg\{ \left[a^2\sin^2\theta+b^2\cos^2\theta - \left(r^2+a^2+b^2\right)\right]\partial_{t}^2 \\
	g^{\mu\nu}\partial_{\mu}\partial_{\nu} = \frac{1}{\Sigma} &\bigg\{ -\Sigma\,\partial_{t}^2 - \frac{\left(r^2+a^2\right)\left(r^2+b^2\right)r_{s}^2}{\Delta}\left(\partial_{t} + \frac{a}{r^2+a^2}\,\partial_{\phi} + \frac{b}{r^2+b^2}\,\partial_{\psi}\right)^2 \\
	& + \left[\frac{1}{\sin^2\theta} - \frac{a^2-b^2}{r^2+a^2}\right]\partial_{\phi}^2 + \left[\frac{1}{\cos^2\theta} - \frac{b^2-a^2}{r^2+b^2}\right]\partial_{\psi}^2 + \frac{\Delta}{r^2}\,\partial_{r}^2 + \partial_{\theta}^2\bigg\} \,,
\ea\ee
while the only additional term needed to construct the massless Klein-Gordon operator involves the Christoffel symbols,
\be
	g^{\mu\nu}\Gamma_{\mu\nu}^{\sigma}\partial_{\sigma} = -\frac{1}{\Sigma}\bigg\{\frac{1}{r}\frac{d}{dr}\left(\frac{\Delta}{r}\right)\,\partial_{r} + \frac{1}{\sin\theta\cos\theta}\frac{d}{d\theta}\left(\sin\theta\cos\theta\right)\,\partial_{\theta}\bigg\} \,.
\ee

To investigate regularity at the future or the past event horizon, one employs advanced (``$+$'') or retarded (``$-$'') coordinates respectively, related to the Boyer-Lindquist coordinates according to
\be\ba
	dt_{\pm} &= dt \pm \frac{\left(r^2+a^2\right)\left(r^2+b^2\right)}{\Delta}dr \,, \\
	d\varphi_{\pm} &= d\phi \pm a\frac{r^2+b^2}{\Delta}dr \,, \\
	dy_{\pm} &= d\psi \pm b\frac{r^2+a^2}{\Delta}dr \,.
\ea\ee
Explicitly, they are given by
\be\ba\label{eq:NullCoordinates}
	t_{\pm} &= t \pm \left\{r + \frac{1}{2} \frac{r_{s}^2}{r_{+}^2-r_{-}^2}\left[r_{+}\ln\left|\frac{r-r_{+}}{r+r_{+}}\right| - r_{-}\ln\left|\frac{r-r_{-}}{r+r_{-}}\right|\right]\right\} \,, \\
	\varphi_{\pm} &= \phi \pm \frac{1}{2}\frac{a}{r_{+}^2-r_{-}^2}\left[\frac{r_{+}^2+b^2}{r_{+}}\ln\left|\frac{r-r_{+}}{r+r_{+}}\right| - \frac{r_{-}^2+b^2}{r_{-}}\ln\left|\frac{r-r_{-}}{r+r_{-}}\right|\right] \,, \\
	y_{\pm} &= \psi \pm \frac{1}{2}\frac{b}{r_{+}^2-r_{-}^2}\left[\frac{r_{+}^2+a^2}{r_{+}}\ln\left|\frac{r-r_{+}}{r+r_{+}}\right| - \frac{r_{-}^2+a^2}{r_{-}}\ln\left|\frac{r-r_{-}}{r+r_{-}}\right|\right] \,,
\ea\ee
where the integration constants have been fixed to ensure the asymptotic behaviors
\be
	t_{\pm}\xrightarrow{r\rightarrow\infty} t \pm r \,,\quad \,\varphi_{\pm}\xrightarrow{r\rightarrow\infty} \phi \,,\quad \,y_{\pm}\xrightarrow{r\rightarrow\infty} \psi \,,
\ee
as well as a smooth extremal limit,
\be\ba
	t_{\pm} &\xrightarrow{r_{-}\rightarrow r_{+}} t \pm \left\{r-\frac{r_{s}}{2}\left[\frac{r_{s}r}{r^2-r_{+}^2} + \frac{r_{s}}{2r_{+}}\log\left|\frac{r-r_{+}}{r+r_{+}}\right|\right]\right\} \,, \\
	\varphi_{\pm} &\xrightarrow{r_{-}\rightarrow r_{+}} \phi \mp \frac{a}{2r_{+}}\left[\frac{b^2+r_{+}^2}{r_{+}^2}\frac{r_{+}r}{r^2-r_{+}^2} + \frac{b^2-r_{+}^2}{2r_{+}^2}\log\left|\frac{r-r_{+}}{r+r_{+}}\right|\right] \,, \\
	y_{\pm} &\xrightarrow{r_{-}\rightarrow r_{+}} \psi \mp \frac{b}{2r_{+}}\left[\frac{a^2+r_{+}^2}{r_{+}^2}\frac{r_{+}r}{r^2-r_{+}^2} + \frac{a^2-r_{+}^2}{2r_{+}^2}\log\left|\frac{r-r_{+}}{r+r_{+}}\right|\right] \,.
\ea\ee
\section{Modified Spherical Harmonics basis}
\label{sec:ApSphericalHarmonics}

In this appendix, we present the basis used for the modified harmonic expansion, naturally associated with the computation of Love numbers for axisymmetric distributions in $d=5$ spacetime dimensions. For the sake of this, we need to perform a harmonic expansion over the $\mathbb{S}^1\times\mathbb{S}^1$ subpart of $\mathbb{S}^3$, appropriate for the general isometry group factor $U\left(1\right)\times U\left(1\right)$ of such configurations. The basis we are looking for is extracted by solving the eigenvalue problem for the Laplace-Beltrami operator on $\mathbb{S}^3$ expressed in the direction cosine angular coordinates appearing in the Myers-Perry black hole line element,
\be\label{eq:LaplaceBeltramiEigenvalue}
	\left[\frac{1}{\sin\theta\cos\theta}\partial_{\theta}\left(\sin\theta\cos\theta\,\partial_{\theta}\right) + \frac{1}{\sin^2\theta}\,\partial_{\phi}^2 + \frac{1}{\cos^2\theta}\,\partial_{\psi}^2\right] \tilde{Y}_{\lambda}\left(\theta,\phi,\psi\right) = \lambda \tilde{Y}_{\lambda}\left(\theta,\phi,\psi\right) \,.
\ee

Let us present first how this is indeed the Laplace-Beltrami operator on $\mathbb{S}^3$. In usual spherical coordinates $\left(r,\vartheta_1,\vartheta_2,\varphi\right)$, the $4$-dimensional spatial position vector with components $\left(x^1,x^2,x^3,x^4\right)$ in Cartesian coordinates is written as
\be\ba
	x^1 &= r\cos\vartheta_1 \,, \\
	x^2 &= r\sin\vartheta_1\cos\vartheta_2 \,, \\
	x^3 &= r\sin\vartheta_1\sin\vartheta_2\cos\varphi \,, \\
	x^4 &= r\sin\vartheta_1\sin\vartheta_2\sin\varphi \,,
\ea\ee
where $x^1$ plays the role of the $3$-dimensional $z$-axis. In these coordinates, $\vartheta_1\in\left[0,\pi\right]$ and $\vartheta_2\in\left[0,\pi\right]$ are two polar angles, while $\varphi\in\left[0,2\pi\right)$ is a periodically identified azimuthal angle. The Laplacian operator acting on scalar functions then reads
\be
	\triangle_{4} = \partial_{1}^2 + \partial_{2}^2 + \partial_{3}^2 + \partial_{4}^2 = \frac{1}{r^3}\partial_{r}\left(r^3\partial_{r}\right) + \frac{1}{r^2}\triangle_{\mathbb{S}^3}^{\left(0\right)} \,,
\ee
with $\triangle_{\mathbb{S}^3}^{\left(0\right)}$ the Laplace-Beltrami operator on $\mathbb{S}^3$ acting on scalar ($s=0$) functions, which in spherical coordinates is given explicitly by
\be
	\triangle_{\mathbb{S}^3}^{\left(0\right)} = \frac{1}{\sin^2\vartheta_1}\left[\partial_{\vartheta_1}\left(\sin^2\vartheta_1\,\partial_{\vartheta_1}\right) + \frac{1}{\sin\vartheta_2}\partial_{\vartheta_2}\left(\sin\vartheta_2\,\partial_{\vartheta_2}\right) + \frac{1}{\sin^2\vartheta_2}\,\partial_{\varphi}^2\right] \,.
\ee
In order to transit to direction cosine coordinates $\left(r,\theta,\phi,\psi\right)$, we split the four Cartesian coordinates $x^{i}$ into two pairs of two and project onto these two planes of rotation,
\be\ba
	x^1 &= r\cos\theta\cos\psi \,, \\
	x^2 &= r\cos\theta\sin\psi \,, \\
	\\
	x^3 &= r\sin\theta\cos\phi \,, \\
	x^4 &= r\sin\theta\sin\phi \,,
\ea\ee
with $\theta\in\left[0,\frac{\pi}{2}\right]$ a direction cosine angle, while $\phi\in\left[0,2\pi\right)$ and $\psi\in\left[0,2\pi\right)$ are two periodically identified azimuthal angles. Then, the Laplace operator on scalar functions has the same form as in spherical coordinates, but with the Laplace-Beltrami operator now identified as
\be
	\triangle_{\mathbb{S}^3}^{\left(0\right)} = \frac{1}{\sin\theta\cos\theta}\partial_{\theta}\left(\sin\theta\cos\theta\,\partial_{\theta}\right) + \frac{1}{\sin^2\theta}\,\partial_{\phi}^2 + \frac{1}{\cos^2\theta}\,\partial_{\psi}^2 \,.
\ee
In particular, the transformation rule between spherical coordinates $\left(\vartheta_1,\vartheta_2,\varphi\right)$ and direction cosine coordinates $\left(\theta,\phi,\psi\right)$ allows to identify $\phi=\varphi$, while $\psi$ and $\theta$ are related to $\vartheta_1$ and $\vartheta_2$ according to
\be
	\sin\theta = \sin\vartheta_1\sin\vartheta_2\,\,\,,\,\,\,\tan\psi = \tan\vartheta_1\cos\vartheta_2 \,.
\ee
The generators of the algebra $\mathfrak{so}\left(4\right)\simeq\mathfrak{so}\left(3\right)\oplus\mathfrak{so}\left(3\right)$ in direction cosine coordinates are more transparent by introducing the sum/difference azimuthal angles $\psi_{\pm}=\psi\pm\phi$ and they are organized in the two commuting $\mathfrak{so}\left(3\right)$'s, labeled by a sign $\sigma=+$ or $\sigma=-$,
\be
	\begin{gathered}
		J_0^{\left(\sigma\right)} = -i\partial_{\sigma} \,, \\
		J_{\pm1}^{\left(\sigma\right)} = e^{\pm i\psi_{\sigma}}\left[\partial_{2\theta} \pm i\cot2\theta\,\partial_{\sigma} \mp\frac{i}{\sin2\theta}\,\partial_{-\sigma}\right] \,, \\
		\left[J_{\pm1}^{\sigma},J_0^{\sigma^{\prime}}\right] = \mp J_{\pm1}^{\left(\sigma\right)}\delta_{\sigma,\sigma^{\prime}} \,\,\,,\,\,\, \left[J_{\pm1}^{\sigma},J_{\mp1}^{\sigma^{\prime}}\right] = \mp 2J_{0}^{\left(\sigma\right)}\delta_{\sigma,\sigma^{\prime}} \,\,\,,\,\,\, \left[J_{\pm1}^{\sigma},J_{\pm1}^{\sigma^{\prime}}\right] = 0 \,.
	\end{gathered}
\ee

Returning to the eigenvalue problem \eqref{eq:LaplaceBeltramiEigenvalue}, this is reduced to a one-dimensional problem after separating the azimuthal angles as
\be
	\tilde{Y}_{\ell m j}\left(\theta,\phi,\psi\right) = S_{\ell m j}\left(\theta\right)\frac{e^{im\phi}}{\sqrt{2\pi}}\frac{e^{ij\psi}}{\sqrt{2\pi}} \,,
\ee
with the azimuthal numbers $m$ and $j$ being integers by virtue of the periodicity of the angles $\phi$ and $\psi$ with period $2\pi$, and $S_{\ell m j}\left(\theta\right)$ satisfying the first order ordinary differential equation
\be
	\left[\frac{1}{\sin\theta\cos\theta}\frac{d}{d\theta}\left(\sin\theta\cos\theta\,\frac{d}{d\theta}\right) - \frac{m^2}{\sin^2\theta} - \frac{j^2}{\cos^2\theta}\right]S_{\ell m j}\left(\theta\right) = -\ell\left(\ell+2\right)S_{\ell m j}\left(\theta\right) \,.
\ee
We remark here that we have set the eigenvalues to $\lambda \equiv -\ell\left(\ell+2\right)$ such that $\ell$ resembles the corresponding orbital quantum number appearing in scalar spherical harmonics on $\mathbb{S}^3$. However, at this point, $\ell$ is not restricted to be a whole number.

This differential equation must now be solved in parallel with the boundary condition of regularity of $S_{\ell m j}\left(\theta\right)$ along the full domain of the direction cosine angle, $\theta\in\left[0,\frac{\pi}{2}\right]$. The unique solution is not hard to extract,
\be
	S_{\ell m j}\left(\theta\right) = N_{\ell m j}\sin^{\left|m\right|}\theta\,\cos^{\left|j\right|}\theta\,{}_2F_1\left(-\frac{\ell-\left|m\right|-\left|j\right|}{2},\frac{\ell+\left|m\right|+\left|j\right|}{2}+1;1+\left|j\right|;\cos^2\theta\right) \,,
\ee
with $N_{\ell m j}$ a normalization constant, while regularity at $\theta=0$ imposes the discretization condition
\be
	\frac{\ell-\left|m\right|-\left|j\right|}{2} \in \mathbb{N}_0 = \left\{0,1,2,\dots\right\} \,.
\ee
We can, thus, assign whole numbers to $\ell$ as in the usual scalar spherical harmonics of $\mathbb{S}^3$, and restrict the domain of the azimuthal numbers $m$ and $j$ such that the above condition is satisfied. This can be achieved, for example, by letting $\left|m\right|\le\ell$ to resemble the corresponding azimuthal number of the usual scalar spherical harmonics, but restricting $j$ to take integer values according to
\be
	j = -\left(\ell-\left|m\right|\right)\,,\,\, -\left(\ell-\left|m\right|\right) + 2\,,\,\, \dots\,,\,\, \left(\ell-\left|m\right|\right) - 2\,,\,\, \left(\ell-\left|m\right|\right) \,,
\ee
where we emphasize the step $2$ in the successively allowed values of $j$.

The eigenfunctions $\tilde{Y}_{\ell m j}$ can be seen to be orthogonal on $\mathbb{S}^3$ in these coordinates using properties of the Jacobi polynomials, while choosing the normalization constant to be
\be
	N_{\ell m j} = \frac{1}{\left(\left|j\right|\right)!}\sqrt{\left(2\ell+2\right)\frac{\left(\frac{\ell+\left|m\right|+\left|j\right|}{2}\right)!\left(\frac{\ell-\left|m\right|+\left|j\right|}{2}\right)!}{\left(\frac{\ell-\left|m\right|-\left|j\right|}{2}\right)!\left(\frac{\ell+\left|m\right|-\left|j\right|}{2}\right)!}} \,,
\ee
ensures the orthonormality identity
\be
	\oint_{\mathbb{S}^3} d\Omega_3\,\tilde{Y}_{\ell m j}^{\ast}\tilde{Y}_{\ell^{\prime}m^{\prime}j^{\prime}} = \delta_{\ell\ell^{\prime}}\delta_{mm^{\prime}}\delta_{jj^{\prime}} \,,
\ee
where asterisks indicate complex conjugation and the $3$-sphere integration measure in the direction cosine coordinates reads
\be
	\oint_{\mathbb{S}^3} d\Omega_3 = \int_{0}^{\frac{\pi}{2}}d\theta\int_{0}^{2\pi}d\phi\int_{0}^{2\pi}d\psi\,\sin\theta\cos\theta \,.
\ee

Two useful properties of these modified spherical harmonic functions are their transformation under complex conjugation, which simply reverses the sign of the azimuthal numbers,
\be
	\tilde{Y}_{\ell m j}^{\ast} = \tilde{Y}_{\ell,-m,-j} \,,
\ee
and under a parity transformation\footnote{In spherical coordinates, parity acts as $\left(\vartheta_1,\vartheta_2,\varphi\right)\rightarrow\left(\pi-\vartheta_1,\pi-\vartheta_2,\pi+\varphi\right)$ which is translated into directed cosine coordinates to $\left(\theta,\phi,\psi\right)\rightarrow\left(\theta,\pi+\phi,\pi+\psi\right)$.} $\left(\theta,\phi,\psi\right)\rightarrow\left(\theta,\pi+\phi,\pi+\psi\right)$, which only adds an overall phase,
\be
	\tilde{Y}_{\ell m j}\left(\theta,\pi+\phi,\pi+\psi\right) = \left(-1\right)^{m+j}\tilde{Y}_{\ell m j}\left(\theta,\psi,\phi\right) \,.
\ee

Let us count how many such basis states exist for a given value of the orbital number $\ell$. This is given by
\be
	\tilde{d}_{\ell} = \sum_{m=-\ell}^{\ell}\left(\sum_{j=-\left(\ell-\left|m\right|\right),2}^{\ell-\left|m\right|}\right) = \sum_{m=-\ell}^{\ell}\left(\ell-\left|m\right|+1\right) = \left(\ell+1\right)^2
\ee
and is exactly the same as the degeneracy of the scalar spherical harmonics\footnote{The scalar spherical harmonics of degree $\ell$ on $\mathbb{S}^{n}$ have a degeneracy,
\be
	d_{\ell}\left(n\right) = \frac{\left(2\ell+n-1\right)\left(\ell+n-2\right)!}{\ell!\left(n-1\right)!} \,.
\ee
For $n=3$, this gives $d_{\ell}\left(3\right)=\left(\ell+1\right)^2$.} on $\mathbb{S}^3$. Subsequently, the modified spherical harmonics $\tilde{Y}_{\ell m j}$ are equivalent to the usual scalar spherical harmonics on $\mathbb{S}^3$. In particular, the scalar spherical harmonics $Y_{\ell,\ell_2,\mu}$ on $\mathbb{S}^3$, with $\left|\mu\right|\le\ell_2\le\ell$ can always be written as a linear combination of the modified spherical harmonics,
\be
	Y_{\ell,\ell_2,\mu}\left(\vartheta_1,\vartheta_2,\varphi\right) = \sum_{m,j}c_{\ell,\ell_2,\mu;m,j}\tilde{Y}_{\ell m j}\left(\theta,\phi,\psi\right) \,,
\ee
and, thus, they form a complete set of $\tilde{d}_{\ell}$ linearly independent and orthonormal unit vectors. We remark here that the azimuthal numbers $\mu$ and $m$ are \textit{not} the same as they appear in $Y_{\ell,\ell_2,\mu}$ and $Y_{\ell m j}$; while $\mu$ and $m$ have the same range in both spherical harmonics bases, their multiplicities do not match. Nevertheless, in the above expansion it is straightforward to see that $c_{\ell,\ell_2,\mu;m,j}=c_{\ell,\ell_2,\mu;j}\delta_{m,\mu}$. In addition, the complex conjugacy relation $Y_{\ell,\ell_2,\mu}^{\ast} = \left(-1\right)^{\mu}Y_{\ell,\ell_2,-\mu}$ further implies
\be
	c_{\ell,\ell_2,\mu;j}^{\ast} = \left(-1\right)^{\mu} c_{\ell,\ell_2,-\mu;-j}\,.
\ee

One useful consequence of this completeness is that the two spherical harmonics bases obey the same addition theorem, expressed in terms of the Gegenabuer polynomials $C_{\ell}^{\left(1\right)}\left(x\right)$ as
\be
	\sum_{m,j}\tilde{Y}_{\ell m j}\left(\mathbf{\Omega}\right)\tilde{Y}_{\ell m j}^{\ast}\left(\mathbf{\Omega}^{\prime}\right) = \sum_{\ell_2,\mu}Y_{\ell,\ell_2,\mu}\left(\mathbf{\Omega}\right)Y_{\ell,\ell_2,\mu}^{\ast}\left(\mathbf{\Omega}^{\prime}\right) = \frac{\ell+1}{2\pi^2}C_{\ell}^{\left(1\right)}\left(\mathbf{\Omega}\cdot\mathbf{\Omega}^{\prime}\right) \,.
\ee

\subsection{Correspondence with STF tensors}
We will now present the $1$-to-$1$ correspondence between $4$-dimensional spatial STF tensors of rank-$\ell$ and the modified spherical harmonics $\tilde{Y}_{\ell m j}$ with the same orbital number $\ell$. There are two key observations that ensure this correspondence. First, the basic STF tensor of rank-$\ell$ $\Omega^{\left\langle L\right\rangle}$, where $\Omega^{i}=\frac{x^{i}}{r}$ are the projectors along the $i$-th spatial direction $x^{i}$, has eigenvalue $-\ell\left(\ell+2\right)$ under the action of the $4$-dimensional flat space Laplace operator, i.e. the same eigenvalue as all $\tilde{Y}_{\ell m j}$ with the same $\ell$.

Second, the number of independent components of a rank-$\ell$ STF tensor in  $n+1$ spatial dimensions is the number of degrees of freedom of a rank-$\ell$ symmetric tensor minus the number of traces that need to be removed,
\be
	d_{\ell}^{\text{STF}}\left(n\right) =
	\begin{pmatrix}
		n+\ell \\
		\ell
	\end{pmatrix} -
	\begin{pmatrix}
		n+\ell-2 \\
		\ell-2
	\end{pmatrix}
	= \frac{\left(2\ell+n-1\right)\left(\ell+n-2\right)!}{\ell!\left(n-1\right)!} \,.
\ee
For $n=3$, this gives that the number of independent components of a $4$-dimensional spatial STF tensor of rank-$\ell$ is equal to $\left(\ell+1\right)^2$, i.e. the same as the number $\tilde{d}_{\ell}$ of basis function $\tilde{Y}_{\ell m j}$ with orbital number $\ell$. Consequently, any $4$-dimensional spatial STF tensor of rank-$\ell$ can be written as a linear combination of $\tilde{Y}_{\ell m j}$ for all the possible values of the azimuthal numbers $m$ and $j$. In particular, for the basic STF tensor $\Omega^{\left\langle L \right\rangle}$,
\be
	\Omega^{\left\langle L \right\rangle} = A_{\ell}\sum_{m,j}\mathcal{Y}^{L}_{\ell m j}\tilde{Y}_{\ell m j}\left(\mathbf{\Omega}\right) \,,
\ee
where $\mathbf{\Omega}$ is a shorthand for the direction cosine coordinates $\left(\theta,\phi,\psi\right)$ associated with the position vector $x^{i}$ from which $\Omega^{i}$ is constructed, the constant STF tensors $\mathcal{Y}^{L}_{\ell m j}$ are given by
\be
	\mathcal{Y}^{L}_{\ell m j} = \frac{1}{A_{\ell}}\oint_{\mathbb{S}^3} d\Omega_{3}\,\Omega^{\left\langle L \right\rangle}\tilde{Y}_{\ell m j}^{\ast}\left(\mathbf{\Omega}\right) \,,
\ee
and $A_{\ell}$ is a real normalization constant chosen such that\footnote{Using the addition theorem of scalar spherical harmonics, which is also an addition theorem for the modified spherical harmonics, this normalization constant can be found to be,
\be
	A_{\ell} = \frac{4\pi^2\ell!}{\left(2\ell+2\right)!!}
\ee}
\be
	\tilde{Y}_{\ell m j}\left(\mathbf{\Omega}\right) = \mathcal{Y}^{L\ast}_{\ell m j}\Omega_{\left\langle L \right\rangle} \,.
\ee

A general $4$-dimensional spatial STF tensor $\mathcal{E}^{L}=\mathcal{E}^{\left\langle L \right\rangle}$ can then be expanded into its modified multipole moments $\mathcal{E}_{\ell m j}$ according to
\be
	\mathcal{E}^{L} = \sum_{m,j}\mathcal{E}_{\ell m j}\mathcal{Y}^{L\ast}_{\ell m j} \,,
\ee
with
\be
	\mathcal{E}_{\ell m j} = A_{\ell} \mathcal{E}_{L}\mathcal{Y}^{L}_{\ell m j} = \mathcal{E}_{L}\oint_{\mathbb{S}^3} d\Omega_{3}\,\Omega^{\left\langle L \right\rangle}\tilde{Y}_{\ell m j}^{\ast}\left(\mathbf{\Omega}\right) \,,
\ee
such that
\be
	\mathcal{E}_{L}\Omega^{L} = \sum_{m,j}\mathcal{E}_{\ell m j}\tilde{Y}_{\ell m j} \,.
\ee

Last, from the complex conjugacy relation of the modified spherical harmonics, we can see that
\be
	\mathcal{Y}^{L\ast}_{\ell m j} = \mathcal{Y}^{L}_{\ell, -m, -j} \,,\quad \mathcal{E}_{\ell m j}^{\ast} = \mathcal{E}_{\ell, -m, -j} \,.
\ee
\section{Source/response split and Schwarzschild limit}
\label{sec:ApSchwarzschildLimit}

In this appendix we demonstrate explicitly the source/response split of the scalar field \eqref{eq:NZRadialSolution} and how the spinless limit recovers the corresponding solution in the $5$-d Schwarzschild black hole background~\cite{Kol:2011vg}. To extract the source and response parts of the solution, we need to identify those components that solve the linearized Klein-Gordon equation and asymptotically behave as
\be
	\begin{gathered}
		R_{\omega\ell m j}\left(\rho\right) = \bar{\mathcal{E}}_{\ell m j}\left(\omega\right)\,\rho^{\hat{\ell}}\left[Z_{\omega\ell m j}^{\text{source}}\left(\rho\right) + k_{\ell m j}\left(\omega\right)\,\left(\frac{\rho_{s}}{\rho}\right)^{2\hat{\ell}+1} Z_{\omega\ell m j}^{\text{response}}\left(\rho\right)\right] \,, \\
		Z_{\omega\ell m j}^{\text{source/response}}\left(\rho\right) \xrightarrow{\rho\rightarrow\infty} 1 \,,
	\end{gathered}
\ee
where $k_{\ell m j}$ are the scalar response coefficients. The source/response split ambiguity that is encountered for the physical values $\ell\in\mathbb{N}$ is bypassed by treating the orbital number $\ell$ as a generic real-valued number $\ell\in\mathbb{R}$. The above asymptotic behaviors then allow to uniquely distinguish between the source and response components of the solution without worrying about ambiguities arising from overlapping series.

The hypergeometric function involved in the near-zone scalar field solution that is regular at the future event horizon \eqref{eq:NZRadialSolution} is expressed in terms of the variable $x=\frac{\rho-\rho_{+}}{\rho_{+}-\rho_{-}}$. Due to a branch cut at $\left|x\right|=1$, this series needs to be analytically continued at large distances according to
\be\ba
	{}_2F_1\left(a,b;c;z\right) &= \frac{\Gamma\left(c\right)\Gamma\left(b-a\right)}{\Gamma\left(b\right)\Gamma\left(c-a\right)}\left(-z\right)^{-a}{}_2F_1\left(a,a-c+1;a-b+1;\frac{1}{z}\right) \\
	&+ \frac{\Gamma\left(c\right)\Gamma\left(a-b\right)}{\Gamma\left(a\right)\Gamma\left(c-b\right)}\left(-z\right)^{-b}{}_2F_1\left(b,b-c+1;b-a+1;\frac{1}{z}\right) \,,
\ea\ee
which is valid for $\left|\text{Arg}\left(-z\right)\right|<\pi$. Applying this prescription for \eqref{eq:NZRadialSolution}, we match the normalization constants $\bar{R}_{\ell m j}\left(\omega\right)$ to
\be
	\bar{R}_{\ell m j}\left(\omega\right) = \bar{\mathcal{E}}_{\ell m j}\left(\omega\right)\left(\rho_{+}-\rho_{-}\right)^{\hat{\ell}}\frac{\Gamma\left(\hat{\ell}+1+i\Gamma^{\left(\sigma\right)}_{+\sigma}\right)\Gamma\left(\hat{\ell}+1+i\Gamma^{\left(\sigma\right)}_{-\sigma}\right)}{\Gamma\left(2\hat{\ell}+1\right)\Gamma\left(1+2iZ_{+}\left(\omega\right)\right)} \,,
\ee
and, then, identify the full source and response components to be\footnote{We remark here the analytic properties of the hypergeometric function ${}_2F_1\left(a,b;c;z\right)$ with respect to its arguments and how this is related to the near-zone scalar field solution \eqref{eq:NZRadialSolution}. In general, ${}_2F_1\left(a,b;c;z\right)$ does not exist when $c$ is a non-positive integer and the analytic continuation formula for large $z$ becomes problematic. Nevertheless, the \textit{regularized} hypergeometric function,
\be
	\mathbf{F}\left(a,b;c,z\right) \equiv \frac{{}_2F_1\left(a,b;c;z\right)}{\Gamma\left(c\right)} \,,
\ee
is well-defined for all values of $c\in\mathbb{C}$ and can be analytically continued at $\left|z\right|>1$ without any issues. This is also the situation with our near-zone scalar field solution \eqref{eq:NZRadialSolution} as the normalization constants $\bar{R}_{\ell m j}\left(\omega\right)$ themselves contain precisely an inverse $\Gamma\left(c\right)$ factor when matched to the field strengths $\bar{\mathcal{E}}_{\ell m j}\left(\omega\right)$. In other words, the solution \eqref{eq:NZRadialSolution} has all the nice analytic properties with respect to all the arguments of the hypergeometric function.}
\be
	\begin{gathered}
		\ba
			Z_{\omega\ell m j}^{\text{source}}\left(\rho\right) &= \left(1-\frac{\rho_{+}}{\rho}\right)^{\hat{\ell}} \left(\frac{\rho-\rho_{-}}{\rho-\rho_{+}}\right)^{\mp i\sigma\tilde{Z}_{-}^{\left(\sigma\right)}\left(\omega\right)} \\
			&\times\,{}_2F_1\left(-\hat{\ell}+i\Gamma^{\left(\sigma\right)}_{\mp\sigma}\left(\omega\right),-\hat{\ell}-i\Gamma^{\left(\sigma\right)}_{\pm\sigma}\left(\omega\right);-2\hat{\ell};\frac{\rho_{+}-\rho_{-}}{\rho_{+}-\rho}\right) \,,
		\ea \\
		\ba
			Z_{\omega\ell m j}^{\text{response}}\left(\rho\right) &= \left(1-\frac{\rho_{+}}{\rho}\right)^{-\hat{\ell}-1} \left(\frac{\rho-\rho_{-}}{\rho-\rho_{+}}\right)^{\mp i\sigma\tilde{Z}_{-}^{\left(\sigma\right)}\left(\omega\right)} \\
			&\times\,{}_2F_1\left(\hat{\ell}+1+i\Gamma^{\left(\sigma\right)}_{\mp\sigma}\left(\omega\right),\hat{\ell}+1-i\Gamma^{\left(\sigma\right)}_{\pm\sigma}\left(\omega\right);2\hat{\ell}+2;\frac{\rho_{+}-\rho_{-}}{\rho_{+}-\rho}\right) \,,
		\ea
	\end{gathered}
\ee
and the scalar response coefficients $k_{\ell m j}\left(\omega\right)$ are given explicitly in \eqref{eq:NZSLNs}.

The two radial functions $\rho^{\hat{\ell}}Z_{\omega\ell m j}^{\text{source}}$ and $\rho^{-\hat{\ell}-1}Z_{\omega\ell m j}^{\text{response}}$ are linearly independent solutions of the near-zone radial Klein-Gordon equation for generic orbital number $\ell$ and have the nice property that they transform into each other under the discrete symmetry transformation $\hat{\ell} \leftrightarrow -\hat{\ell}-1$ of the near-zone Klein-Gordon equation,
\be
	\rho^{\hat{\ell}}Z_{\omega\ell m j}^{\text{source}} \xleftrightarrow{\hat{\ell} \leftrightarrow -\hat{\ell}-1} \rho^{-\hat{\ell}-1}Z_{\omega\ell m j}^{\text{response}} \,.
\ee
%The linear independence can be directly seen from the Wronskian,
%\be
%	\mathscr{W}\left\{\rho^{\hat{\ell}}Z_{\ell m j}^{\text{source}},\rho^{-\hat{\ell}-1}Z_{\ell m j}^{\text{response}},\rho\right\} = -\frac{2\hat{\ell}+1}{\left(\rho-\rho_{+}\right)\left(\rho-\rho_{-}\right)}\,e^{2\pi\Gamma_{-}}
%\ee
%which never vanishes for generic $\hat{\ell}$. However, when $2\hat{\ell}$ is an integer, the above two solutions become degenerate and need to be treated more carefully. This is closely related to the source/response split ambiguity issue that gets lifted when we analytically continue the orbital number to range over the field of real numbers.
%\todo[inline]{Used equation 15.10.7 in NIST}
%
%We point out here that, instead of finding the solution regular at the black hole future event horizon and analytically continuing to large distances, we could instead use a generic linear combination of the source and response solutions above, with arbitrary response coefficients $k_{\ell m j}\left(\omega\right)$, which are eventually fixed by imposing the ingoing boundary condition at the event horizon.

We now consider the spinless limit of our solution and compare it with the already known results obtained in~\cite{Kol:2011vg} in the static limit. To do this, we need to extract the corresponding behavior of the parameters $Z_{+}\left(\omega\right)$ and $\tilde{Z}^{\left(\sigma\right)}_{-}\left(\omega\right)$, which are given in explicitly in \eqref{eq:NHbc} and \eqref{eq:NZZetaMinus}. First of all, as the spin parameters $a$ and $b$ are sent to zero, the inner and outer horizons have the asymptotic behaviors
\be
	\rho_{+} \xrightarrow{a,b\rightarrow0} \rho_{s}\left[1 + \mathcal{O}\left(a^2,b^2\right)\right] \,\,\,,\,\,\,\rho_{-} \xrightarrow{a,b\rightarrow0} \frac{a^2b^2}{\rho_{s}}\left[1+\mathcal{O}\left(a^2,b^2\right)\right] \,.
\ee
As such,
\be
	Z_{+}\left(\omega\right) \xrightarrow{a,b\rightarrow0} \frac{\omega r_{s}}{2} + \mathcal{O}\left(a,b\right) \,,\quad \tilde{Z}^{\left(\sigma\right)}_{-}\left(\omega\right) \xrightarrow{a,b\rightarrow0} \sigma\frac{\omega r_{s}}{2} + \mathcal{O}\left(a,b\right) \,.
\ee
We see, thus, that both $Z_{+}\left(\omega\right)$ and $\tilde{Z}^{\left(\sigma\right)}_{-}\left(\omega\right)$ become independent of the azimuthal numbers $m$ and $j$ in the spinless limit, as is appropriate for the spherically symmetric $5$-d Schwarzschild black hole background geometry. Furthermore, $\Gamma^{\left(\sigma\right)}_{\pm\sigma}\left(\omega\right)=\omega r_{s} \left(1\mp1\right)/2 + \mathcal{O}\left(a,b\right)$ and the near-zone solution in the spinless limit becomes
\be\ba
	R_{\omega\ell m j} &\xrightarrow{a,b\rightarrow0} \bar{\mathcal{E}}_{\ell m j}\left(\omega\right)\,\rho_{s}^{\hat{\ell}}\,\frac{\Gamma\left(\hat{\ell}+1\right)\Gamma\left(\hat{\ell}+1+i\omega r_{s}\right)}{\Gamma\left(2\hat{\ell}+1\right)\Gamma\left(1+i\omega r_{s}\right)} \\
	&\times \left(1-\frac{\rho_{s}}{\rho}\right)^{i\omega r_{s}/2}{}_2F_1\left(\hat{\ell}+1,-\hat{\ell};1+2i\omega r_{s};1-\frac{\rho}{\rho_{s}}\right) \,.
\ea\ee
Using the Pfaff transformation,
\be
	{}_2F_1\left(a,b;c;z\right) = \left(1-z\right)^{-b}{}_2F_1\left(c-a,b;c;\frac{z}{z-1}\right) \,,
\ee
which is valid for $\left|\text{Arg}\left(1-z\right)\right|<\pi$, this can be rewritten as
\be\ba
	R_{\omega\ell m j} &\xrightarrow{a,b\rightarrow0} \bar{\mathcal{E}}_{\ell m j}\left(\omega\right)\,\rho_{s}^{\hat{\ell}}\,\frac{\Gamma\left(\hat{\ell}+1\right)\Gamma\left(\hat{\ell}+1+i\omega r_{s}\right)}{\Gamma\left(2\hat{\ell}+1\right)\Gamma\left(1+i\omega r_{s}\right)} \\
	&\times \left(1-\frac{\rho_{s}}{\rho}\right)^{i\omega r_{s}/2}\left(\frac{\rho_{s}}{\rho}\right)^{-\hat{\ell}}{}_2F_1\left(-\hat{\ell},-\hat{\ell}+2i\omega r_{s};1+2i\omega r_{s};1-\frac{\rho_{s}}{\rho}\right) \,.
\ea\ee
In the $\omega\rightarrow0$ limit, this is in complete agreement, up to an overall conventional constant factor, with the corresponding result in equation (4.6) of~\cite{Kol:2011vg}. The scalar response coefficients \eqref{eq:NZSLNs} also have a smooth spinless limit that can be easily shown to agree with equation (4.7) of~\cite{Kol:2011vg} when $\omega=0$ after employing the Legendre duplication formula for the Gamma function,
\be
	\Gamma\left(z\right)\Gamma\left(z+\frac{1}{2}\right) = 2^{1-2z}\sqrt{\pi}\,\Gamma\left(2z\right) \,.
\ee
\section{Derivation of $\SL$ generators}
\label{sec:ApSL2RGenerators}

In Section \ref{sec:SL2R} we presented the existence of near-zone truncations of the massless Klein-Gordon equation in the background of the $5$-d Myers-Perry black hole equipped with an $\SL$ symmetry structure. In this appendix, we will sketch the derivation of the vector fields generating these symmetries. We will do this by starting with a generic ansatz for the form of the $\SL$ generators and require that the associated Casimir operators yield operators that preserve the characteristic exponents of the full equations of motion in the vicinity of the black hole event horizon. We will end up with an infinite number of $\SL$ algebras, most of which are not consistent near-zone truncations and also not globally defined.

The upshot of using this approach is that finding the most general truncation that preserves the near-horizon characteristic exponent also ensures that we will find all the possible near-zone truncations admitting $\SL$ symmetries as a subset. As we will see, there will be two towers of near-zone $\SL$ symmetries controlled by an arbitrary parameter which spontaneously breaks the $\SL$ symmetry down to $U\left(1\right)$. The only possible globally defined near-zone $\SL$ symmetries will then correspond to setting this symmetry breaking parameter to zero and will precisely correspond to the Love symmetries presented in Section \ref{sec:SL2R}. We will also investigate the situations where the $\SL$ symmetry of the truncations preserving the near-horizon characteristic exponents can be enhanced to full $2$-d conformal structure $\SL\times\SL$.

\subsection{Truncated radial operators preserving the near-horizon characteristic exponents}
The full massless Klein-Gordon operator in the background of the $5$-d Myers-Perry black hole has been introduced in Section \ref{sec:TLNsComputation} and the corresponding radial and angular operators are given in \eqref{eq:FullEOM}. We wish to explore the possibility of truncating the radial operator such that we preserve the characteristic exponents as we approach the event horizon at $\rho=\rho_{+}$. The most general such truncation has the form
\be
	\mathbb{O}_{\text{trunc}} =\partial_{\rho}\,\Delta\,\partial_{\rho} - \frac{\left(\rho_{+}-\rho_{-}\right)^2}{4\Delta}\beta^2\left(\partial_{t}+\Omega_{+}\,\partial_{+}+\Omega_{-}\,\partial_{-}\right)^2 + \delta\gamma^{\mu\nu}\partial_{\mu}\partial_{\nu} + \delta\gamma^{\mu}\,\partial_{\mu} \,,
\ee
where $\delta\gamma^{\mu\nu}$ and $\delta\gamma^{\mu}$ are terms that are subleading in the vicinity of the event horizon, $\Delta=\left(\rho-\rho_{+}\right)\left(\rho-\rho_{-}\right)$ is the discriminant function for the $5$-d Myers-Perry black hole and $\beta=\rho_{s}r_{+}/\left(\rho_{+}-\rho_{-}\right)$ is its inverse Hawking temperature. We note here that we are working in sum/difference azimuthal angles, $\psi_{\pm}=\psi\pm\phi$, with $\Omega_{\pm}=\Omega_{\psi}\pm\Omega_{\phi}$ the angular velocities along these two directions, and we are using the notation $\partial_{\pm}\equiv\partial_{\psi_{\pm}} = \left(\partial_{\psi}\pm\partial_{\phi}\right)/2$. The subleading terms $\delta\gamma^{\mu\nu}$ and $\delta\gamma^{\mu}$ that preserve the background $\mathbb{R}_{t}\times U\left(1\right)_{\phi}\times U\left(1\right)_{\psi}$ symmetries will then have the generic form
\be
	\begin{gathered}
		\delta\gamma^{tt} = f_{tt}\left(\rho\right) \,,\quad \delta\gamma^{t\psi_{\pm}} = \Omega_{\pm}f_{t\psi_{\pm}}\left(\rho\right) \,, \\
		\delta\gamma^{\psi_{\pm}\psi_{\pm}} = \Omega_{\pm}^2f_{\psi_{\pm}\psi_{\pm}}\left(\rho\right) \,,\quad \delta\gamma^{\psi_{+}\psi_{-}} = \Omega_{+}\Omega_{-}f_{\psi_{+}\psi_{-}}\left(\rho\right) \,, \\
		\delta\gamma^{t\rho} = \Delta f_{t\rho}\left(\rho\right) \,,\quad \delta\gamma^{\psi_{\pm}\rho} = \Omega_{\pm}\Delta f_{\psi_{\pm}\rho	}\left(\rho\right) \,,\quad \delta\gamma^{\rho\rho} = \Delta^2 f_{\rho\rho}\left(\rho\right) \,,
	\end{gathered}
\ee
\be
	\begin{gathered}
		\delta\gamma^{t} = f_{t}\left(\rho\right) \,,\quad \delta\gamma^{\psi_{\pm}} = \Omega_{\pm}f_{\psi_{\pm}}\left(\rho\right) \,,\quad \delta\gamma^{\rho} = \Delta f_{\rho}\left(\rho\right) \,,
	\end{gathered}
\ee
with all $f_{\mu\nu}$ and $f_{\mu}$ being radial functions that are regular as $\rho\rightarrow\rho_{+}$. We remark here that we allow for time-reversal violating terms. To simplify the problem, we can perform $\rho$-dependent coordinate transformations to eliminate as many as possible of the above terms. In particular, introducing coordinates $(\tilde{t},\tilde{\rho},\tilde{\psi}_{+},\tilde{\psi}_{-})$, related to $\left(t,\rho,\psi_{+},\psi_{-}\right)$ according to
\be
	\begin{gathered}
		\frac{d\tilde{\rho}}{\sqrt{\Delta\left(\tilde{\rho}\right)}} = \frac{d\rho}{\sqrt{\Delta\left(\rho\right)+\delta\gamma^{\rho\rho}\left(\rho\right)}} \,, \\
		d\tilde{t} = dt - \frac{\delta\gamma^{t\rho}\left(\rho\right)}{\Delta\left(\rho\right)+\delta\gamma^{\rho\rho}\left(\rho\right)}d\rho \,,\quad d\tilde{\psi}_{\pm} = d\psi_{\pm} - \frac{\delta\gamma^{\psi_{\pm}\rho}\left(\rho\right)}{\Delta\left(\rho\right)+\delta\gamma^{\rho\rho}\left(\rho\right)}d\rho \,,
	\end{gathered}
\ee
we can set
\be
	\delta\tilde{\gamma}^{\tilde{\rho}\tilde{\rho}} = 0 \,,\quad \delta\tilde{\gamma}^{\tilde{t}\tilde{\rho}} = 0 \,,\quad \delta\tilde{\gamma}^{\tilde{\psi}_{\pm}\tilde{\rho}} = 0 \,.
\ee
We will adopt these coordinates henceforth and drop the tildes to ease our notation.

\subsection{Solving the $\SL$ constraints}
With this setup for the generic truncation of the radial operator, we now explore the existence of three operators, $L_0$, $L_{+1}$ and $L_{-1}$, generating the $\SL$ algebra,
\be\label{eq:AlgebraConstraints}
	\left[L_{m},L_{n}\right] = \left(m-n\right)L_{m+n} \,,\quad m,n=0,\pm1 \,,
\ee
and whose Casimir precisely recovers a truncation of the radial operator of the type we just described,
\be\label{eq:CasimirConstraints}
	\mathcal{C}_2 = L_0^2 - \frac{1}{2}\left(L_{+1}L_{-1}+L_{-1}L_{+1}\right) \equiv \mathbb{O}_{\text{trunc}} \,.
\ee
Representations of $\SL$ are labeled by $\mathcal{C}_2$ and $L_0$. Using the $\mathbb{R}_{t}\times U\left(1\right)_{\phi}\times U\left(1\right)_{\psi}$ isometry of the background, we therefore make the following generic ansatz for the $\SL$ generators
\be
	\begin{gathered}
		L_0 = -\left(\beta_{t}\,\partial_{t} + \beta_{+}\Omega_{+}\,\partial_{+} + \beta_{-}\Omega_{-}\,\partial_{-}\right) \,, \\
		L_{\pm1} = \tilde{G}_{\pm}\left(x\right)\partial_{\rho} + \tilde{K}_{\pm}\left(x\right)\partial_{t} + \tilde{H}_{\pm}^{\left(+\right)}\left(x\right)\Omega_{+}\,\partial_{+} + \tilde{H}_{\pm}^{\left(-\right)}\left(x\right)\Omega_{-}\,\partial_{-} \,,
	\end{gathered}
\ee
where all components of the $L_0$ vector field are constants and all components of the $L_{\pm1}$ vector fields are spacetime functions $\tilde{X}_{\pm}\left(x\right)=\tilde{X}_{\pm}\left(t,\rho,\psi_{+},\psi_{-}\right)$. The algebra constraints \eqref{eq:AlgebraConstraints} and the Casimir constraints \eqref{eq:CasimirConstraints} will now be solved to fix the exponents $\beta_{\pm}$ and $\beta_{t}$, the functions $\tilde{X}_{\pm}\left(x\right)$ and the subleading terms $\delta\gamma^{\mu\nu}\left(\rho\right)$ and $\delta\gamma^{\mu}\left(\rho\right)$ appearing in the truncation of the radial operator.

We start with the algebra constraint $\left[L_{\pm1},L_0\right]=\pm L_{\pm1}$,
\be
	\left(\beta_{t}\,\partial_{t} + \beta_{+}\Omega_{+}\,\partial_{+} + \beta_{-}\Omega_{-}\,\partial_{-}\right)\tilde{X}_{\pm}\left(t,\rho,\psi_{+},\psi_{-}\right) = \pm \tilde{X}_{\pm}\left(t,\rho,\psi_{+},\psi_{-}\right) \,.
\ee
This can be used to eliminate the explicit $t$-dependence by introducing
\be
	\hat{\psi}_{\pm} = \psi_{\pm} - \frac{\beta_{\pm}}{\beta_{t}}\Omega_{\pm}t \quad \Rightarrow \quad \tilde{X}_{\pm}\left(t,\rho,\psi_{+},\psi_{-}\right) = e^{\pm t/\beta_{t}}X_{\pm}(\rho,\hat{\psi}_{+},\hat{\psi}_{-}) \,.
\ee
It is therefore favorable to work in the $(t,\rho,\hat{\psi}_{+},\hat{\psi}_{-})$ coordinates instead of $\left(t,\rho,\psi_{+},\psi_{-}\right)$ to solve the constraints. The generators in these coordinates are given by
\be
	\begin{gathered}
		L_0 = -\beta_{t}\,\partial_{t} \,, \\
		\ba
			L_{\pm1} &= e^{\pm t/\beta_{t}}\bigg[G_{\pm}(\rho,\hat{\psi}_{+},\hat{\psi}_{-})\partial_{\rho} + K_{\pm}(\rho,\hat{\psi}_{+},\hat{\psi}_{-})\partial_{t} \\
			&+ \hat{H}_{\pm}^{\left(+\right)}(\rho,\hat{\psi}_{+},\hat{\psi}_{-})\Omega_{+}\,\hat{\partial}_{+} + \hat{H}_{\pm}^{\left(-\right)}(\rho,\hat{\psi}_{+},\hat{\psi}_{-})\Omega_{-}\,\hat{\partial}_{-} \bigg] \,,
		\ea
	\end{gathered}
\ee
with $\hat{H}_{\pm}^{\left(i\right)} = H_{\pm}^{\left(i\right)} - \frac{\beta_{i}}{\beta_{t}}K_{\pm}$.

Moving forward, the Casimir constraints \eqref{eq:CasimirConstraints} imply that products of components of the $L_{\pm1}$ vector fields are independent of the azimuthal angles. We, thus, write $X_{\pm}(\rho,\hat{\psi}_{+},\hat{\psi}_{-}) = e^{\pm A(\hat{\psi}_{+},\hat{\psi}_{-})}\chi_{\pm}\left(\rho\right)$. In fact, taking derivatives with respect to $\hat{\psi}_{\pm}$ of the various Casimir and algebra constraints reveals that the dependence on the azimuthal angles in the exponential must be linear,
\be
	X_{\pm}(\rho,\hat{\psi}_{+},\hat{\psi}_{-}) = e^{\pm (\tau_{+}\hat{\psi}_{+} + \tau_{-}\hat{\psi}_{-})}\chi_{\pm}\left(\rho\right) \,.
\ee
The $\rho\rho$-component of the Casimir constraints \eqref{eq:CasimirConstraints} and the $\rho$-component of the last algebra constraint $\left[L_{+1},L_{-1}\right]=2L_0$ then completely fix the $\rho$-components of $L_{\pm1}$ to be, up to automorphisms,
\be
	g_{\pm}\left(\rho\right) = \mp\sqrt{\Delta} \,.
\ee
The $t\rho$- and $\psi_{\pm}\rho$-components of the Casimir constraints then imply
\be
	k_{+}\left(\rho\right) = k_{-}\left(\rho\right) \equiv k\left(\rho\right) \,,\quad \hat{h}_{+}^{\left(i\right)}\left(\rho\right) = \hat{h}_{-}^{\left(i\right)}\left(\rho\right) \equiv \hat{h}^{\left(i\right)}\left(\rho\right) \,.
\ee
The remaining algebra constraints from $\left[L_{+1},L_{-1}\right]=2L_0$ become
\begin{subequations}
	\be\label{eq:AC1}
		\sqrt{\Delta}\,k^{\prime} + \left(\frac{k}{\beta_{t}}+\sum_{i=+,-}\tau_{i}\Omega_{i}\hat{h}^{\left(i\right)}\right)k = \beta_{t} \,,
	\ee
	\be\label{eq:AC2}
		\sqrt{\Delta}\,\hat{h}^{\left(i\right)\prime} + \left(k+\sum_{j=+,-}\tau_{j}\Omega_{j}\hat{h}^{\left(j\right)}\right)\hat{h}^{\left(i\right)} = 0 \,,
	\ee
\end{subequations}
while the remaining Casimir constraints read
\begin{subequations}
	\begin{align}
		\label{eq:CC1} k^2 -1 &= \frac{\left(\rho_{+}-\rho_{-}\right)^2}{4\Delta}\beta^2 - f_{tt}\left(\rho\right) \,, \\
		\label{eq:CC2} k\hat{h}^{\left(\pm\right)} &= \frac{\left(\rho_{+}-\rho_{-}\right)^2}{4\Delta}\beta^2\left(1-\frac{\beta_{\pm}}{\beta_{t}}\right) - \left(f_{t\psi_{\pm}} - \frac{\beta_{\pm}}{\beta_{t}}f_{tt}\right) \,, \\
		\label{eq:CC3} \left[\hat{h}^{\left(\pm\right)}\right]^2 &= \frac{\left(\rho_{+}-\rho_{-}\right)^2}{4\Delta}\beta^2\left(1-\frac{\beta_{\pm}}{\beta_{t}}\right)^2 - \left(f_{\psi_{\pm}\psi_{\pm}} - 2\frac{\beta_{\pm}}{\beta_{t}}f_{t\psi_{\pm}} + \frac{\beta_{\pm}^2}{\beta_{t}^2}f_{tt}\right) \,, \\
		\begin{split}\label{eq:CC4}
			\hat{h}^{\left(+\right)}\hat{h}^{\left(-\right)} &= \frac{\left(\rho_{+}-\rho_{-}\right)^2}{4\Delta}\beta^2\left(1-\frac{\beta_{+}}{\beta_{t}}\right)\left(1-\frac{\beta_{-}}{\beta_{t}}\right) \\
			&\quad - \left(f_{\psi_{+}\psi_{-}} - \frac{\beta_{+}}{\beta_{t}}f_{t\psi_{+}}-\frac{\beta_{-}}{\beta_{t}}f_{t\psi_{-}} + \frac{\beta_{+}\beta_{-}}{\beta_{t}^2}f_{tt}\right) \,.
		\end{split}
	\end{align}
\end{subequations}

Let us sketch how to solve these. The algebra constraints can be solved for the functions $k\left(\rho\right)$ and $\hat{h}^{\left(\pm\right)}\left(\rho\right)$. The integration constants associated with the differential equations \eqref{eq:AC1}-\eqref{eq:AC2} are then fixed by the near-horizon behaviors of these functions as dictated by the Casimir constraints \eqref{eq:CC1}-\eqref{eq:CC4}. These near-horizon behaviors also result in a relation between the constants $\beta_{t}$, $\beta_{\pm}$ and $\tau_{\pm}$,
\be\label{eq:SL2Rbetas}
	\beta_{t}\left(1-\beta\sum_{i=+,-}\tau_{i}\Omega_{i}\right) = \beta\left(1-\sum_{i=+,-}\tau_{i}\beta_{i}\Omega_{i}\right) \,.
\ee
The final expressions of the generators $L_0$, $L_{\pm1}$ after translating back into $\left(t,\rho,\psi_{+},\psi_{-}\right)$ coordinates are
\be\label{eq:SL2RGenericGenerators}
	\begin{gathered}
		L_0 = -\left(\beta_{t}\,\partial_{t} + \beta_{+}\Omega_{+}\,\partial_{+} + \beta_{-}\Omega_{-}\,\partial_{-}\right) \,, \\
		\ba
			L_{\pm1} &= \exp\left\{\pm\left[t/\beta + \tau_{+}\left(\psi_{+}-\Omega_{+}t\right) + \tau_{-}\left(\psi_{-}-\Omega_{-}t\right)\right]\right\} \\
			&\quad\times\left[\mp\sqrt{\Delta}\,\partial_{\rho} - \partial_{\rho}\left(\sqrt{\Delta}\right)L_0 + \frac{\rho_{+}-\rho_{-}}{2\sqrt{\Delta}}\left(\beta K+L_0\right) \right] \,,
		\ea
	\end{gathered}
\ee
with the associated Casimir given by
\be\label{eq:SL2RGenericCasimir}
	\mathcal{C}_2 = \partial_{\rho}\,\Delta\,\partial_{\rho} - \frac{\left(\rho_{+}-\rho_{-}\right)^2}{4\Delta}\beta^2K^2 + \frac{\rho_{+}-\rho_{-}}{\rho-\rho_{-}}L_0\left(L_0+\beta K\right) \,,
\ee
where $K=\partial_{t} + \Omega_{+}\,\partial_{+} + \Omega_{-}\,\partial_{-}$ is the Killing vector relative to which the event horizon is a Killing horizon. Interestingly, by working in the advanced (retarded) null coordinates \eqref{eq:NullCoordinates}, one can then check that the above vectors fields are automatically regular at the future (past) event horizon. We also remind here that one has the freedom of performing the coordinate transformations
\be
	\rho\rightarrow\rho+\Delta^2g_{\rho}\left(\rho\right) \,,\quad t\rightarrow t+g_{t}\left(\rho\right) \,,\quad \psi_{\pm}\rightarrow\psi_{\pm}+g_{\psi_{\pm}}\left(\rho\right) \,,
\ee
for arbitrary radial functions $g_{\rho}\left(\rho\right)$, $g_{t}\left(\rho\right)$, $g_{\psi_{+}}\left(\rho\right)$ and $g_{\psi_{-}}\left(\rho\right)$ that are regular at the event horizon; these give rise to non-zero subleading contributions $\delta\gamma^{\rho\rho}$, $\delta\gamma^{t\rho}$ and $\delta\gamma^{\psi_{\pm}\rho}$ respectively.

Let us now start discussing the properties of these truncations of the radial operators that are equipped with $\SL$ symmetries as demonstrated above. First of all, we notice that the $\SL$ symmetry generated by \eqref{eq:SL2RGenericGenerators} is in general spontaneously broken down to $U\left(1\right)$, generated by $L_0$, from the periodic identification of the azimuthal coordinates when $\tau_{\pm}\ne0$. An interesting remark here is that the $\tau_{\pm}\ne0$ generators can be obtained from the globally defined ones, with $\tau_{\pm}=0$, via temporal translations involving the co-rotating azimuthal angles,
\be
	L_{m}^{\left(\tau_{\pm}=0\right)} \xrightarrow{t\rightarrow t + \beta\sum_{i=+,-}\tau_{i}\left(\psi_{i}-\Omega_{i}t\right)} L_{m}^{\left(\tau_{\pm}\ne0\right)} \,.
\ee

Furthermore, the Casimir operator \eqref{eq:SL2RGenericCasimir} has the property of preserving the characteristic exponents near the event horizon by construction. However, it does not in general capture any near-zone truncation of the radial operator. For this to happen, the Casimir operator must exactly reproduce all the static terms in the radial operator as well. This additional requirement gives two infinite towers of near-zone $\SL$'s controlled by the parameters $\tau_{\pm}$. The two towers can be labeled by a sign $\sigma=+,-$ and correspond to fixing the parameters $\beta_{\pm}$ to
\be
	\beta_{\sigma} = \beta \,,\quad \beta_{-\sigma} = 0 \,\quad \text{for $\sigma=+$ OR $\sigma=-$} \,.
\ee
These near-zone $\SL$'s are generated by the vector fields
\be\label{eq:SL2RNZGenerators}
	\begin{gathered}
		L_0^{\left(\sigma\right)} = -\left(\beta_{t}^{\left(\sigma\right)}\,\partial_{t} + \beta\Omega_{\sigma}\,\partial_{\sigma}\right) \,, \\
		\ba
			L_{\pm1}^{\left(\sigma\right)} &= \exp\left\{\pm\left[t/\beta + \tau_{+}\left(\psi_{+}-\Omega_{+}t\right)+\tau_{-}\left(\psi_{-}-\Omega_{-}t\right)\right]\right\} \\
			&\quad\times\bigg[\mp\sqrt{\Delta}\,\partial_{\rho} + \partial_{\rho}\left(\sqrt{\Delta}\right)\left(\beta_{t}^{\left(\sigma\right)}\,\partial_{t} + \beta\Omega_{\sigma}\,\partial_{\sigma}\right) \\
			&\quad\quad + \frac{\rho_{+}-\rho_{-}}{2\sqrt{\Delta}}\left[\left(\beta-\beta_{t}^{\left(\sigma\right)}\right)\partial_{t}+ \beta\Omega_{-\sigma}\,\partial_{-\sigma}\right] \bigg] \,,
		\ea
	\end{gathered}
\ee
and the associated Casimir operator is given by
\be\ba\label{eq:SL2RNZCasimir}
	\mathcal{C}_2^{\left(\sigma\right)} &= \partial_{\rho}\,\Delta\,\partial_{\rho} - \frac{\left(\rho_{+}-\rho_{-}\right)^2}{4\Delta}\beta^2\left(\partial_{t}+\Omega_{+}\,\partial_{+}+\Omega_{-}\,\partial_{-}\right)^2 \\
	&\quad- \frac{\rho_{+}-\rho_{-}}{\rho-\rho_{-}}\left(\beta_{t}^{\left(\sigma\right)}\,\partial_{t} + \beta\Omega_{\sigma}\,\partial_{\sigma}\right)\left[\left(\beta-\beta_{t}^{\left(\sigma\right)}\right)\partial_{t}+\beta\Omega_{-\sigma}\partial_{-\sigma}\right] \,,
\ea\ee
where $\beta_{t}^{\left(\sigma\right)}$, $\tau_{+}^{\left(\sigma\right)}$ and $\tau_{-}^{\left(\sigma\right)}$ are related according to
\be
	\beta_{t}^{\left(\sigma\right)}\left(1-\beta\tau_{+}^{\left(\sigma\right)}\Omega_{+}-\beta\tau_{-}^{\left(\sigma\right)}\Omega_{-}\right) = \beta\left(1-\beta\tau_{\sigma}^{\left(\sigma\right)}\Omega_{\sigma}\right) \,.
\ee
Supplemented with the near-zone-approximation-preserving temporal translations $t\rightarrow t + g_{t}\left(\rho\right)$, this exhausts all the possible near-zone $\SL$ symmetries.

If we want these near-zone $\SL$ symmetries to be globally defined, one must further impose $\tau_{\pm}=0$, in which case we must have $\beta_{t}^{\left(\sigma\right)}=\beta$ and we end up with the fact that the most general globally defined near-zone $\SL$ symmetries are, up to $\rho$-dependent temporal translations, the Love symmetries presented in Section \ref{sec:SL2R}.

\subsection{Extension to $\SL\times\SL$}
We will finish with a short investigation on the possibility of extending the above-found $\SL$ symmetries, for which the radial operator truncations preserve the characteristic exponents near the event horizon, into the full $2$-d global conformal structure $\SL\times\SL$.

Consider, therefore, two general such $\SL$ symmetries generated by vector fields $L_{m}$ and $\bar{L}_{m}$ of the form \eqref{eq:SL2RGenericGenerators}. They are characterized by parameters $\{\beta_{t},\beta_{\pm},\tau_{\pm}\}$ and $\{\bar{\beta}_{t},\bar{\beta}_{\pm},\bar{\tau}_{\pm}\}$ respectively, with each set of parameters being subject to the relation \eqref{eq:SL2Rbetas}. By working out the requirement that
\be
	\left[L_{m},\bar{L}_{n}\right] = 0 \,,\quad m,n=0,\pm1 \,,
\ee
we extract the following summarizing condition
\be
	L_0 + \bar{L}_0 = -\beta K \,,
\ee
that is, $\bar{\beta}_{t} = \beta - \beta_{t}$ and $\bar{\beta}_{\pm} = \beta - \beta_{\pm}$. The associated Casimirs turn out to be exactly the same and can be written in the suggestive form
\be
	\mathcal{C}_2 = \bar{\mathcal{C}}_2 = \partial_{\rho}\,\Delta\,\partial_{\rho} - \frac{\rho_{+}-\rho_{-}}{\rho-\rho_{+}}\left(\frac{L_0+\bar{L}_0}{2}\right)^2 + \frac{\rho_{+}-\rho_{-}}{\rho-\rho_{-}}\left(\frac{L_0-\bar{L}_0}{2}\right)^2 \,.
\ee
In a $\text{CFT}_2$ interpretation, this shows that the characteristic exponents near the outer and inner horizons are the (squares of half of the) scaling dimension and spin-weight of the spacetime scalar field under the action of the $\text{CFT}_2$ dilaton and Lorentz generators respectively. In this language, the above truncations of the radial operator seek to preserve the scaling dimension but allow to approximate the $\text{CFT}_2$ spin-weight. The remaining generators can similarly be written as
\be\ba
	L_{\pm1} &= e^{\pm\left[t/\beta + \tau_{+}\left(\psi_{+}-\Omega_{+}t\right) + \tau_{-}\left(\psi_{-}-\Omega_{-}t\right)\right]} \\
	&\quad\times\left[\mp\sqrt{\Delta}\,\partial_{\rho} - \sqrt{\frac{\rho-\rho_{-}}{\rho-\rho_{+}}}\frac{L_0+\bar{L}_0}{2} - \sqrt{\frac{\rho-\rho_{+}}{\rho-\rho_{-}}}\frac{L_0-\bar{L}_0}{2}\right] \,, \\
	\bar{L}_{\pm1} &= e^{\pm\left[t/\beta + \bar{\tau}_{+}\left(\psi_{+}-\Omega_{+}t\right) + \bar{\tau}_{-}\left(\psi_{-}-\Omega_{-}t\right)\right]} \\
	&\quad\times\left[\mp\sqrt{\Delta}\,\partial_{\rho} - \sqrt{\frac{\rho-\rho_{-}}{\rho-\rho_{+}}}\frac{L_0+\bar{L}_0}{2} + \sqrt{\frac{\rho-\rho_{+}}{\rho-\rho_{-}}}\frac{L_0-\bar{L}_0}{2}\right] \,.
\ea\ee

Last, for the case of near-zone $\SL\times\SL$'s, there are two towers of such enhancements labeled by a sign $\sigma=+,-$. They correspond to $\left(\beta_{\sigma},\beta_{-\sigma}\right)=\left(\beta,0\right)$ and, thus, $\left(\bar{\beta}_{\sigma},\bar{\beta}_{-\sigma}\right)=\left(0,\beta\right)$. One of the outcomes of the current analysis is then that near-zone $\SL\times\SL$'s can \textit{never} be globally defined. The best one can do is to have near-zone $\SL\times\SL$ symmetries spontaneously broken down to $\SL\times U\left(1\right)$ from the periodic identification of the azimuthal angles. These are precisely the enhancements of the Love symmetries presented in Section \ref{sec:Properties}. We remark here that the construction of near-zone $\SL\times\SL$ symmetries described above also contains the Kerr/CFT proposal for $5$-dimensional rotating black holes in~\cite{Krishnan:2010pv} as a special case, corresponding to a different near-zone truncation with $\beta_{t}=\beta\frac{r_{+}-r_{-}}{2r_{+}}$ and $\bar{\beta}_{t}=\beta\frac{r_{+}+r_{-}}{2r_{+}}$ and which has the unique property of preserving the characteristic exponent at the inner horizon as well.

\addcontentsline{toc}{section}{References}
\bibliographystyle{JHEP}
\bibliography{TLNsAll_References_Inspirehep}

\end{document}